\documentclass[twocolumn,superscriptaddress,floatfix,prb, aps]{revtex4-1}
\usepackage{graphicx,amsfonts,amssymb,amsmath,hyperref,hypcap,enumerate}
\usepackage{xcolor}
\usepackage{soul}
\usepackage{scalerel}
\usepackage{adjustbox,stmaryrd}

\usepackage{cancel}

\hypersetup{  colorlinks=true, linkcolor=blue, citecolor=red, urlcolor=blue  }

\newcommand{\RNum}[1]{\uppercase\expandafter{\romannumeral #1\relax}}
\newcommand{\sub}[1]{\scaleto{\text{#1}}{4pt}}
\newcommand{\mb}[1]{\mathbf{#1}}
\newcommand{\bs}[1]{\boldsymbol{#1}}

\begin{document}

\title{Theory of ARPES in Graphene-Based Moir\'e Superlattices}
\author{Jihang Zhu}
\affiliation{Department of Physics, University of Texas at Austin, Austin TX 78712}
\author{Jingtian Shi} 
\affiliation{Department of Physics, University of Texas at Austin, Austin TX 78712} 
\author{Allan H. MacDonald}
\affiliation{Department of Physics, University of Texas at Austin, Austin TX 78712}


\begin{abstract}
Graphene-based moir\'e superlattices are now established as an interesting platform for strongly-correlated many-electron physics, and have
so far been characterized mainly by transport and scanning tunneling microscopy (STM)  measurements.
Motivated by recent experimental progress, 
we present a theoretical model study whose aim is to assess the potential of angle-resolved photoemission 
spectroscopy (ARPES) to resolve some of the many open issues in these systems.  The theory is developed specifically
for graphene on hexagonal boron nitride (G/hBN) and twisted bilayer graphene (TBG) moir\'e superlattices, but is readily generalized to any system with active degrees of freedom
in graphene sheets.  
\end{abstract}

\maketitle

\section{Introduction}
A large body of theoretical and experimental work\cite{graphene_Novoselov_2004, graphene_Geim_2007} over the past decade 
has achieved a thorough understanding of most single-layer and few-layer graphene film properties.  
Progress in this field has been aided by success in reducing disorder effects to very low levels and
by the identification of hexagonal boron nitride (hBN),\cite{hBN_sub_Dean_2010, hBN_sub_Dean_2012, hBN_sub_Xue_2011} with its large band gaps and atomically smooth surfaces, 
as the substrate of choice.  
The recent discovery of superconducting, correlated insulating and orbital magnetic states in magic-angle\cite{tBLG_Bistritzer_2011}
twisted bilayer graphene (MATBG) \cite{tBLG_SC_Cao_2018, tBLG_Insu_Cao_2018, tBLG_Lu_2019, tBLG_Yankowitz_2019, tBLG_Sharpe_2019, tBLG_Serlin_2020} 
has now added strongly-correlated-electron behavior to the physics that can be 
explored in graphene multi-layers.
MATBG's strong-correlation physics is a consequence of unusual flat-band behavior 
near a discrete set of magic twist angles.\cite{tBLG_Bistritzer_2011}  The flat bands emerge from 
interference between intralayer and interlayer hopping processes that are 
individually strong.  The residual dispersion in these bands is important for understanding 
physical properties but, because it results from a delicate cancellation, 
is difficult to predict reliably on the basis of theoretical considerations alone.
The difficulty of quantitative theoretical modeling is 
heightened by the large number of carbon atoms ($\sim 10^4$) per 
superlattice unit cell, by the important role of interactions in reshaping the
moir\'e superlattice bands,\cite{tBLG_Guinea_2018, tBLG_MingHF, tBLG_Ming_weakfield} by the critical importance of non-local 
exchange interactions,\cite{tBLG_MingHF} and by a tendency toward spin and/or valley flavor symmetry breaking\cite{tBLG_MingHF, tBLG_Ming_weakfield, tBLG_Po_2018, tBLG_Bultinck_2019, tBLG_You_2019}
that is still incompletely understood. Because ARPES directly probes the momentum-dependence of the 
one-particle electronic Green's function, it is uniquely positioned to guide
progress toward a quantitative understanding of MATBG properties.  

ARPES has become an indispensable tool for studies of strongly interacting\cite{ARPES_Damascelli_2003, ARPES_Vishik_2010, ARPES_Lu_2012}
and topological materials,\cite{ARPES_3DTI_Xia_2009} and has been applied successfully to single-layer and multilayer epitaxial graphene samples formed on the surface of silicon carbide.\cite{ARPES_BLG_Ohta_2006, ARPES_Ohta_2007, ARPES_Sprinkle_2009, ARPES_Bostwick_2007, Bostwick_2007, ARPES_Zhou_2007, ARPES_Bostwick_2006, ARPES_Coletti_2013, ARPES_G_Hwang_2011, ARPES_G_Gierz_2011, ARPES_TBGonSiC_Ohta, ARPES_BLG_Kim} The typical photon beam spot size of conventional ARPES experiments is $\sim 25-100 \mu$m,\cite{ARPES_2017} larger than or roughly equal to the
$\sim 1-100\mu$m size of typical MATBG samples prepared by mechanical exfoliation of two-dimensional (2D) crystals.
Applying the power of ARPES to MATBG physics
requires either access to the nano length scale in ARPES, or larger moir\'e samples. 
Recent progress in nano-ARPES\cite{nanoARPES_Dudin_2010, nanoARPES_Bostwick_2012, nanoARPES_Avila_2013, nanoARPES_Avila_2013_2, nanoARPES_Avila_2013_3, nanoARPES_Avila_2014, nanoARPES_Coy_2015} may provide the necessary opening and has been implemented to mechanically exfoliated van der Waals heterostructures.\cite{nanoARPES_BLG_Joucken, nanoARPES_G_Chen, nanoARPES_WS2_Katoch, nanoARPES_Joucken, nanoARPES_Nguyen, nanoARPES_Wang, nanoARPES_tBLG_Simone} 
Preliminary applications of nano-ARPES to G/hBN\cite{nanoARPES_Wang} and TBG moir\'e superlattices\cite{nanoARPES_tBLG_Simone, nanoARPES_tBLG_Utama, nanoARPES_Razado, ARPES_tBLG_Ohta} have been reported recently.

The ARPES spectra of graphene moir\'e systems have been studied previously using both 
tight-binding model \cite{ARPES_theo_Amorim, ARPES_theo_tTL_Amorim} and 
continuum model approaches.\cite{ARPES_theo_Anshuman,ARPES_theo_GhBN_Falko} 
In this paper we use an accurate continuum model to compute theoretical ARPES spectra
of both G/hBN and MATBG with the goal of informing the interpretation of future ARPES experiments, either nano-ARPES studies of MATBG samples similar to those that are currently available or conventional ARPES studies of large area MATBG samples which could become 
available in the future.
We find that key parameters of low-energy effective models, like the size of mass term that 
expresses broken inversion symmetry in G/hBN and the G/G interlayer intra-sublattice and inter-sublattice tunneling parameters, can be inferred from 
ARPES momentum distributions.  Although a complete treatment of the role of interactions 
lies out of the scope of this paper, we do comment on the ability of ARPES to measure flat-band 
shape renormalization by electron-electron interactions, and the broken spin and/or valley flavor symmetries 
thought to occur at fractional flat band filling.

This paper is organized as follows. In section \ref{section2} we discuss the general theory of ARPES 
in graphene-based moir\'e superlattices described by $\mb{k} \cdot \mb{p}$ continuum models.  
In sections \ref{section3} and \ref{section4} we focus on two prototypical moir\'e superlattice systems, G/hBN 
in which ARPES can be used to determine the important inversion symmetry breaking mass parameter, and TBG in which 
ARPES can characterize strain relaxation within the moir\'e pattern and identify when the magic angle is reached.
In the latter case, important parameters can be identified by performing measurements of momentum space distributions at energies well away from the 
flat bands that do not require extremely precise energy resolution.  
In section \ref{section5}, we discuss ARPES momentum distributions at van Hove singularity (VHS) energies in TBG, which can be revealed in both large and small twist angle 
regimes.  Finally in Section~\ref{summary} we conclude with a general discussion of some of the issues that could 
be clarified if accurate ARPES measurements become a possibility.  

\section{ARPES in graphene-based moir\'e superlattices}\label{section2}

The ARPES intensity $ I(\mb{p},E)$ is proportional to the transition probability from a Bloch 
initial state with crystal momentum $\mb{k}$ and energy $E$ 
to a photoelectron final state with momentum $\mb{p}$ and kinetic energy $E_{\text{kin}}$. Energy conservation guarantees $E_{\text{kin}} = \hbar \omega + E - \phi$, where $\hbar \omega$ is the photon energy and $\phi$ is the work function. The initial state energy $E$ is relative to the Fermi energy. In non-interacting 
electron models the ARPES spectrum of a 2D solid is non-zero only if one of the occupied band states at momentum $\mb{k}$,
where $\mb{k}$ is the in-plane projection $\mb{p}_{\parallel}$ reduced to the 2D Brillouin zone (BZ), has energy $E$.
The intensity of the peak produced by an occupied band state at a given  
extended zone momentum replica depends on the Bloch state wavefunction.
This dependence is particularly simple when all the states of interest 
are linear combinations of carbon $\pi$-orbitals on different lattice sites,
as we now explain. 

The moir\'e superlattice period of G/hBN multilayers depends on both the lattice constant mismatch and twist angle between the graphene and hBN layers, whereas the moir\'e superlattice period of 
TBG depends only on twist angle. In both G/hBN and TBG cases we will assume near perfect alignment so that the moir\'e modulation has a long wavelength.
Since we are interested in electronic states at energies near the Dirac point we can use 
$\mb{k} \cdot \mb{p}$ continuum models\cite{tBLG_Bistritzer_2011} in which $\pi$-orbital 
envelope function spinors satisfy effective Schrodinger equations.  The number of components of the 
envelope function spinors is two (for the two honeycomb sublattices) times the number of active graphene
layers in the moir\'e heterojunction. 
At low energies the correction to the Dirac Hamiltonians of isolated graphene layers can be approximated by a 
sublattice and position-dependent terms that have the periodicity of the moir\'e pattern.
For example, these have been detailed for the G/hBN and TBG cases discussed below 
in Refs.\onlinecite{tBLG_Bistritzer_2011, GhBN_Jeil_2014, GhBN_Jeil_2017, GhBN_Jeil_2015}.
In the TBG case, the moir\'e superlattice is defined mainly by
the spatial pattern of interlayer tunneling, whereas in the G/hBN case the 
moir\'e superlattice is defined by the spatial pattern of sublattice-dependent 
energies and inter-sublattice tunneling.  

Specializing to the case in which a single graphene layer is active, the initial electronic states prior to photoemission are moir\'e band eigenstates $|\xi, n,\mb{k} \rangle$,
two-component sublattice spinors that have a Bloch state plane-wave expansion: 
\begin{equation}
    |\xi,n,\mb{k} \rangle = \sum\limits_{\alpha, \mb{g}} \psi^{\xi}_{n\alpha \mb{g}}(\mb{k}) |\mb{k} + \mb{g},\alpha \rangle.
\end{equation}
Here $\xi=\pm$ is a valley index, $n$ is a band index, $\mb{g}$ is a moir\'e reciprocal lattice vector,
$|\mb{k}, \alpha \rangle$ is a graphene $\pi$-orbital state with definite sublattice $\alpha=\text{A},\text{B}$ and momentum $\mb{k}$.
In the calculations below we cut-off the momentum expansion at
$\mb{g} \in \{\mb{0}, \mb{g}_1, \dots, \mb{g}_6\}$ for G/hBN, where 
$ \mb{g}_1, \dots, \mb{g}_6$ are the six first-shell moir\'e reciprocal lattice vectors. 
For TBG case, we include three shells of moir\'e reciprocal lattice vectors, i.e. $|\mb{g}_{\text{max}}|=3g$ where $g$ is the length of the primitive reciprocal lattice vector.

When multilayer graphene is probed using high-energy photon beams, in the soft x-ray regime for example, the photoemission final state is well approximated as free-electron \cite{final_state_Strocov} and
photoelectron scattering and diffraction effects can be ignored.
This approximation is justified because i) the crystal potential is relatively small compared
to the photoelectron's kinetic energy,\cite{Damascelli_ARPES_2004} ii)
the scattering cross section is small for light atoms\cite{Puschnig_2009} and 
iii) $\pi$-orbitals in graphene form delocalized itinerant band states.\cite{Medjanik_2017} Indeed, the free-electron final state approximation has worked very well in previous studies.\cite{ARPES_theo_Amorim, ARPES_theo_Anshuman, ARPES_Puschnig, ARPES_Shirley} Note that the photon energy should be high but not too high, because high-photon-energy decreases the energy resolution and momentum resolution. The neglected final state effects\cite{ARPES_G_Gierz_2011,final_state_Ayria, final_state_Barrett, final_state_Strocov_2006} can be important at low photon energies ($\lesssim 50$ eV), but are out of the scope of this paper.

By generalizing the established theory\cite{ARPES_Eli2008} of monolayer graphene sheet ARPES intensity
summarized in Appendix \ref{appendixA}, where matrix element effects that are dependent on experimental geometry are ignored and a free-electron final state is assumed, we obtain the following expression for the 
dependence of the ARPES signal on the initial Bloch state energy $E$ and photoelectron momentum $\mb{p}$: 
\begin{equation}
\begin{split}
\label{eq:arpesgeneral}
I(\mb{p},E) 
\propto& \sum\limits_{\xi,n,\mb{k}} \big|\langle \mb{p}|\xi, n,\mb{k} \rangle \big|^2 \delta(E-\varepsilon^\xi_{n\mb{k}}) \\
\propto& 
\big| \phi(\mb{p})\big|^2  \sum\limits_{\xi,n,\mb{k}}
\delta(E-\varepsilon^\xi_{n\mb{k}}) \\
& \Big| \sum\limits_{\alpha,\mb{g}} \psi^\xi_{n\alpha \mb{g}}(\mb{k}) 
e^{-i\mb{G} \cdot \bs\tau_\alpha}
\delta_{\mb{p}_{\parallel},\mb{k}+\mb{g}+\mb{G}}
\Big|^2,
\end{split}
\end{equation}
where $\phi(\mb{p}) = \int d^3 \mb{r} e^{-i\mb{p} \cdot \mb{r}} \phi(\mb{r})$ is the Fourier transform of the atomic $\pi$-orbital and $\mb{G}$ is a reciprocal lattice vector of an isolated graphene layer.
Equation~(\ref{eq:arpesgeneral}) ignores a factor related to photon polarization.
A given photoelectron momentum $\mb{p}$ picks a specific $\mb{G}$, 
valley wavevector $\mb{K}_{\xi}$, and moir\'e reciprocal lattice vector $\mb{g}$ to 
map $\mb{k}=\mb{p}_{\parallel}-\mb{g}-\mb{G}$ into the moir\'e Brillouin zone (MBZ). 
Below we assume that $\mb{p}_{\parallel}$ is near the 
$\mb{K}_{+}=(4\pi/3a,0)$, where $a$ is graphene's lattice constant, Eq.~(\ref{eq:arpesgeneral})
simplifies to 
\begin{equation}
    I(\mb{p},E) \propto \big| \phi(\mb{p}) \big|^2  \sum\limits_{n,\mb{k}}
    \Big| \sum\limits_{\alpha,\mb{g}} \psi_{n\alpha \mb{g}}^+(\mb{k}) \delta_{\mb{p}_{\parallel}, \mb{k} + \mb{g}} \Big|^2 \delta(E-\varepsilon_{n\mb{k}}^+). 
\label{Eq_GhBN_G0}
\end{equation}

Photon polarization effects\cite{ARPES_G_Hwang_2011, Adotv_Ismail} add a momentum-dependent weighting factor 
and can alter momentum distribution 
function anisotropy.  When they are taken into account using the dipole approximation, 
as summarized in Appendix \ref{appendixA}, we obtain 
\begin{equation}\label{Eq_I_Adotv}
\begin{split}
    I(\mb{p},E) &\propto \sum\limits_{\xi,n,\mb{k}} \big| \langle \mb{p}| \mb{A} \cdot \hat{\mb{v}}|\xi,n,\mb{k} \rangle \big|^2 \delta(E-\varepsilon^\xi_{n\mb{k}}) \\
    &= \mb{A} \cdot \sum\limits_{\xi,n,\mb{k}} \big| \langle \mb{p}| \bs{\nabla}_\mb{k} H |\xi,n,\mb{k} \rangle \big|^2 \delta(E-\varepsilon^\xi_{n\mb{k}}),
\end{split}
\end{equation}
where the free-electron final state is projected to the Bloch basis
\begin{equation}
    |\mb{p}\rangle = \sum\limits_{\mb{k},\mb{g},\alpha} \delta_{\mb{p}_{\parallel},\mb{k}+\mb{g}+\mb{G}} e^{i\mb{G} \cdot \bs{\tau}_\alpha} \phi^*(\mb{p}) |\mb{k}+\mb{g},\alpha \rangle.
\end{equation}

In multilayer systems, the out-of-plane momentum component $p_z$ of the photoelectron 
controls interlayer interferences, which is absent in 
the single active layer G/hBN case, but included in the TBG case in section \ref{section4}.

\section{Graphene on hBN}\label{section3}

\begin{figure}
\includegraphics[width=\linewidth]{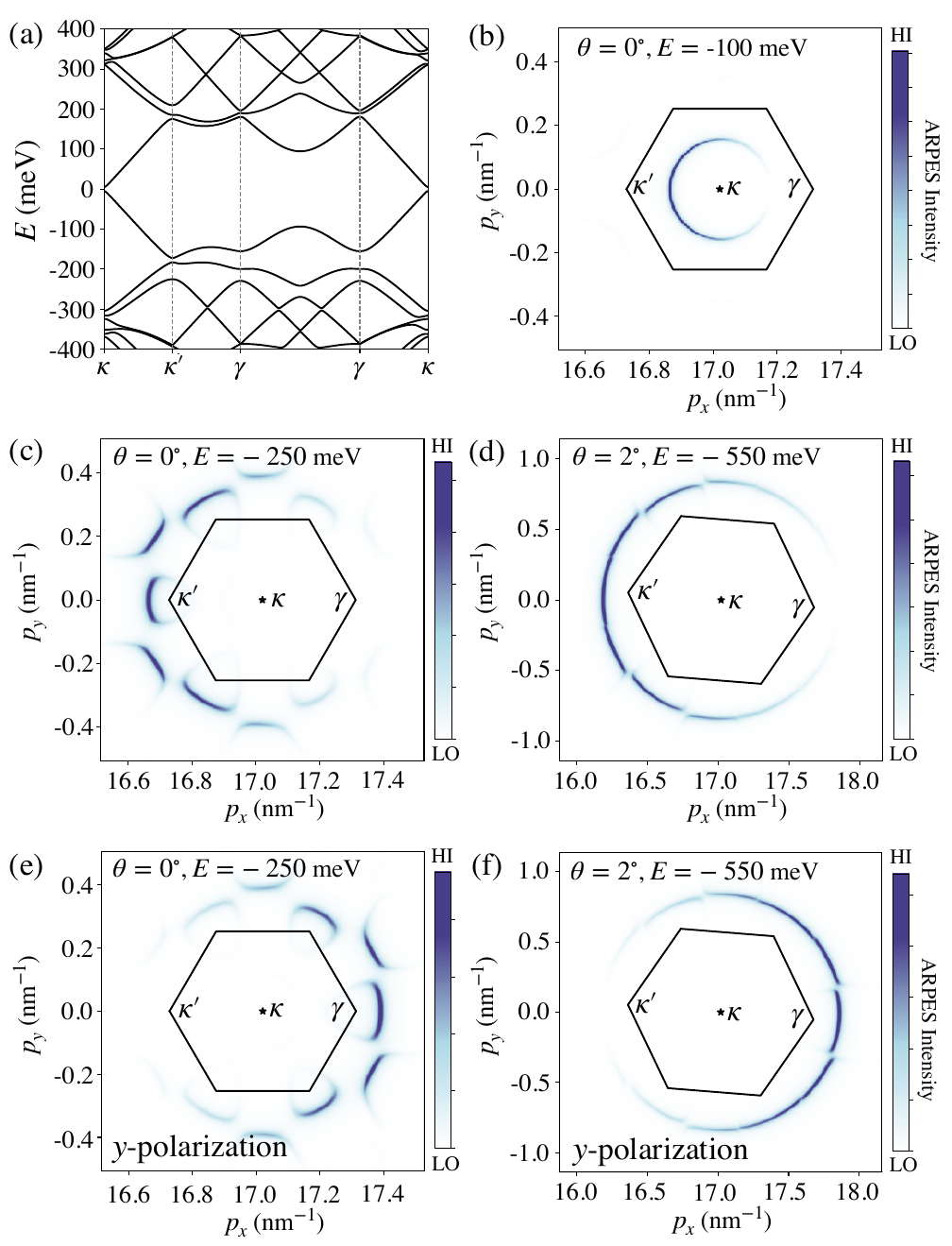}
\vspace{-10pt}
\caption {\small
(a) Moir\'e band structure of G/hBN with twist angle $\theta=0^\circ$, 
calculated using the \textit{ab initio} relaxed first harmonic parameters from Ref.\onlinecite{GhBN_Jeil_2017}.
(b-d) Constant-energy ARPES momentum distributions near BZ corner $\mb{K}_+$ calculated using Eq.~(\ref{Eq_GhBN_G0}) in which a factor related to photon polarization is dropped.
(b) $\theta=0^\circ$ at energy $E=-100$ meV, where hBN substrate has little effect on the energy bands;
(c) $\theta=0^\circ$ at energy $E=-250$ meV, where the 
hBN substrate has a large effect on the energy bands of graphene; 
(d) $\theta=2^\circ$ at $E=-550$meV. 
hBN's effect is negligible when the graphene and hBN layers are away from alignment.
(e-f) Constant-energy ARPES momentum distributions calculated for $y$-polarized light using 
Eq.~(\ref{Eq_I_Adotv}). The $x$-polarized light yields ARPES contours identical to those calculated
in (c-d) using Eq.~(\ref{Eq_GhBN_G0}). Photons with $y$-polarization rotate the anisotropy by $\pi$ compared to photons with $x$-polarization. In (b-f), the hexagon is the MBZ.
}\label{fig_G_hBN}
\end{figure}

The moir\'e band structure of graphene on aligned hBN is illustrated in Fig.~\ref{fig_G_hBN}(a).
These bands were calculated from a continuum model\cite{GhBN_Jeil_2017}
that accounts for lattice relaxation. 
In this model low energy states in graphene are most strongly modified by the substrate hBN layer when
the two layers are aligned ($\theta=0^\circ$).
In this case the inversion symmetry breaking in the presence of hBN opens a 
gap with size $\sim 7$ meV\cite{GhBN_Jeil_2017} at charge neutrality and a gap 
between the highest-energy valence band and remote valence bands. Both gaps are apparent in transport measurements.\cite{GhBN_gap_Hunt, GhBN_gap_Woods, SDC_Ponomarenko} 
Figures~\ref{fig_G_hBN}(b-d) show the corresponding ARPES momentum distribution functions near BZ corner $\mb{K}_+$, using Eq.~(\ref{Eq_GhBN_G0}) in which a factor related to photon polarization is dropped,
calculated at an energy near the middle of the highest valence band and 
at an energy below the energy gap separating this band from lower energy states.  
For the aligned ($\theta=0^\circ$) case, the 
hBN substrate has little effect (Fig.~\ref{fig_G_hBN}(b)) on the ARPES spectrum except at energies that are close to the
induced gaps on the hole-side (Fig.~\ref{fig_G_hBN}(c)).  In Fig.~\ref{fig_G_hBN}(b) in particular, the constant energy surface is still well inside the
MBZ and the ARPES momentum distribution is similar to the 
circular constant-energy surface of monolayer graphene\cite{ARPES_Bostwick_2007, ARPES_G_Hwang_2011, ARPES_G_Gierz_2011} shown in Appendix \ref{appendixA}.
At a lower energy illustrated in Fig.~\ref{fig_G_hBN}(c), Bragg scattering by 
moir\'e reciprocal lattice vectors thoroughly mixes isolated layer momentum eigenstates and 
this is reflected in the momentum distribution functions.
The avoided crossings that are apparent in Fig.~\ref{fig_G_hBN}(c) are sometimes referred to as secondary Dirac cones.\cite{SDC_Ponomarenko, SDC_Park, SDC_Jung, SDC_Ortix, SDC_Wallbank, G_hBN_SDC_2016, SDC_Yankowitz}  
When the two layers are not accurately aligned, as in the $\theta = 2^\circ$ case illustrated in Fig.~\ref{fig_G_hBN}(d), the unperturbed energy at the MBZ
boundary is large, increasing the range of energy over which the ARPES momentum 
distribution is not strongly altered by hBN. This result agrees with previous ARPES observations.\cite{nanoARPES_Wang}

The momentum distribution functions in Fig.~\ref{fig_G_hBN} are anisotropic as a function of momentum direction.  
These dark corridor\cite{ARPES_G_Gierz_2011} anisotropies are well known from previous ARPES studies of epitaxial graphene systems\cite{ARPES_Shirley, ARPES_Eli2008, Bostwick_2007, ARPES_theo_Falko, ARPES_G_Soren, ARPES_G_Hwang_2011, ARPES_G_Gierz_2011, ARPES_Puschnig} and result from interference between photoemissions from two 
honeycomb sublattices.  
The ARPES intensity anisotropy also has a photon-polarization dependence\cite{ARPES_G_Gierz_2011, ARPES_G_Hwang_2011, ARPES_polarization_Liu, ARPES_review_Moser}
that is ignored when Eq.~(\ref{Eq_GhBN_G0}) is used for the momentum distribution function, highlighted in Figs.~\ref{fig_G_hBN}(e-f) which illustrate momentum distributions calculated for the case of 
$y$-polarized light using Eq.~(\ref{Eq_I_Adotv}). For momenta near $\mb{K}_+$, the ARPES momentum distribution contour with $x$-polarized light calculated using Eq.~(\ref{Eq_I_Adotv}) is identical to 
the result obtained using Eq.~(\ref{Eq_GhBN_G0}) and shown in Fig.~\ref{fig_G_hBN}(b-d).
This is a consequence of the Dirac Hamiltonian property: $\bs{\nabla}_\mb{k}H_0 = \hbar v_{\sub{F}}(\sigma_x, \sigma_y)$.
The same observation applies for the TBG model discussed 
in section \ref{section4}, in which interlayer tunneling is momentum independent.
The constant-energy ARPES anisotropies of G/hBN using $x$- and $y$-polarized light are analogous to the monolayer graphene case shown in Appendix~\ref{appendixA}, where in both cases photons with $y$-polarization rotate the anisotropy by $\pi$ compared to photons with $x$-polarization.
The photon-polarization dependent ARPES measurements have been implemented to determine the signs of intralayer and interlayer tunneling parameters in monolayer graphene and Bernal-stacked bilayer graphene,\cite{ARPES_G_Hwang_2011} as described in Appendix \ref{appendixB}.

\begin{figure}
\includegraphics[width=\linewidth]{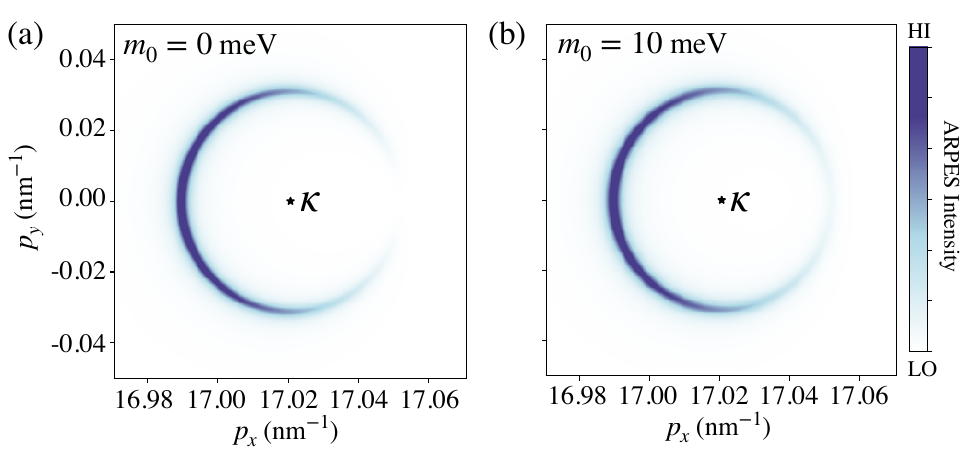}
\vspace{-10pt}
\caption{\small
Constant-energy ARPES maps of $0^\circ$-twist G/hBN (a) at $E=-20$ meV for $m_0=0$ meV; (b) at $E=-23$ meV for $m_0=10$ meV, which produces a band gap $\Delta_{\text{gap}} \sim 23$ meV at charge neutrality.
The mass parameter $m_0$ weakens the anisotropy.
}\label{fig_G_hBN_mass}
\end{figure}

This anisotropy of graphene sheet ARPES can be used
to measure one of the key parameters of G/hBN systems, the mass parameter $m_0$,   
as illustrated in Fig.~\ref{fig_G_hBN_mass}. 
The mass parameter characterizes
the strength of sublattice symmetry breaking\cite{Gierz_G_subsym} in graphene and plays a key role in the 
appearance of the quantized anomalous Hall effect.\cite{tBLG_Sharpe_2019, tBLG_Serlin_2020, QAHE_Polshyn, QAHE_Stepanov, QAHE_theo_Bultinck, QAHE_theo_Ya-Hui, QAHE_theo_Ya-Hui_PRR, QAHE_theo_Song, QAHE_theo_XiaoDi, QAHE_theo_Wolf, QAHE_theo_Jianpeng, QAHE_theo_Repellin, QAHE_theo_Fengcheng} 
It has contributions both from single-particle physics and from interacting self-energies, and 
in the latter case can be spin/valley-flavor dependent.\cite{GhBN_Jeil_2014, GhBN_Jeil_2017, GhBN_Jeil_2015, tBLG_MingHF,tBLG_Ming_weakfield} 
It influences the photoemission by concentrating the quasiparticle 
states more on one sublattice, thereby weakening sublattice interference and the resulting anisotropy of the APRES 
signal.  When a mass term $m_0$ is added to the isolated layer Dirac Hamiltonian, 
the eigenvector becomes 
\begin{equation}\label{mass_anisotropy}
    \Psi^\xi(\mb{q}) \propto 
    \begin{pmatrix}
    e^{-i\xi \theta_{\mb{q}}/2} \\
    \Big(-\frac{m_0}{v_{\scaleto{\text{F}}{3pt}} q}+s\sqrt{1+\frac{m_0^2}{v_{\scaleto{\text{F}}{3pt}}^2 q^2}}\Big) e^{i\xi \theta_{\mb{q}}/2}
    \end{pmatrix},
\end{equation}
where $\mb{q}$ is momentum measured from the Dirac point, and $s=+1(-1)$ denotes conduction(valence) band.
Near $\mb{K}_{+}$, the ARPES signal is the square of the sum of the sublattice components of the quasiparticle wavefunctions.
As shown in Fig.~\ref{fig_G_hBN_mass}, the anisotropy is noticeably weaker for $m_0=10$ meV (Fig.~\ref{fig_G_hBN_mass}(b)) than for 
$m_0=3.62$ meV\cite{GhBN_Jeil_2017} (Fig.~\ref{fig_G_hBN_mass}(a)). By comparing the contrast ratio between the weakest and strongest photoemission intensity
on the Fermi contour, it should be possible to measure this key parameter. 

ARPES momentum distribution functions are influenced both by all details of the single-particle Hamiltonian 
and by electron-electron interaction effects.  Comparing ARPES spectra with theoretical model 
calculations like those illustrated in Fig.~\ref{fig_G_hBN}(b-f) sheds light on both single-particle and 
interaction corrections, although they might be difficult to separate.  In the case of G/hBN heterojunctions,
the questions that ARPES can answer are mostly quantitative in character.  
We therefore turn now to the case in which 
ARPES has the greatest potential to answer key qualitative questions, namely the case of 
TBG heterojunctions, especially close to the magic twist angles.  

\section{Graphene on Graphene} \label{section4}

Bilayer graphene moir\'e superlattices are formed by a relative twist between different graphene sheets.  For our TBG calculations, we assume the second layer is twisted clockwise by $\theta$ 
with respect to the first layer.
The ARPES momentum distribution function calculations in this section are based on a  
low-energy continuum moir\'e Hamiltonian of small-twist-angle TBG.\cite{tBLG_Bistritzer_2011}  ARPES measurements have the potential to 
validate and refine these models, and to identify important interaction effects.  By diagonalizing the 
continuum model Hamiltonian the miniband Bloch wavefunctions can be expanded in the form
\begin{equation}
\begin{split}
    |\xi,n,\mb{k} \rangle 
    &= \sum\limits_{l,\alpha,\mb{g}}  \psi^\xi_{n l \alpha \mb{g}}(\mb{k})|\mb{k}+\mb{g}, l \alpha \rangle \\
     &= \frac{1}{\sqrt{N}} \sum\limits_{l, \alpha,\mb{g}, \mb{R}_1}  \psi^\xi_{n l \alpha \mb{g}}(\mb{k}) e^{i(\mb{k}+\mb{g}) \cdot (\mb{R}_l+\bs\tau_{l\alpha})}|\mb{R}_l, \alpha \rangle,
\end{split}
\end{equation}
where $l=1,2$ label layers and $\alpha=$A,B label sublattices. The coordinates of carbon atoms in two layers are related by $\mb{R}_2=\mathcal{R}_{-\theta}(\mb{R}_1-\bs{\tau}) + \mb{d}$, $\bs{\tau}_{2\alpha} = \mathcal{R}_{-\theta} \bs{\tau}_{1\alpha}$, and $\mathcal{R}$ is the rotation operator.
As in the single active layer case, we employ a free electron final state approximation and ignore photon polarization effects to obtain the following expression for the photoemission transition amplitudes: 
\begin{widetext}
\begin{equation}\label{Eq_TBG}
\begin{split}
    \langle \mb{p}|\xi,n,\mb{k} \rangle \propto \phi(\mb{p})
    \sum\limits_{\alpha,\mb{g}}
    \Big[ \psi^\xi_{n1\alpha \mb{g}}(\mb{k})
    \delta_{\mb{p}_\parallel, \mb{k}+\mb{g}+\mb{G}_1} e^{-i\mb{G}_1 \cdot \bs{\tau}_{1\alpha}} e^{-ip_z z_1}
    + \psi^\xi_{n2\alpha \mb{g}}(\mb{k})
    \delta_{\mb{p}_\parallel, \mb{k}+\mb{g}+\mb{G}_2} e^{-i[\mb{G}_1 \cdot (\bs{\tau}_{1\alpha}-\bs{\tau}) + \mb{G}_2 \cdot \mb{d}]} e^{-ip_z z_2}
    \Big],
\end{split}
\end{equation}
\end{widetext}
Here $\mb{G}_1$ is a reciprocal lattice vector of the first layer and $\mb{G}_2 = \mathcal{R}_{-\theta} \mb{G}_1$ is the corresponding reciprocal lattice vector of the second layer. We take $z_1=d/2$ and $z_2=-d/2$, where $d=0.34$ nm is the adjacent layer distance. For initial AB-stacking, $\bs\tau=\bs\tau_{\sub{1B}}=(0,a/\sqrt{3})$.

\begin{figure}
\includegraphics[width=\linewidth]{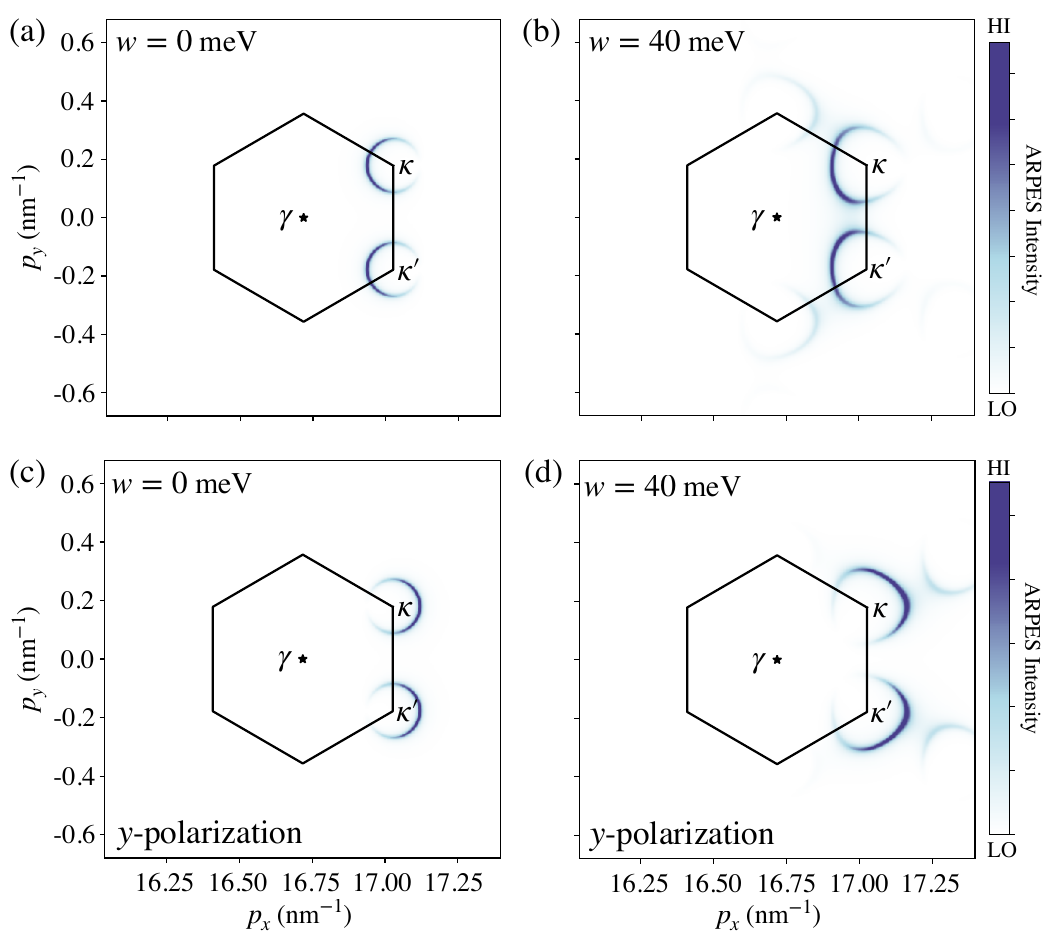}
\vspace{-10pt}
\caption{\small
Constant-energy ARPES momentum distribution of $1.2^\circ$-TBG with $v_{\sub{F}}=10^6$ m/s at energy $E=-60$ meV (a) without interlayer tunneling and (b) with interlayer tunneling strength $w=40$ meV.   (The experimental tunneling strength is thought to be close to 
$w \sim 110$ meV. Since the continuum model 
moir\'e bands depend only on the ratio of $w$ to the twist angle,
these results also apply to TBG with a realistic interlayer tunneling amplitude 
at a twist angle $\sim 3 ^\circ$ after rescaling of momentum measured from the Dirac point.)
As the interlayer tunneling is turned on, the ARPES signal begins to reflect the
altered wavefunctions and dispersions of the moir\'e minibands. 
Band flattening leads to more rapid dependence of the momentum distribution function on energy.
(c-d) corresponds to (a-b) respectively using $y$-polarized light.
}\label{fig_tBLG}
\end{figure}

\begin{figure*}
\includegraphics[scale=1.1]{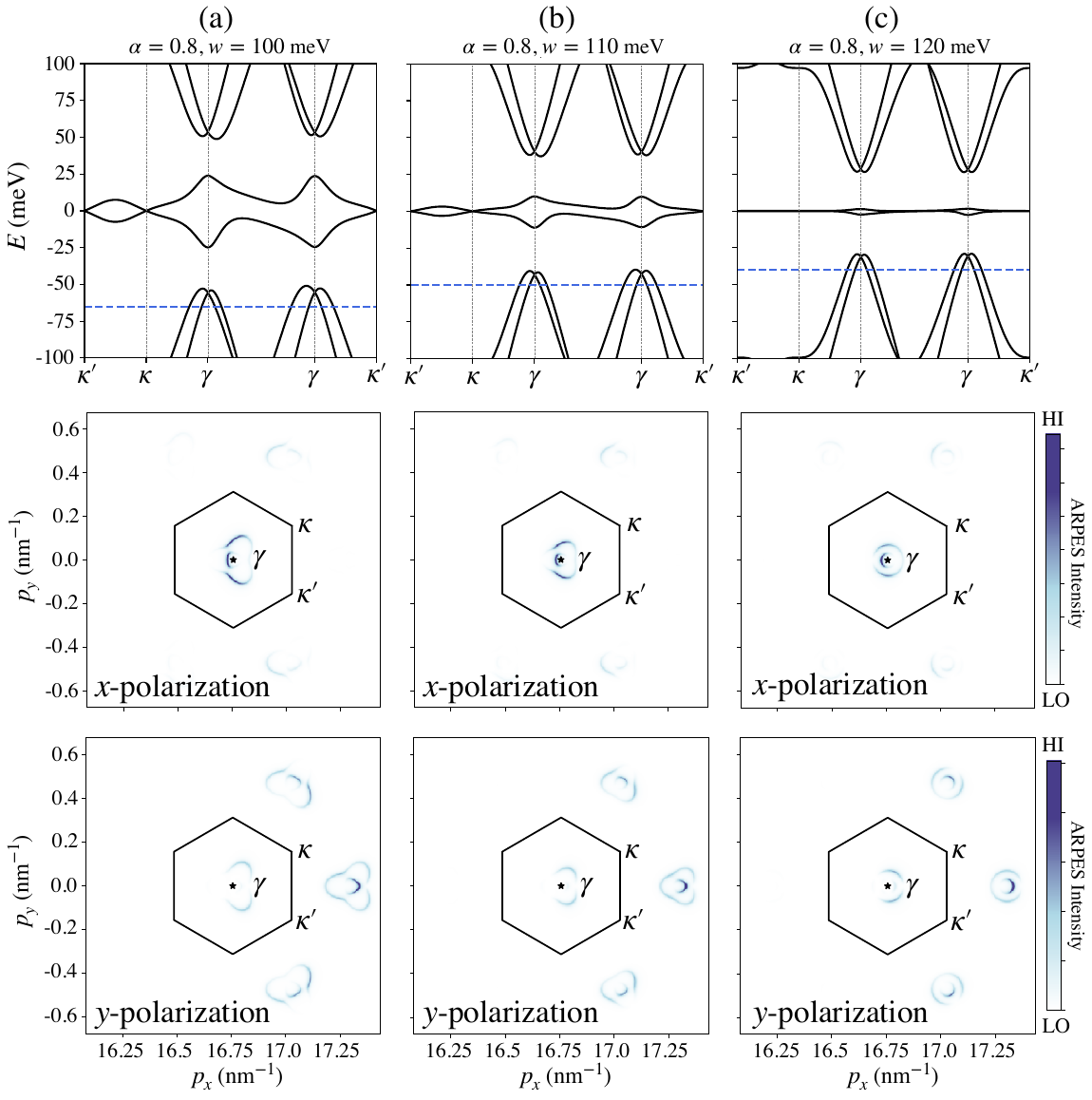}
\vspace{-10pt}
\caption{\small
Band structures and constant-energy ARPES momentum distributions for 
$1.05^\circ$-TBG with tunneling ratio $\alpha=w^{\sub{AA}}/w^{\sub{AB}}=0.8$, and tunneling strength $w=w^{\sub{AB}}$.
The momentum space maps are calculated at the energy near the top of the remote
valence bands specified by the blue dashed line in the band structure
plots. (a) Tunneling strength $w=100$ meV, (b) $w=110$ meV, (c) $w=120$ meV.
}\label{fig_tBLG_w1}
\end{figure*}

Taking photon polarization effects into account, the ARPES intensity is the same as in Eq.~(\ref{Eq_I_Adotv}) 
with the free-electron final state projected to the Bloch state basis:
\begin{equation}
\begin{split}
|\mb{p} \rangle = &\phi^*(\mb{p}) \sum\limits_{\alpha,\mb{g}} \Big[ \delta_{\mb{p}_\parallel,\mb{k}+\mb{g}+\mb{G}_1} e^{i\mb{G}_1 \cdot \bs{\tau}_{1\alpha}} e^{ip_zd/2}
|\mb{k}+\mb{g},1\alpha \rangle \\
&+\delta_{\mb{p}_\parallel,\mb{k}+\mb{g}+\mb{G}_2} e^{i[\mb{G}_1 \cdot (\bs{\tau}_{1\alpha}-\bs{\tau}) + \mb{G}_2 \cdot \mb{d}]} e^{-ip_zd/2}
|\mb{k}+\mb{g},2\alpha \rangle \Big].
\end{split}
\end{equation}

Each photoelectron momentum $\mb{p}$ picks a specific valley $\xi$, a reciprocal lattice vector $\mb{G}_1$ and a moir\'e reciprocal lattice vector $\mb{g}$ to map $\mb{k}$ into the first MBZ. The ARPES contour becomes complex, depending on $\bs{\tau}$ and $\mb{d}$, for $\mb{G}_2=\mathcal{R}_{-\theta}\mb{G}_1 \neq 0$.
We therefore focus only on photoelectron momenta $\mb{p}$ near $\mb{K}_+ = (4\pi/3a,0)$, the most intense signal comes from $\mb{G}_1=\mb{G_1}^\prime= \mb{G}'_2=0$. 
Thus, the ARPES intensity is proportional to
\begin{equation}
\begin{split}
    I(\mb{p},E) \propto \sum\limits_{n,\mb{k}} \Big|& \sum\limits_{\alpha,\mb{g}} \delta_{\mb{p}_\parallel, \mb{k}+\mb{g}} \big(\psi^+_{n1\alpha \mb{g}}(\mb{k}) e^{-ip_zd/2} \\
    + &\psi^+_{n2\alpha \mb{g}}(\mb{k}) e^{ip_zd/2} \big) \Big|^2 \delta(E-\varepsilon^+_{n\mb{k}}).
\end{split}
\end{equation}

Interlayer interference becomes important for large photon energies because $p_zd$ is not negligible, which is the case we are considering in order to employ the free-electron final state approximation. For a $100$ eV photon, the out-of-plane momentum of the photoelectron emitted near the BZ corner is $p_z \sim 5$ \AA$^{-1}$. The ARPES intensity $I(\mb{p},E)$ depends periodically on $p_z$ and thereby on photon energy, 
in analogy to the bilayer graphene case illuminated in Appendix \ref{appendixB}. We will ignore the photon-energy dependence of ARPES intensity calculations in the remaining part of the paper.

The principle elements of the TBG photoemission signal near valley $\mb{K}_+$ are illustrated in
Fig.~\ref{fig_tBLG}.  
When the two graphene layers are artificially decoupled, 
the individual layer Dirac cones are displaced in momentum space and centered 
on the displaced BZ corners, $\bs{\kappa} = \mathcal{R}_{\theta/2} \mb{K}_+$ and $\bs{\kappa}' = \mathcal{R}_{-\theta/2} \mb{K}_+$,
of two layers.  As shown in Fig.~\ref{fig_tBLG}(a), two Dirac cones appear at $\bs{\kappa}$, which is the first layer Dirac point, and at 
$\bs{\kappa}'$, which is the second layer Dirac point. 
As illustrated in Fig.~\ref{fig_tBLG}(b), 
when interlayer tunneling $w$ is turned on the circular constant energy 
surfaces of the decoupled layers are distorted, and replicas displaced by moir\'e reciprocal lattice vectors
appear that have different matrix elements.  The interlayer tunneling strength $w=40$ meV chosen in Fig.~\ref{fig_tBLG}(b) corresponds to the moderate coupling strength present 
above the first magic twist angle. All TBG calculations in this paper take the Fermi velocity to be $v_{\sub{F}} = 10^6$ m/s. The appropriate value of 
$w$, including its many-body renormalization, plays a key role in TBG electronic properties.  
These figures show that if the twist angle is known,
a numerical value of $w$ can be estimated from ARPES momentum distribution functions.

The anisotropies of the ARPES momentum distribution functions
around $\bs{\kappa}$ and $\bs{\kappa}'$ in Fig.~\ref{fig_tBLG}(a-b) can be understood in 
terms of interference of patterns sourced from two sublattices in each layer: 
\begin{equation}
    I_l(\mb{p}) \propto \cos^2\Big(
    -\frac{\mb{p} \cdot (\bs\tau_{l\sub{B}} - \bs\tau_{l{\sub{A}}})}{2} + \frac{\xi}{2} (\theta^l_{\mb{q}}-\theta_l) + \frac{\pi (1-s)}{4}
    \Big).
\end{equation}
The pattern is analogous to the monolayer graphene case illustrated in 
Appendix \ref{appendixA}, except that two graphene layers here have a relative twist. 
$\theta^l_{\mb{q}}$ is the angle of momentum $\mb{q}$ measured from the Dirac point of layer $l$,
$\theta_l$ is the twist angle of layer $l$ ($\theta_1=\theta/2$, $\theta_2=-\theta/2$). 
In the first layer for example, $\bs\tau_{{\sub{1A}}} = (0,0)$, $\bs\tau_{{\sub{1B}}} = \mathcal{R}_{\theta/2}\bs\tau_{\sub{B}} = e^{i(\pi+\theta)/2} a/\sqrt{3}$ and $\mb{p} = \mb{K}_1^+ + \mb{q}$. Then
\begin{equation}
\begin{split}
    \mb{p} \cdot (\bs\tau_{1{\sub{B}}} - \bs\tau_{{\sub{1A}}}) &= (\mb{K}_1^+ + \mb{q}) \cdot \bs\tau_{{\sub{1B}}} 
    = \mb{q} \cdot \bs\tau_{{\sub{1B}}}.
\end{split}
\end{equation}
Thus for the valence band in valley $+$: $\xi=1$, $s = -1$, 
and the minimum of intensity occurs when $\theta_{\mb{q}}-\theta/2 - \mb{q} \cdot \bs\tau_{{\sub{1B}}} = 0$: 
\begin{equation}\label{min_PE}
     \theta_{\mb{q}} -\frac{\theta}{2} = \frac{qa}{\sqrt{3}} \sin(\theta_{\mb{q}}-\frac{\theta}{2}).
\end{equation}
Equation~(\ref{min_PE}) has the solution $\theta_{\mb{q}}=\theta/2$ if $q \ll |\mb{G}|$. 
The anisotropy of photoemission discussed above for the monolayer case is reoriented by 
the graphene layer twists, providing a handle to measure twist angles from ARPES spectra.

The ARPES momentum distributions with $y$-polarized light corresponding to Fig.~\ref{fig_tBLG}(a-b) are shown in Fig.~\ref{fig_tBLG}(c-d). Comparing Figs.~\ref{fig_tBLG}(a) and \ref{fig_tBLG}(c), the photon-polarization dependent anisotropy as a result of the interference between intralayer sublattices bears resemblance to that of monolayer graphene (Appendix \ref{appendixA}). In addition, comparing Figs.~\ref{fig_tBLG}(b) and \ref{fig_tBLG}(d), we see that 
the interference between interlayer sublattices rotates in the $y$-polarization case  and shifts the overall minibands anisotropies.

\begin{figure*}
\includegraphics[scale=1.1]{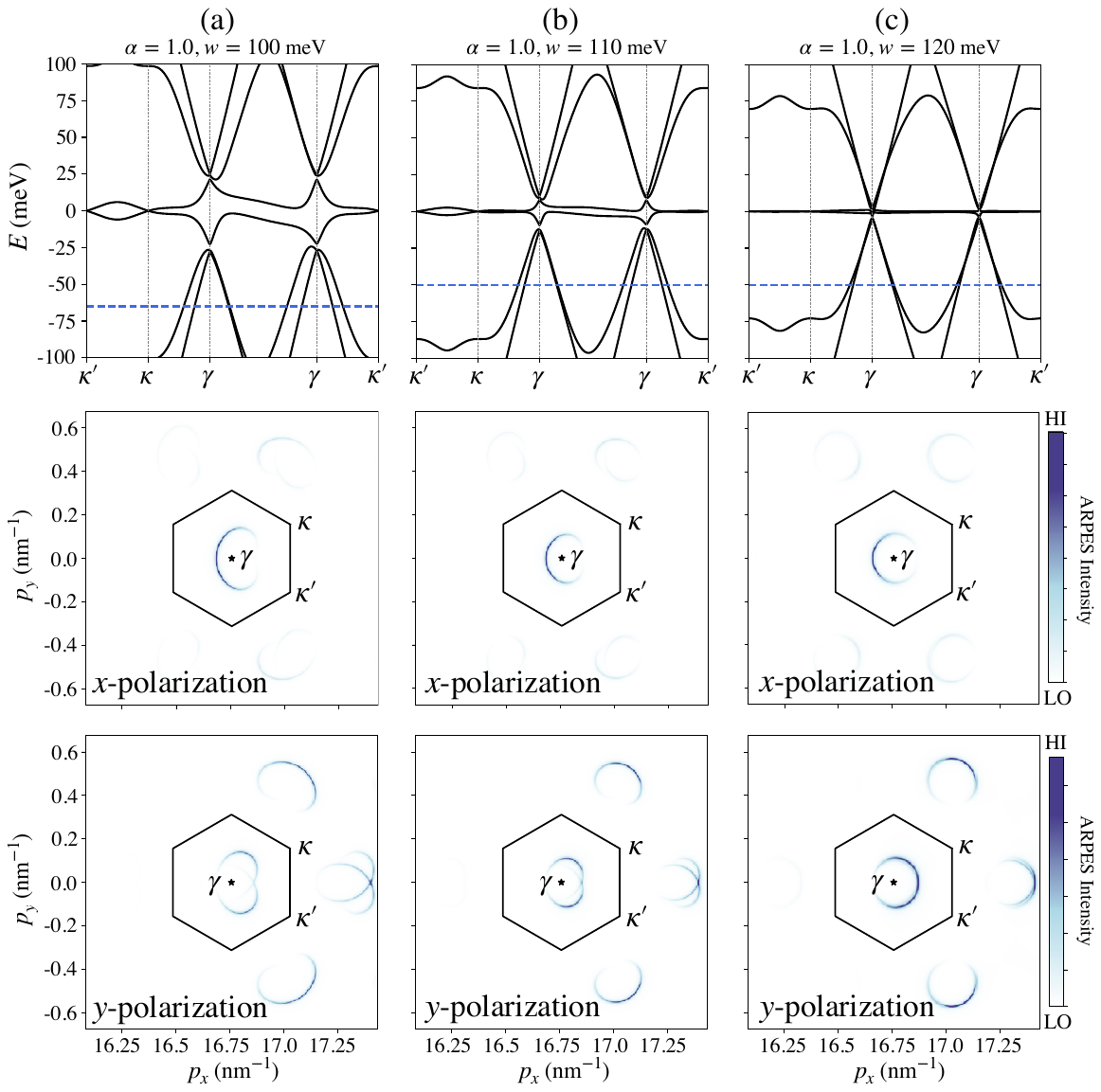}
\vspace{-10pt}
\caption{\small
Band structures and constant-energy ARPES momentum distributions of $1.05^\circ$-TBG with tunneling ratio $\alpha=w^{\sub{AA}}/w^{\sub{AB}}=1$, and tunneling strength $w=w^{\sub{AB}}$.  The momentum distribution functions were evaluated at 
energy levels indicated by the blue dashed lines in the band structures.
(a) Tunneling strength $w=100$ meV, (b) $w=110$ meV, (c) $w=120$ meV.
}\label{fig_tBLG_w2}
\end{figure*}

At the first magic twist angle, interlayer tunneling dominates the physics.
The ARPES signal at energies in the flat bands is discussed at length in the following 
section, but there is a strong influence not only on the flat bands 
but also on the remote bands, whose quasiparticles wavefunctions have non-trivial 
momentum space structure manifested by complex momentum distribution functions like
those illustrated in Fig.~\ref{fig_tBLG_w1}.
This figure highlights the dependence on an 
important phenomenological parameter often used in continuum models of TBG, 
the ratio of the interlayer tunneling amplitude between $\pi$-orbitals on the same sublattice $w^{\sub{AA}}$ 
to the tunneling amplitude between $\pi$-orbitals on different sublattices $w^{{\sub{AB}}}$.  
These amplitudes are equal by symmetry when strain relaxation of the twisted bilayers is 
neglected,\cite{tBLG_Bistritzer_2011} and important 
strain  features can be captured\cite{Uchida_tBLG_corrugation, Koshino_tBLG_corrugation, tBLG_corrugation_2015, tBLG_corrugation_2016,  tBLG_corrugation_Jain}
by letting $w^{\sub{AA}}$ be smaller than $w^{\sub{AB}}$.  
The correction accounts partially\cite{tBLG_relaxation_Kaxiras, tBLG_relaxation_Koshino}
for strain and corrugation effects, neglected in simple bilayer models.
The ratio $\alpha=w^{\sub{AA}}/w^{\sub{AB}}$ is used as a parameter in the calculations below. 
Tight-binding model estimates\cite{Koshino_tBLG_corrugation} suggest that $\alpha \approx 0.8$,
but this estimate should be checked experimentally.  $\alpha$ might also be 
altered by electron-electron interaction effects.  
Figures~\ref{fig_tBLG_w1},\ref{fig_tBLG_w2} compare band structures of $1.05^\circ$-TBG, and 
momentum distribution functions calculated at energy levels away from flat interval
for different tunneling strengths $w=w^{\sub{AB}}$ and for tunneling ratios $\alpha=0.8$ 
(Fig.~\ref{fig_tBLG_w1}) and  $\alpha=1$ (Fig.~\ref{fig_tBLG_w2}). 
The energies at which the momentum distributions are calculated 
are indicated in the band structure plots by blue dashed lines. 
As in the G/hBN\cite{SDC_Park, second_Dirac_point2012, second_Dirac_point2013} case,
there are secondary Dirac cones at the moir\'e $\gamma$ point indicated in Fig.~\ref{fig_tBLG_w1} at 
which isolated layer bands are degenerate.  We see in Fig.~\ref{fig_tBLG_w2} that
the signature of the secondary Dirac cones becomes less prominent as $\alpha \to 1$, providing a handle to choose
the best values of this parameter.
The proximity of the magic twist angle, which depends on the product of $\theta$ and $w$, can 
also be detected by examining the remote bands, as illustrated in 
Figs.~\ref{fig_tBLG_w1},\ref{fig_tBLG_w2}.

Theory\cite{tBLG_MingHF, C3_strain_Bi, C3_Shang, C3_Markus} and scanning probe experiments\cite{TBG_STM_Jiang, TBG_STM_Kerelsky, TBG_STM_Choi, TBG_STM_Xie}
suggest that broken $C_3$ rotational symmetry is common when the Fermi level is in the middle of the flat bands 
of MATBG or when the strain induced by substrate is considered.  The constant energy maps in
Fig.~\ref{fig_tBLG_w1} and Fig.~\ref{fig_tBLG_w2} retain $C_3$ rotational symmetry,
but because of matrix element effects the intensity does not.
By using the polarized light, described in Appendix \ref{appendixA} and Eq.~(\ref{Eq_I_Adotv}), the full shape of constant-energy ARPES contours can be seen as shown in Figs~\ref{fig_tBLG_w1},\ref{fig_tBLG_w2}.

\begin{figure*}
\includegraphics[scale=0.9]{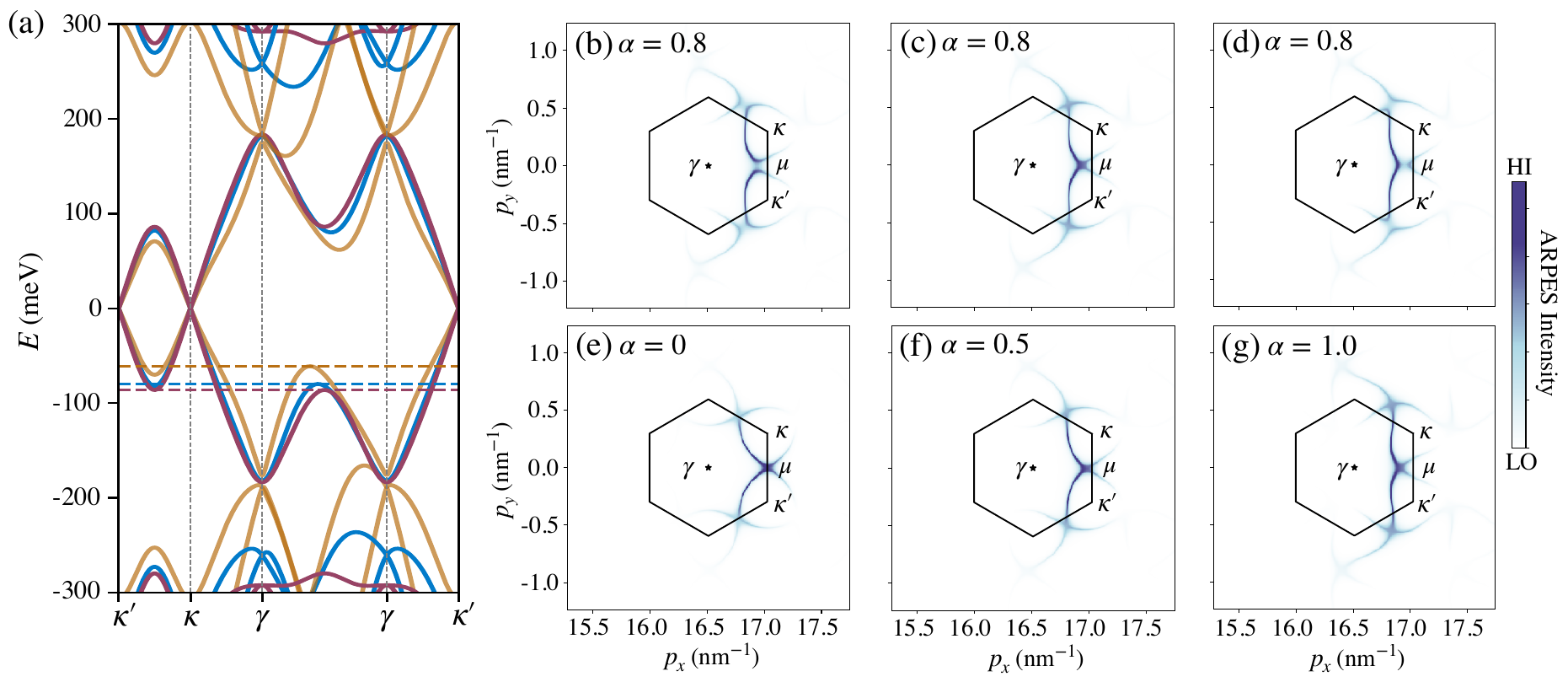}
\vspace{-10pt}
\caption{\small
(a) Band structures of $2^\circ$-TBG with tunneling ratio $\alpha=0$ (red), $\alpha=0.5$ (blue) and $\alpha=1$ (yellow). The colored dashed lines mark the corresponding valence band VHS energies.
(b-d) Constant-energy ARPES maps of $2^\circ$-TBG with $\alpha=0.8$ for three energy levels near the valence band VHS. 
(b) At $2$ meV above the valence band VHS, (c) at the valence band VHS energy, 
(d) at $2$ meV below the valence band VHS. 
(e-g) Constant-energy ARPES maps of $2^\circ$-TBG at the valence band VHS for different tunneling ratios. (e) $\alpha = 0$, (f) $\alpha = 0.5$, (g) $\alpha = 1.0$. The VHSs are always on the $\gamma-\mu$ high symmetry lines and strong lattice relaxation (small $\alpha$) moves the VHS towards the $\mu$ points of the MBZ. 
When $\alpha=0$, VHSs are exactly on $\mu$ points. All of these calculations were performed
with tunneling strength $w=110$ meV.
}\label{tBLG_vhs}
\end{figure*}

\section{van Hove singularities}\label{section5}

So far we have discussed momentum distribution functions measured at energies outside
the flat bands.  The most powerful experimental information will come from 
measurements within partially occupied flat bands, although these will also require the most precise energy resolution.  The dispersion that remains within the flat bands
near the magic angle, where they attain their
minimum width, is very sensitive to details of the single-particle band structure 
calculations, including especially filling-factor dependent band 
renormalizations\cite{TBG_STM_Kerelsky, TBG_STM_Xie, TBG_STM_Choi, TBG_STM_Jiang, TBG_STM_Dillon} due to 
mean-field Hartree and exchange interactions.\cite{tBLG_MingHF, tBLG_Ming_weakfield, tBLG_Guinea_2018, TBG_VHS}
It is also known that the flat band spectrum is very sensitive to the strain parameter $\alpha$.  Below we calculate for reference ARPES momentum distribution functions at selected energies within the flat bands when the interaction effects are neglected.  
These calculations are most likely to be relevant when the bilayer is surrounded by nearby conducting layers,
for example gate layers, that screen Coulomb interactions strongly.

When interactions are neglected the most prominent feature of the flat bands 
are the van Hove singularities (VHSs) that occur at Lifshitz phase transition energies,\cite{VHS_Wu2021} 
which in the past have been studied mainly outside of the flat-band regime. 
When they are weak compared to the flat band width, the influence of interactions is prominent only for Fermi energies close to VHSs where they can lead to competing broken symmetry states.\cite{VHS_kohn_1965, VHS_Markiewicz_1997, VHS_Gonzalez_2008, VHS_Fleck_1997, VHS_Rice_1975, VHS_Valenzuela_2008, VHS_Makogon_2011, VHS_Nandkishore_2012, VHS_Li_2012}
Tuning the Fermi level across a VHS,
generally leads to a change in Fermi surface topology.  
The band filling factors at which VHSs occur in MATBG are strongly sensitive to band structure details that are not always accurately known, and 
could be identified by performing gate-voltage dependent ARPES measurements.  
For example, the continuum model band structures in Fig.~\ref{tBLG_vhs}(a),
calculated at $\theta= 2^\circ$ and  $\alpha=0, 0.5$ and $1$,
have valence band van Hove singularities at energies marked by dashed lines.
At this twist angle there are three VHSs along the $\gamma-\mu$ lines in the MBZ. 
Because of the change in constant-energy surface topology from $\gamma$-centered 
electron pockets at energies below the VHS to $\kappa$- and $\kappa'$-centered hole pockets
above the VHS, ARPES momentum distribution functions can distinguish whether a constant energy surface is below or above the VHS energy, as illustrated in Fig.~\ref{tBLG_vhs}(b-d).
When $\alpha=0$, i.e. the interlayer tunnelling between the same sublattice $w^{\text{AA}}=0$, the VHS is exactly at the $\mu$ point. As $\alpha$ increases, the VHS position
moves away from the $\mu$ point along the $\gamma-\mu$ lines
as illustrated in Fig.~\ref{tBLG_vhs}(e-g).

\begin{figure*}
\includegraphics[scale=0.8]{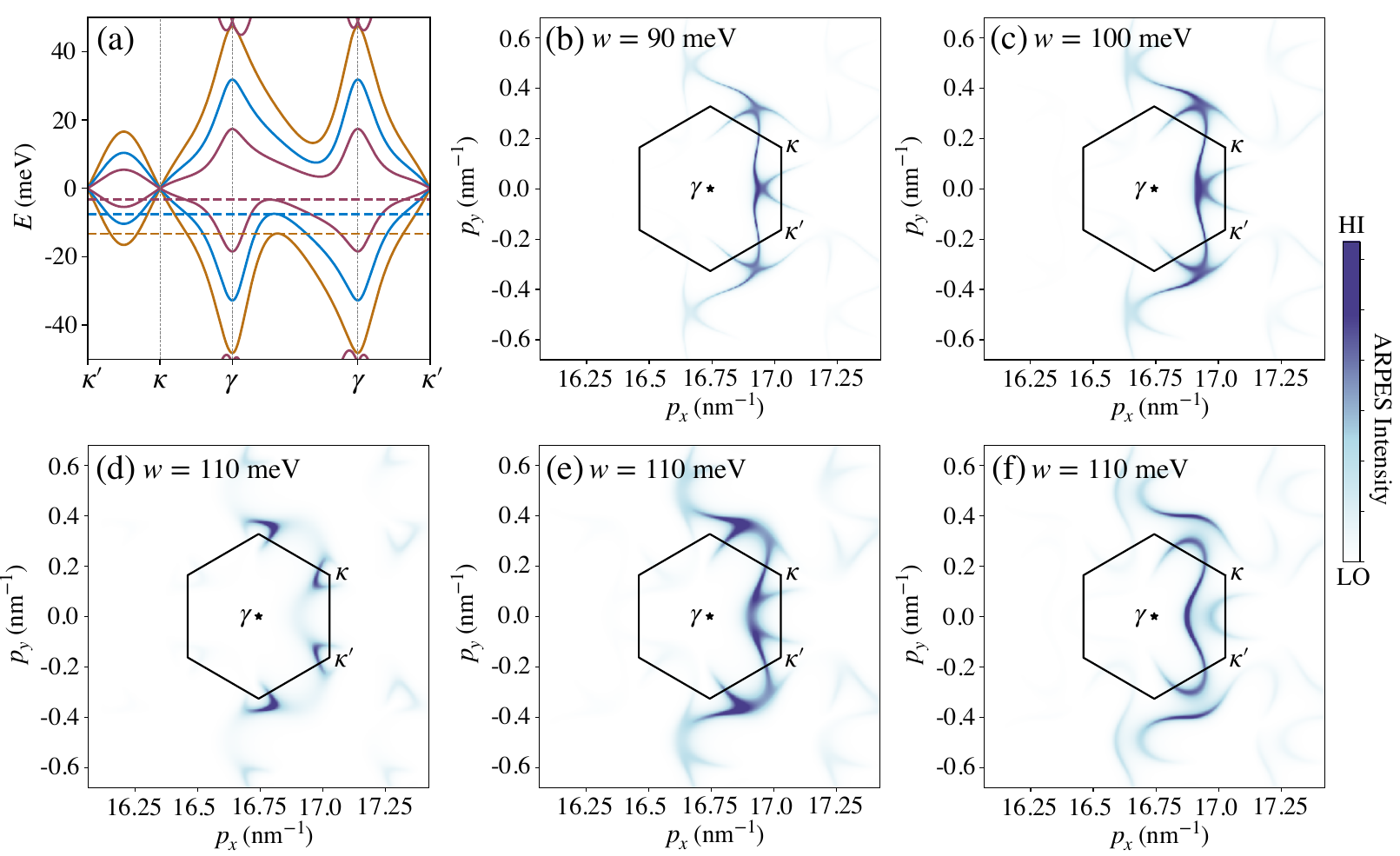}
\vspace{-10pt}
\caption{\small
(a) Band structure of $1.1^\circ$-TBG with $\alpha=0.8$ and three different tunneling strengths $w=90$ meV(yellow), $w=100$ meV(blue) and $w=110$ meV(red). 
The colored dashed lines mark the corresponding valence band VHSs, which are not on the $\kappa-\kappa'$ line. 
(b,c,e) Constant-energy maps momentum distribution functions at 
the valence band VHS energies for $1.1^\circ$-TBG with $\alpha=0.8$. (b) $w=90$ meV, (c) $w=100$ meV and (e) $w=110$ meV. 
(d-f) Constant-energy maps near the valence band VHS of $1.1^\circ$-TBG with $\alpha=0.8$ and $w=110$ meV. The momentum distribution functions (d) at $1$ meV above the valence band VHS, (e) at the valence band VHS, (f) and at $1$ meV below the valence band VHS illustrates how the constant energy surface topology changes.
}\label{MATBG_vhs}
\end{figure*}

At smaller twist angle near the magic angle regime, for example $1.1^\circ$, the flat band energy scales are reduced, as shown in Fig.~\ref{MATBG_vhs}(a), 
but the valence band constant energy surface topology, as shown in Fig.~\ref{MATBG_vhs}(d-f), remains similar as larger twist angles. 
Near the magic angle, each VHS on the $\gamma-\mu$ line splits into two VHSs.\cite{TBG_VHS, LiangFu_VHS}
In Fig.~\ref{MATBG_vhs}(b,c,e), we fix twist angle to be $1.1^\circ$ while tuning the tunneling strength $w$. 
Increasing $w$ plays the same role as decreasing twist angle in the low-energy continuum model. In Fig.~\ref{MATBG_vhs}(e), the VHSs start to split.

\section{Discussion} \label{summary}
In this paper we have analyzed how valence band ARPES momentum distribution functions depend on graphene moir\'e superlattice band Hamiltonians.  For G/hBN the 
critical parameter is the value of the mass parameter $m_0$ which expresses the degree to which inversion symmetry in the graphene sheet is violated by interaction with the substrate.
We point out that $m_0$ parameter, thought to be key to the quantum anomalous Hall effect,
can be extracted from measurements of the anisotropy of the momentum space distribution maps at energies close to the charge neutrality.
Since momentum-space anisotropy decreases when $m_0$ is larger than the conduction-valence
band splitting at $m_0=0$ (see Fig.~\ref{fig_G_hBN_mass}), 
finer momentum-space resolution will be needed to identify smaller values of $m_0$.
The reduced anisotropy is due to weaker interference between honeycomb sublattices with increasing $m_0$. For TBG moir\'{e} superlattices, the important strain-dependent parameter $\alpha$
that characterizes the ratio of intra-sublattice to inter-sublattice tunneling between layers is available from measurements deep in the valence band, which do not require exceptional energy resolution.  For this reason we 
expect that nano-ARPES performed on moir\'{e} superlattice samples, which are typically less
than 100 $\mu$m in size, can provide important information about moir\'{e} superlattice 
electronic structure, and guide us toward accurate parameter values for low-energy model even before 
extreme energy resolution is achieved.

That said, the full potential impact of ARPES in understanding MATBG will
be realized only if sufficient energy resolution can be achieved in momentum-resolved spectra 
taken with partially occupied flat bands. Existing results from STM\cite{TBG_STM_Xie, TBG_STM_Dillon, TBG_STM_Kerelsky, TBG_STM_Jiang, TBG_STM_Choi}
suggest that useful
results will require an energy resolution scale that is small, perhaps very small, compared 
to the $\sim 40$ meV width the flat bands broaden to when partially occupied.
Key questions that need to be answered, and can potentially be answered by ARPES, include the following:
i) Is the valence band minimum at $\gamma$ as it is in single-particle theory, or elsewhere in the 
MBZ; ii) Are large Fermi surface reconstructions associated with broken spin and/or valley symmetries at 
both integer and fractional moir\'e band fillings as 
suggested by weak-field Hall transport measurements? iii) 
Do the broken spin and/or valley symmetries thought to be necessary for interaction-induced insulating 
states persist to non-integer band filling factors, including those where superconductivity is observed?
iv) Finally, are there
well-defined Fermi surfaces at metallic filling factors with large quasiparticle normalization factors, and
if so, what is their shape.  The history of progress in advancing ARPES techniques
over recent decades suggests that we be optimistic about their application to graphene-based 
moir\'e superlattices.

\begin{acknowledgments}
This research was primarily supported by the National Science Foundation through the Center for Dynamics and Control of Materials: an NSF MRSEC under Cooperative Agreement No. DMR-1720595. 
We acknowledge helpful interactions with Dan Dessau, Eli Rotenberg and Simon Moser.
\end{acknowledgments}

\appendix
\section{ARPES in monolayer graphene}\label{appendixA}
Accurate calculations of photoemission matrix elements are 
often challenging.  In the free-electron final state approximation, the
photoemission process promotes an electron 
with crystal momentum $\mb{k}$ from the a Bloch state of the 
target material to a free-space state with momentum $\mb{p}$.
The ejected electron is called a photoelectron. 
In a $\pi$-orbital tight-binding model, the initial state of this photoemission process
is a Bloch state with $\pi$-orbital amplitudes on both sublattices of
monolayer graphene's 2D honeycomb lattices.  Using a $\mb{k} \cdot \mb{p}$
description of low energy states in the graphene sheet's $\pi$-band, the initial Bloch state's are 
labelled by valley $\xi=\pm$ and band $n=\text{c}$ (conduction) or $\text{v}$ (valence):
\begin{equation}
|\xi,n,\mb{k}\rangle = \sum\limits_{\alpha=\text{A,B}} \psi^\xi_{n\alpha}(\mb{k}) |\mb{k},\alpha \rangle,
\end{equation}
where $\mb{k}$ is the full momentum measured from the Brillouin zone (BZ) center $\Gamma$. 
The transition amplitude to the final free-particle state is 
\begin{equation}\label{def_phi}
    \langle \mb{p}|\xi, n,\mb{k} \rangle 
    = \frac{1}{\sqrt{N}} \sum\limits_{\mb{R}, \alpha} \psi^\xi_{n\alpha}(\mb{k}) e^{i(\mb{k}-\mb{p}_{\parallel}) \cdot (\mb{R}+\bs\tau_\alpha)} \phi(\mb{p}),
\end{equation}
where $\phi(\mb{p}) = \int d^3\mb{r} e^{-i\mb{p} \cdot (\mb{r} - \mb{R} - \bs\tau_\alpha)} \phi(\mb{r} - \mb{R} - \bs\tau_\alpha)$ is the Fourier transform of atomic $\pi$-orbital on sublattice 
$\alpha$ at lattice vector $\mb{R}$ and $\mb{p}_{\parallel}$ is the in-plane projection of 3D momentum $\mb{p}$.
Dropping factors that 
depend on the photon polarization and measuring energy relative to a convenient zero, it follows that 
the ARPES intensity
\begin{widetext}
\begin{equation}\label{eqn_A3}
I(\mb{p},E) 
\propto \sum\limits_{\xi,n,\mb{k}} \big|\langle \mb{p}|\xi, n,\mb{k} \rangle \big|^2 \delta(E-\varepsilon^\xi_{n\mb{k}}) 
\propto \big| \phi(\mb{p}) \big|^2 \sum\limits_{\xi,n,\mb{k}} \Big| \sum\limits_{\alpha} \psi^\xi_{n\alpha}(\mb{k})
\delta_{\mb{p}_{\parallel},\mb{k}+\mb{G}}
e^{-i\mb{G} \cdot \bs\tau_\alpha} \Big|^2 \delta(E-\varepsilon^\xi_{n\mb{k}}),
\end{equation}
\end{widetext}
where $\mb{G}$ is a reciprocal lattice vector of graphene. 
For each photoelectron momentum $\mb{p}$, the most intense signal comes from the closest extended-zone valley. 
The photoemission process picks a specific valley and a specific $\mb{G}$ to map $\mb{k}$ into the first BZ. 

\begin{figure}
\includegraphics[width=\linewidth]{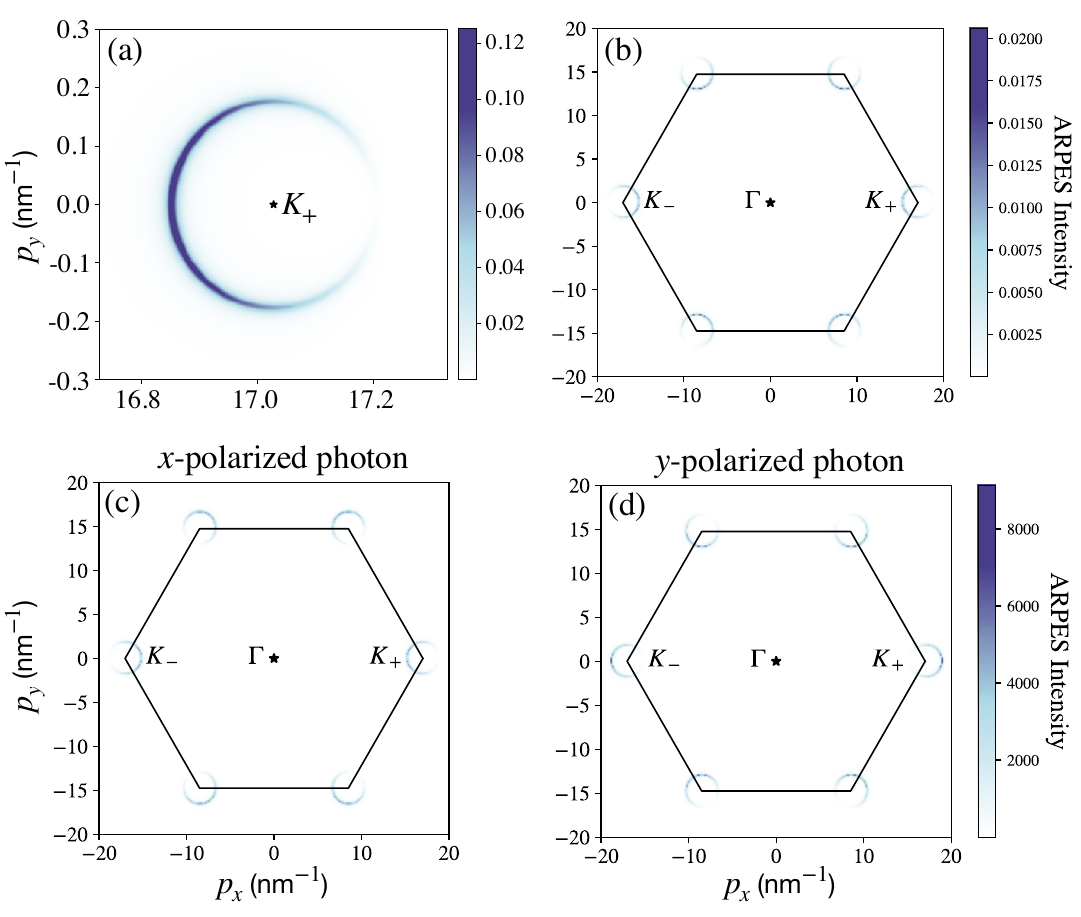}
\vspace{-10pt}
\caption{\small
Constant-energy ARPES maps of monolayer graphene, ignoring photon's polarization effects, at (a) $E=-120$ meV near $\mb{K}_+$; (b) $E=-1200$ meV for photoelectron momentum $\mb{p}$ in the range of the first BZ (hexagon). Constant-energy ARPES maps of monolayer graphene at $E=-1200$ meV using (c) $x$-polarized light and (d) $y$-polarized light.
}\label{fig_monoG}
\end{figure}

Constant-energy photoemission maps at $E=-120$ meV and $E=-1200$meV are shown in 
Figs.~\ref{fig_monoG}(a-b) for $\mb{p}_{\parallel}$ near $\mb{K}_+=(4\pi/3a,0)$ (a) and 
over the full first BZ (b). The anisotropy of ARPES signal in Fig.~\ref{fig_monoG} can be understood 
as a two-source interference pattern from two sublattices,\cite{ARPES_Eli2008}
\begin{equation}
    I(\mb{p}) \propto \cos^2\Big(
    -\frac{\mb{p}_{\parallel} \cdot (\bs\tau_{\sub{B}} - \bs\tau_{\sub{A}})}{2} + \frac{\xi \theta_{\mb{q}}}{2} + \frac{\pi (1-n)}{4}
    \Big),
\end{equation}
$n=+1(-1)$ denotes conduction(valence) band. $\bs\tau_{\sub{A}} = (0,0)$, $\bs\tau_{\sub{B}} = (0,a/\sqrt{3})$. $\theta_{\mb{q}}$ is the angle of wave vector $\mb{q}$ measured from BZ corners.

The anisotropy is also reflected by directly substituting eigenvectors of the Dirac Hamiltonian in Eq.(\ref{eqn_A3}):
\begin{equation}
    \psi_s^\xi(\mb{q}) = \frac{1}{\sqrt{2}}
    \begin{pmatrix}
        e^{-i\xi\theta_{\mb{q}}/2} \\
        n e^{i\xi\theta_{\mb{q}}/2}
    \end{pmatrix},
\end{equation}
to obtain
\begin{widetext}
\begin{equation}
\begin{split}
    I(\mb{p})
    &\propto \Big| e^{-i\xi\theta_{\mb{q}}/2} e^{i\mb{G} \cdot (\bs\tau_{\sub{B}}-\bs\tau_{\sub{A}})/2} + n e^{i\xi\theta_{\mb{q}}/2} e^{-i\mb{G} \cdot (\bs\tau_{\sub{B}}-\bs\tau_{\sub{A}})/2} \Big|^2 \delta_{\mb{p}_\parallel, \mb{q} + \mb{K}_\xi + \mb{G}} \\
    &\propto \cos^2\Big( \frac{-\mb{G} \cdot (\bs\tau_{\sub{B}}-\bs\tau_{\sub{A}})}{2} + \frac{\xi\theta_{\mb{q}}}{2} + \frac{\pi(1-n)}{4}
    \Big) \delta_{\mb{p}_\parallel, \mb{q} + \mb{K}_\xi + \mb{G}}
\end{split}
\end{equation}
\end{widetext}

When the photon's polarization $\mb{A}$ is explicitly taken into account, 
the ARPES intensity is proportional to
$ |\langle \mb{p} | \mb{A} \cdot \hat{\mb{v}} | \xi, n, \mb{k} \rangle|^2 $, where $\hat{\mb{v}} = \bs\nabla_{\mb{k}} H$ \cite{Adotv_Ismail} is the velocity operator and $|\mb{p} \rangle$ is the free-electron final state projected to the Bloch state basis
\begin{equation}
\begin{split}
    |\mb{p} \rangle 
    &= \sum\limits_{\mb{k},\alpha} |\mb{k},\alpha \rangle \langle \mb{k},\alpha|\mb{p}\rangle \\
    &= \sum\limits_{\mb{k},\alpha} \delta_{\mb{p}_\parallel,\mb{k}+\mb{G}} e^{i\mb{G} \cdot \bs{\tau}_\alpha} \phi^*(\mb{p}) |\mb{k},\alpha \rangle
\end{split}
\end{equation}
Specifically, the constant-energy ARPES intensity is
\begin{widetext}
\begin{equation}\label{Eq_PE_G_xy}
I(\mb{p},E) 
\propto \sum\limits_{\xi,n,\mb{k}} \big|\langle \mb{p}|\mb{A} \cdot \hat{\mb{v}} |\xi, n,\mb{k} \rangle \big|^2 \delta(E-\varepsilon^\xi_{n\mb{k}})
=  \sum\limits_{\xi,n,\mb{k}} \big|\mb{A} \cdot \langle \mb{p}|\bs{\nabla}_\mb{k} H |\xi, n,\mb{k} \rangle \big|^2 \delta(E-\varepsilon^\xi_{n\mb{k}})
\end{equation}
\end{widetext}

Fig.~\ref{fig_monoG}(c-d) plot the constant-energy ARPES signals 
using $x$- and $y$-polarized light respectively. The anisotropies can be understood by substituting $\bs{\nabla}_{\mb{k}}H \propto (\xi\sigma_x,\sigma_y)$ in Eq.~(\ref{Eq_PE_G_xy}), which gives
\begin{widetext}
\begin{equation}
\begin{split}
    &I^{x\text{-pol}}(\mb{p})
    \propto \cos^2\Big( \frac{\mb{G} \cdot (\bs\tau_{\sub{B}}-\bs\tau_{\sub{A}})}{2} + \frac{\xi\theta_{\mb{q}}}{2} + \frac{\pi(1-n)}{4}
    \Big) \delta_{\mb{p}_\parallel, \mb{q} + \mb{K}_\xi + \mb{G}} \\
    &I^{y\text{-pol}}(\mb{p})
    \propto \cos^2\Big( \frac{\mb{G} \cdot (\bs\tau_{\sub{B}}-\bs\tau_{\sub{A}})}{2} + \frac{\xi\theta_{\mb{q}}}{2} + \frac{\pi(1+n)}{4}
    \Big) \delta_{\mb{p}_\parallel, \mb{q} + \mb{K}_\xi + \mb{G}}
\end{split}
\end{equation}
\end{widetext}

\section{ARPES in bilayer graphene}\label{appendixB}
We comment here on the importance of reaching a consensus on the signs of hopping amplitudes in graphene multilayers. For Bernal-stacked bilayer graphene, multiple studies adopted interlayer hoppings with wrong signs\cite{BLG_spectrum_Koshino, BLG_t3_Iorsh, BLG_t3_Park, BLG_t3_Gruneis, BLG_t3_Jolie, BLG_t3_McCann, BLG_t3_Cserti, ARPES_Eli2008, BLG_t3_Jung_2011, BLG_t3_Nilsson, BLG_t3_Castro_2010} in the $\pi$-orbital tight-binding model. As Ref.\onlinecite{BLG_sign_Jeil} clarified, not only the magnitudes but also the signs of hopping parameters play a crucial role in electronic properties. We will show that the signs of intralayer and interlayer hoppings can be identified by careful ARPES measurements.

Using the four-component spinor basis, $\Psi_{\mb{k}}= (c_{1{\sub{A}}}, c_{1{\sub{B}}}, c_{2{\sub{A}}}, c_{2{\sub{B}}})^T$, with layer ($1,2$) and sublattice ($\text{A},\text{B}$) degrees of freedom, the Hamiltonian of Bernal-stacked bilayer graphene is
\begin{widetext}
\begin{equation}\label{Eq_H_BLG}
    H(\mb{k})=
    \begin{pmatrix}
        \varepsilon_{{\sub{1A}}} & t_0 f(\mb{k}) & t_4 f(\mb{k})  & t_3 f^*(\mb{k}) \\
        t_0 f^*(\mb{k}) & \varepsilon_{{\sub{1B}}} & t_1 & t_4 f(\mb{k}) \\
        t_4 f^*(\mb{k}) & t_1 & \varepsilon_{{\sub{2A}}} & t_0 f(\mb{k}) \\
        t_3 f(\mb{k}) & t_4 f^*(\mb{k}) & t_0 f^*(\mb{k}) & \varepsilon_{{\sub{2B}}}
    \end{pmatrix},
\end{equation}
\end{widetext}
where
\begin{equation}
    f(\mb{k})=\sum\limits_{j=1}^3 e^{i\mb{k} \cdot \bs{\delta}_j},
\end{equation}
$\bs{\delta}_j$ is the position of B sublattice relative to A sublattice.
$t_0$ is the intralayer nearest-neighbor (NN) hopping parameter, $t_1$ is the interlayer hopping between dimer sites and $t_3$ and $t_4$ are interlayer next-nearest-neighbor (NNN) hopping parameters between non-dimer sites:
\begin{equation}
\begin{split}
    t_0 &= \langle \mb{R}_{\sub{1A}} | \mathcal{H} | \mb{R}_{\sub{1B}} \rangle 
    = \langle \mb{R}_{\sub{2A}} | \mathcal{H} | \mb{R}_{\sub{2B}} \rangle \\
    t_1 &= \langle \mb{R}_{\sub{1B}}| \mathcal{H} | \mb{R}_{\sub{2A}} \rangle \\
    t_3 &= \langle \mb{R}_{\sub{1A}}| \mathcal{H} | \mb{R}_{\sub{2B}} \rangle \\
    t_4 &= \langle \mb{R}_{\sub{1A}}| \mathcal{H} | \mb{R}_{\sub{2A}} \rangle 
    = \langle \mb{R}_{\sub{1B}}| \mathcal{H} | \mb{R}_{\sub{2B}} \rangle \\
\end{split}
\end{equation}
$|\mb{R}_{\alpha} \rangle$ is localized Wannier orbitals. $t_0$ is related to the Fermi velocity by $v_{\sub{F}} = \sqrt{3}a|t_0|/2\hbar$, $t_1$ and $t_3$ determine the amplitude and orientation of trigonal warping and $t_4$ introduces particle-hole asymmetry.

Ref.\onlinecite{BLG_sign_Jeil} ascertained that, using the maximally localized Wannier wave function method, $t_0$ is negative and $t_1$, $t_3$ and $t_4$ are positive. The negative sign of $t_0$ and positive sign of $t_1$ have been testified by polarization-dependent ARPES measurements in Ref.\onlinecite{ARPES_G_Hwang_2011}.

\begin{figure*}
\includegraphics[width=\linewidth]{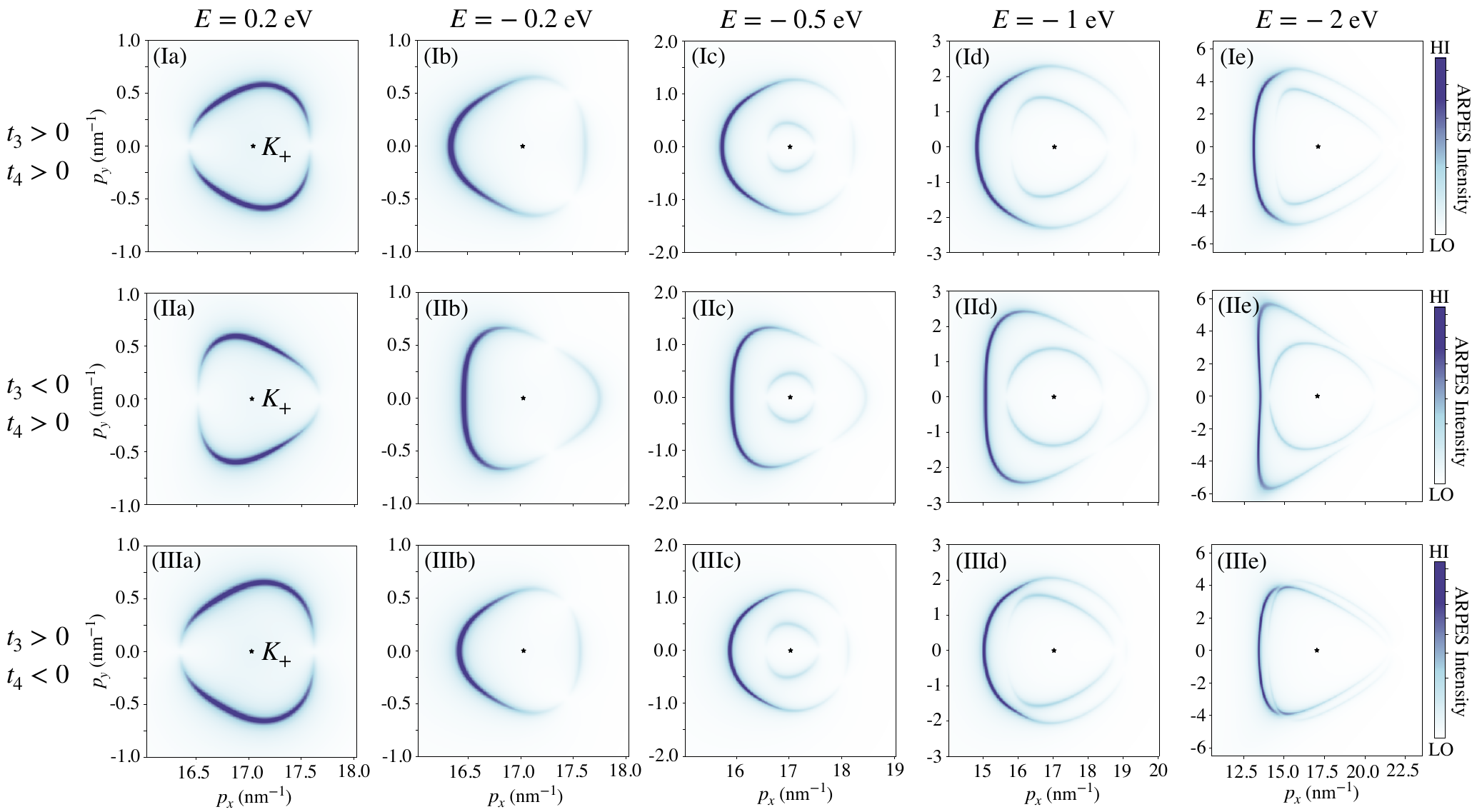}
\vspace{-10pt}
\caption {\small
Constant-energy ARPES momentum distributions, using $x$-polarized light, near $\mb{K}_+$ at various energies indicated at the top of each column. We use \textit{ab initio} tight-binding parameters in Ref.\onlinecite{BLG_sign_Jeil}, $t_0=-2.61$ eV, $t_1=0.361$ eV, $|t_3|=0.283$ eV and $|t_4|=0.138$ eV. (Ia-Ie) $t_3>0, t_4>0$; (IIa-IIe) $t_3<0, t_4>0$; (IIIa-IIIe) $t_3>0, t_4<0$. Note that we use a power-law normalized colorbar.
}\label{fig_BLG}
\end{figure*}

\begin{figure}
\includegraphics[width=\linewidth]{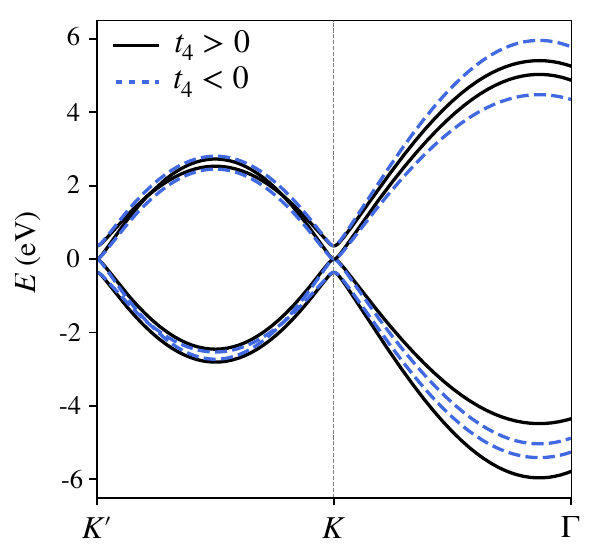}
\vspace{-10pt}
\caption {\small
Band structure of Bernal-stacked bilayer graphene with positive $t_4$ (solid black line) and negative $t_4$ (blue dashed line).
}\label{fig_band_BLG}
\end{figure}

\begin{figure*}
\includegraphics[width=\linewidth]{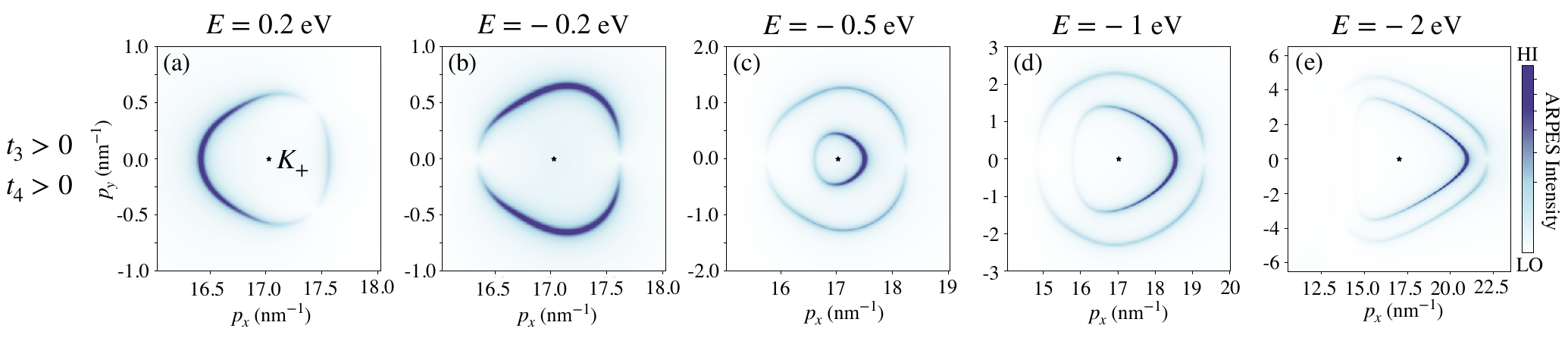}
\vspace{-10pt}
\caption {\small
Constant-energy ARPES momentum distributions, using $y$-polarized light, near $\mb{K}_+$ at various energies. The tight-binding parameters are the same as in Fig.~\ref{fig_BLG}(Ia-Ie). Note that we use a power-law normalized colorbar.
}\label{fig_vy_BLG}
\end{figure*}

The signs of $t_3$ and $t_4$ can also be determined by photon-polarization-dependent ARPES using Eq.~(\ref{Eq_PE_G_xy}). Figure~\ref{fig_BLG} show constant-energy ARPES contours near valley $\mb{K}_+=(4\pi/3a,0)$, using $x$-polarized\cite{ARPES_G_Hwang_2011} beam, for different signs of $t_3$ and $t_4$ at various energies. For positive $t_3$ (Fig.~\ref{fig_BLG}(Ia-Ie,\RNum{3}a-\RNum{3}e)), the trigonal warping orientations of the highest valence band and lowest conduction band are inverted as the Fermi level is tuned away from the charge neutrality, while the trigonal warping orientations of the lowest valence band and highest conduction band stay the same. For negative $t_3$ (Fig.~\ref{fig_BLG}(\RNum{2}a-\RNum{2}e)), the trigonal warping orientations of the highest valence band and lowest conduction band stay invariant as tuning the Fermi level, and the trigonal warpings of the lowest valence band and highest conduction band are less evident.
The opposite sign of $t_4$ interchanges conduction and valence bands, as shown in the band structure in Fig.~\ref{fig_band_BLG}. Two conduction bands intersect for positive $t_4$ and two valence bands intersect for negative $t_4$.
Figure~\ref{fig_vy_BLG} show constant-energy ARPES contours with $y$-polarized light. By comparing Fig.~\ref{fig_BLG} and Fig.~\ref{fig_vy_BLG} with ARPES experiments,\cite{ARPES_G_Hwang_2011, ARPES_BLG_Ohta_2006}  it is inferred that $t_0<0$, $t_1>0$, $t_3>0$ and $t_4>0$. This result can also be found in recent scanning tunnelling microscopy experiment.\cite{BLG_STM_Joucken}

\begin{figure*}
\includegraphics[width=0.8\linewidth]{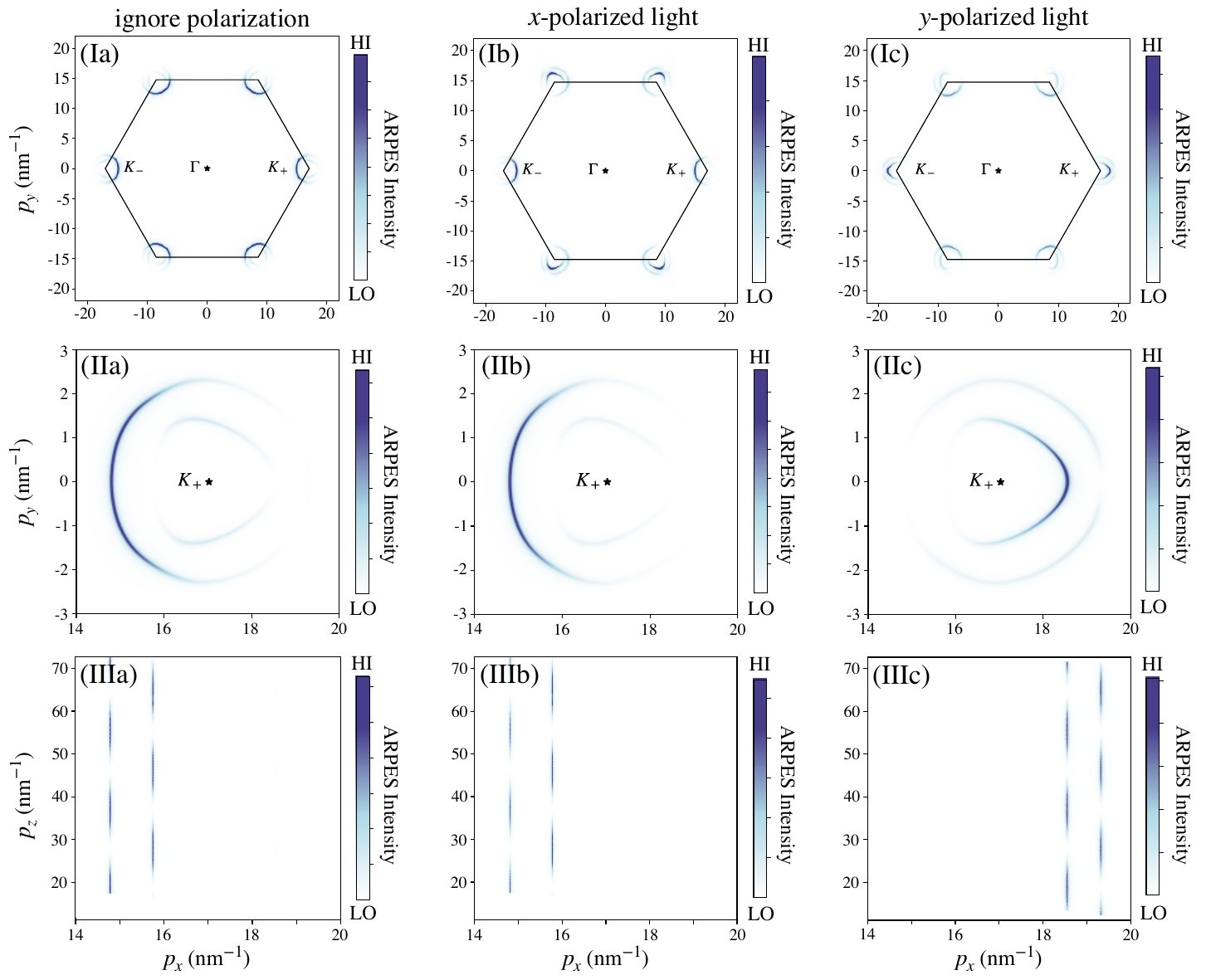}
\vspace{-10pt}
\caption {\small
Constant-energy ARPES momentum distributions. Columns a,b and c are calculated ignoring photon's polarization, with $x$-polarized light and with $y$-polarized light respectively. (Ia,Ib,Ic) are near the first BZ. (IIa, IIb, IIc) are near the BZ corner $\mb{K}_+$.  (IIIa, IIIb, IIIc) are near $\mb{K}_+$ and calculated with fixed $p_y=0$, photon's energy $\hbar \omega$ ranges from 20 to 210 eV. $p_z$ is related to $\hbar \omega$ by Eq.~(\ref{Eq_Ephoton_to_pz}). All figures are calculated with $t_0=-2.61$ eV, $t_1=0.361$ eV, $t_3=0.283$ eV and $t_4=0.138$ eV. Note that the scales of colorbars are different.
}\label{fig_BLG_pz}
\end{figure*}

With the correct hopping signs discussed above, Fig.~\ref{fig_BLG_pz} shows constant-energy ARPES momentum distributions near the first BZ (Fig.~\ref{fig_BLG_pz} (Ia-Ic)) and zoom-in figures near valley $\mb{K}_+$ (Fig.~\ref{fig_BLG_pz} (IIa-IIc)). Figures in columns a,b and c are calculated ignoring photon polarization, and 
with $x$-polarized light and with $y$-polarized light respectively.
In multilayer systems, including the bilayer graphene we are discussing here and the TBG in section \ref{section4} and \ref{section5} in the main text, interference between orbitals in different layers results in photon-energy-dependent ARPES signal. Interlayer interference becomes important when the photon energy is large
enough that $p_zd \sim 1$, where $d$ is the adjacent layer distance and $p_z$ is the $z$-component of photoelectron momentum. With this consideration, the ARPES intensity becomes
\begin{widetext}
\begin{equation}
    I(\mb{p},E) \propto \big| \phi(\mb{p}) \big|^2 \sum\limits_{n,\mb{k}} \Big| \sum\limits_{\alpha} \psi_{n\alpha}(\mb{k}) \delta_{\mb{p}_\parallel,\mb{k}} e^{-ip_z \hat{z} \cdot \bs{\tau}_{\alpha}} \Big|^2 \delta(E-\varepsilon_{n\mb{k}}),
\end{equation}
\end{widetext}
$\psi_{n\alpha}(\mb{k})$ is the eigenvector of Hamiltonian Eq.~(\ref{Eq_H_BLG}) and $\alpha$ represents layer and sublattice indices. In accordance with Hamiltonian Eq.~(\ref{Eq_H_BLG}),
\begin{equation}
\begin{split}
    &\bs{\tau}_{\sub{1A}} = (0,0,\frac{d}{2}),
    \bs{\tau}_{\sub{1B}} = (0,\frac{a}{\sqrt{3}},\frac{d}{2}),\\
    &\bs{\tau}_{\sub{2A}} = (0,\frac{a}{\sqrt{3}},-\frac{d}{2}),
    \bs{\tau}_{\sub{2B}} = (0,\frac{2a}{\sqrt{3}},-\frac{d}{2}).
\end{split}
\end{equation}
$p_z$ is related to photoelectron's kinetic energy $E_{\text{kin}}$ by
\begin{equation}
    \frac{\hbar^2}{2m}(p^2_\parallel + p_z^2) = E_{\text{kin}} = \hbar \omega + E - \phi,
\end{equation}
where $\hbar \omega$ is photon energy, $E$ is the initial Bloch state energy measured relative to the Fermi energy and $\phi$ is the work function. For large enough photon's energy $\hbar \omega$,
\begin{equation}\label{Eq_Ephoton_to_pz}
    p_z \approx \sqrt{\frac{2m \hbar \omega}{\hbar^2} - p^2_\parallel}
\end{equation}
Fixed $p_y=0$, Fig.~\ref{fig_BLG_pz}(IIIa-IIIc) show ARPES intensities near the BZ corner $\mb{K}_+$ as a function of $p_z$ and $p_x$. Photon's energy ranges from $20$ to $210$ eV in Fig.~\ref{fig_BLG_pz}(IIIa-IIIc).

\bibliographystyle{apsrev4-1}
\bibliography{ARPES}

\begin{thebibliography}{136}%
\makeatletter
\providecommand \@ifxundefined [1]{%
 \@ifx{#1\undefined}
}%
\providecommand \@ifnum [1]{%
 \ifnum #1\expandafter \@firstoftwo
 \else \expandafter \@secondoftwo
 \fi
}%
\providecommand \@ifx [1]{%
 \ifx #1\expandafter \@firstoftwo
 \else \expandafter \@secondoftwo
 \fi
}%
\providecommand \natexlab [1]{#1}%
\providecommand \enquote  [1]{``#1''}%
\providecommand \bibnamefont  [1]{#1}%
\providecommand \bibfnamefont [1]{#1}%
\providecommand \citenamefont [1]{#1}%
\providecommand \href@noop [0]{\@secondoftwo}%
\providecommand \href [0]{\begingroup \@sanitize@url \@href}%
\providecommand \@href[1]{\@@startlink{#1}\@@href}%
\providecommand \@@href[1]{\endgroup#1\@@endlink}%
\providecommand \@sanitize@url [0]{\catcode `\\12\catcode `\$12\catcode
  `\&12\catcode `\#12\catcode `\^12\catcode `\_12\catcode `\%12\relax}%
\providecommand \@@startlink[1]{}%
\providecommand \@@endlink[0]{}%
\providecommand \url  [0]{\begingroup\@sanitize@url \@url }%
\providecommand \@url [1]{\endgroup\@href {#1}{\urlprefix }}%
\providecommand \urlprefix  [0]{URL }%
\providecommand \Eprint [0]{\href }%
\providecommand \doibase [0]{http://dx.doi.org/}%
\providecommand \selectlanguage [0]{\@gobble}%
\providecommand \bibinfo  [0]{\@secondoftwo}%
\providecommand \bibfield  [0]{\@secondoftwo}%
\providecommand \translation [1]{[#1]}%
\providecommand \BibitemOpen [0]{}%
\providecommand \bibitemStop [0]{}%
\providecommand \bibitemNoStop [0]{.\EOS\space}%
\providecommand \EOS [0]{\spacefactor3000\relax}%
\providecommand \BibitemShut  [1]{\csname bibitem#1\endcsname}%
\let\auto@bib@innerbib\@empty
\bibitem [{\citenamefont {Novoselov}\ \emph {et~al.}(2004)\citenamefont
  {Novoselov}, \citenamefont {Geim}, \citenamefont {Morozov}, \citenamefont
  {Jiang}, \citenamefont {Zhang}, \citenamefont {Dubonos}, \citenamefont
  {Grigorieva},\ and\ \citenamefont {Firsov}}]{graphene_Novoselov_2004}%
  \BibitemOpen
  \bibfield  {author} {\bibinfo {author} {\bibfnamefont {K.~S.}\ \bibnamefont
  {Novoselov}}, \bibinfo {author} {\bibfnamefont {A.~K.}\ \bibnamefont {Geim}},
  \bibinfo {author} {\bibfnamefont {S.~V.}\ \bibnamefont {Morozov}}, \bibinfo
  {author} {\bibfnamefont {D.}~\bibnamefont {Jiang}}, \bibinfo {author}
  {\bibfnamefont {Y.}~\bibnamefont {Zhang}}, \bibinfo {author} {\bibfnamefont
  {S.~V.}\ \bibnamefont {Dubonos}}, \bibinfo {author} {\bibfnamefont {I.~V.}\
  \bibnamefont {Grigorieva}}, \ and\ \bibinfo {author} {\bibfnamefont {A.~A.}\
  \bibnamefont {Firsov}},\ }\href {\doibase 10.1126/science.1102896} {\bibfield
   {journal} {\bibinfo  {journal} {Science}\ }\textbf {\bibinfo {volume}
  {306}},\ \bibinfo {pages} {666} (\bibinfo {year} {2004})}\BibitemShut
  {NoStop}%
\bibitem [{\citenamefont {Geim}\ and\ \citenamefont
  {Novoselov}(2007)}]{graphene_Geim_2007}%
  \BibitemOpen
  \bibfield  {author} {\bibinfo {author} {\bibfnamefont {A.~K.}\ \bibnamefont
  {Geim}}\ and\ \bibinfo {author} {\bibfnamefont {K.~S.}\ \bibnamefont
  {Novoselov}},\ }\href {\doibase 10.1038/nmat1849} {\bibfield  {journal}
  {\bibinfo  {journal} {Nature Materials}\ }\textbf {\bibinfo {volume} {6}},\
  \bibinfo {pages} {183} (\bibinfo {year} {2007})}\BibitemShut {NoStop}%
\bibitem [{\citenamefont {Dean}\ \emph {et~al.}(2010)\citenamefont {Dean},
  \citenamefont {Young}, \citenamefont {Meric}, \citenamefont {Lee},
  \citenamefont {Wang}, \citenamefont {Sorgenfrei}, \citenamefont {Watanabe},
  \citenamefont {Taniguchi}, \citenamefont {Kim}, \citenamefont {Shepard},\
  and\ \citenamefont {Hone}}]{hBN_sub_Dean_2010}%
  \BibitemOpen
  \bibfield  {author} {\bibinfo {author} {\bibfnamefont {C.~R.}\ \bibnamefont
  {Dean}}, \bibinfo {author} {\bibfnamefont {A.~F.}\ \bibnamefont {Young}},
  \bibinfo {author} {\bibfnamefont {I.}~\bibnamefont {Meric}}, \bibinfo
  {author} {\bibfnamefont {C.}~\bibnamefont {Lee}}, \bibinfo {author}
  {\bibfnamefont {L.}~\bibnamefont {Wang}}, \bibinfo {author} {\bibfnamefont
  {S.}~\bibnamefont {Sorgenfrei}}, \bibinfo {author} {\bibfnamefont
  {K.}~\bibnamefont {Watanabe}}, \bibinfo {author} {\bibfnamefont
  {T.}~\bibnamefont {Taniguchi}}, \bibinfo {author} {\bibfnamefont
  {P.}~\bibnamefont {Kim}}, \bibinfo {author} {\bibfnamefont {K.~L.}\
  \bibnamefont {Shepard}}, \ and\ \bibinfo {author} {\bibfnamefont
  {J.}~\bibnamefont {Hone}},\ }\href {\doibase 10.1038/nnano.2010.172}
  {\bibfield  {journal} {\bibinfo  {journal} {Nature Nanotechnology}\ }\textbf
  {\bibinfo {volume} {5}},\ \bibinfo {pages} {722} (\bibinfo {year}
  {2010})}\BibitemShut {NoStop}%
\bibitem [{\citenamefont {Dean}\ \emph {et~al.}(2012)\citenamefont {Dean},
  \citenamefont {Young}, \citenamefont {Wang}, \citenamefont {Meric},
  \citenamefont {Lee}, \citenamefont {Watanabe}, \citenamefont {Taniguchi},
  \citenamefont {Shepard}, \citenamefont {Kim},\ and\ \citenamefont
  {Hone}}]{hBN_sub_Dean_2012}%
  \BibitemOpen
  \bibfield  {author} {\bibinfo {author} {\bibfnamefont {C.~R.}\ \bibnamefont
  {Dean}}, \bibinfo {author} {\bibfnamefont {A.}~\bibnamefont {Young}},
  \bibinfo {author} {\bibfnamefont {L.}~\bibnamefont {Wang}}, \bibinfo {author}
  {\bibfnamefont {I.}~\bibnamefont {Meric}}, \bibinfo {author} {\bibfnamefont
  {G.-H.}\ \bibnamefont {Lee}}, \bibinfo {author} {\bibfnamefont
  {K.}~\bibnamefont {Watanabe}}, \bibinfo {author} {\bibfnamefont
  {T.}~\bibnamefont {Taniguchi}}, \bibinfo {author} {\bibfnamefont
  {K.}~\bibnamefont {Shepard}}, \bibinfo {author} {\bibfnamefont
  {P.}~\bibnamefont {Kim}}, \ and\ \bibinfo {author} {\bibfnamefont
  {J.}~\bibnamefont {Hone}},\ }\href {\doibase
  https://doi.org/10.1016/j.ssc.2012.04.021} {\bibfield  {journal} {\bibinfo
  {journal} {Solid State Communications}\ }\textbf {\bibinfo {volume} {152}},\
  \bibinfo {pages} {1275 } (\bibinfo {year} {2012})}\BibitemShut {NoStop}%
\bibitem [{\citenamefont {Xue}\ \emph {et~al.}(2011)\citenamefont {Xue},
  \citenamefont {Sanchez-Yamagishi}, \citenamefont {Bulmash}, \citenamefont
  {Jacquod}, \citenamefont {Deshpande}, \citenamefont {Watanabe}, \citenamefont
  {Taniguchi}, \citenamefont {Jarillo-Herrero},\ and\ \citenamefont
  {LeRoy}}]{hBN_sub_Xue_2011}%
  \BibitemOpen
  \bibfield  {author} {\bibinfo {author} {\bibfnamefont {J.}~\bibnamefont
  {Xue}}, \bibinfo {author} {\bibfnamefont {J.}~\bibnamefont
  {Sanchez-Yamagishi}}, \bibinfo {author} {\bibfnamefont {D.}~\bibnamefont
  {Bulmash}}, \bibinfo {author} {\bibfnamefont {P.}~\bibnamefont {Jacquod}},
  \bibinfo {author} {\bibfnamefont {A.}~\bibnamefont {Deshpande}}, \bibinfo
  {author} {\bibfnamefont {K.}~\bibnamefont {Watanabe}}, \bibinfo {author}
  {\bibfnamefont {T.}~\bibnamefont {Taniguchi}}, \bibinfo {author}
  {\bibfnamefont {P.}~\bibnamefont {Jarillo-Herrero}}, \ and\ \bibinfo {author}
  {\bibfnamefont {B.~J.}\ \bibnamefont {LeRoy}},\ }\href {\doibase
  10.1038/nmat2968} {\bibfield  {journal} {\bibinfo  {journal} {Nature
  Materials}\ }\textbf {\bibinfo {volume} {10}},\ \bibinfo {pages} {282}
  (\bibinfo {year} {2011})}\BibitemShut {NoStop}%
\bibitem [{\citenamefont {Bistritzer}\ and\ \citenamefont
  {MacDonald}(2011)}]{tBLG_Bistritzer_2011}%
  \BibitemOpen
  \bibfield  {author} {\bibinfo {author} {\bibfnamefont {R.}~\bibnamefont
  {Bistritzer}}\ and\ \bibinfo {author} {\bibfnamefont {A.~H.}\ \bibnamefont
  {MacDonald}},\ }\href {\doibase 10.1073/pnas.1108174108} {\bibfield
  {journal} {\bibinfo  {journal} {Proceedings of the National Academy of
  Sciences}\ }\textbf {\bibinfo {volume} {108}},\ \bibinfo {pages} {12233}
  (\bibinfo {year} {2011})}\BibitemShut {NoStop}%
\bibitem [{\citenamefont {Cao}\ \emph {et~al.}(2018{\natexlab{a}})\citenamefont
  {Cao}, \citenamefont {Fatemi}, \citenamefont {Fang}, \citenamefont
  {Watanabe}, \citenamefont {Taniguchi}, \citenamefont {Kaxiras},\ and\
  \citenamefont {Jarillo-Herrero}}]{tBLG_SC_Cao_2018}%
  \BibitemOpen
  \bibfield  {author} {\bibinfo {author} {\bibfnamefont {Y.}~\bibnamefont
  {Cao}}, \bibinfo {author} {\bibfnamefont {V.}~\bibnamefont {Fatemi}},
  \bibinfo {author} {\bibfnamefont {S.}~\bibnamefont {Fang}}, \bibinfo {author}
  {\bibfnamefont {K.}~\bibnamefont {Watanabe}}, \bibinfo {author}
  {\bibfnamefont {T.}~\bibnamefont {Taniguchi}}, \bibinfo {author}
  {\bibfnamefont {E.}~\bibnamefont {Kaxiras}}, \ and\ \bibinfo {author}
  {\bibfnamefont {P.}~\bibnamefont {Jarillo-Herrero}},\ }\href {\doibase
  10.1038/nature26160} {\bibfield  {journal} {\bibinfo  {journal} {Nature}\
  }\textbf {\bibinfo {volume} {556}},\ \bibinfo {pages} {43} (\bibinfo {year}
  {2018}{\natexlab{a}})}\BibitemShut {NoStop}%
\bibitem [{\citenamefont {Cao}\ \emph {et~al.}(2018{\natexlab{b}})\citenamefont
  {Cao}, \citenamefont {Fatemi}, \citenamefont {Demir}, \citenamefont {ang},
  \citenamefont {Tomarken}, \citenamefont {Luo}, \citenamefont
  {Sanchez-Yamagishi}, \citenamefont {Watanabe}, \citenamefont {Taniguchi},
  \citenamefont {Kaxiras}, \citenamefont {Ashoori},\ and\ \citenamefont
  {Jarillo-Herrero}}]{tBLG_Insu_Cao_2018}%
  \BibitemOpen
  \bibfield  {author} {\bibinfo {author} {\bibfnamefont {Y.}~\bibnamefont
  {Cao}}, \bibinfo {author} {\bibfnamefont {V.}~\bibnamefont {Fatemi}},
  \bibinfo {author} {\bibfnamefont {A.}~\bibnamefont {Demir}}, \bibinfo
  {author} {\bibfnamefont {S.}~\bibnamefont {ang}}, \bibinfo {author}
  {\bibfnamefont {S.~L.}\ \bibnamefont {Tomarken}}, \bibinfo {author}
  {\bibfnamefont {J.~Y.}\ \bibnamefont {Luo}}, \bibinfo {author} {\bibfnamefont
  {J.~D.}\ \bibnamefont {Sanchez-Yamagishi}}, \bibinfo {author} {\bibfnamefont
  {K.}~\bibnamefont {Watanabe}}, \bibinfo {author} {\bibfnamefont
  {T.}~\bibnamefont {Taniguchi}}, \bibinfo {author} {\bibfnamefont
  {E.}~\bibnamefont {Kaxiras}}, \bibinfo {author} {\bibfnamefont {R.~C.}\
  \bibnamefont {Ashoori}}, \ and\ \bibinfo {author} {\bibfnamefont
  {P.}~\bibnamefont {Jarillo-Herrero}},\ }\href {\doibase 10.1038/nature26154}
  {\bibfield  {journal} {\bibinfo  {journal} {Nature}\ }\textbf {\bibinfo
  {volume} {556}},\ \bibinfo {pages} {80} (\bibinfo {year}
  {2018}{\natexlab{b}})}\BibitemShut {NoStop}%
\bibitem [{\citenamefont {Lu}\ \emph {et~al.}(2019)\citenamefont {Lu},
  \citenamefont {Stepanov}, \citenamefont {Yang}, \citenamefont {Xie},
  \citenamefont {Aamir}, \citenamefont {Das}, \citenamefont {Urgell},
  \citenamefont {Watanabe}, \citenamefont {Taniguchi}, \citenamefont {Zhang},
  \citenamefont {Bachtold}, \citenamefont {MacDonald},\ and\ \citenamefont
  {Efetov}}]{tBLG_Lu_2019}%
  \BibitemOpen
  \bibfield  {author} {\bibinfo {author} {\bibfnamefont {X.}~\bibnamefont
  {Lu}}, \bibinfo {author} {\bibfnamefont {P.}~\bibnamefont {Stepanov}},
  \bibinfo {author} {\bibfnamefont {W.}~\bibnamefont {Yang}}, \bibinfo {author}
  {\bibfnamefont {M.}~\bibnamefont {Xie}}, \bibinfo {author} {\bibfnamefont
  {M.~A.}\ \bibnamefont {Aamir}}, \bibinfo {author} {\bibfnamefont
  {I.}~\bibnamefont {Das}}, \bibinfo {author} {\bibfnamefont {C.}~\bibnamefont
  {Urgell}}, \bibinfo {author} {\bibfnamefont {K.}~\bibnamefont {Watanabe}},
  \bibinfo {author} {\bibfnamefont {T.}~\bibnamefont {Taniguchi}}, \bibinfo
  {author} {\bibfnamefont {G.}~\bibnamefont {Zhang}}, \bibinfo {author}
  {\bibfnamefont {A.}~\bibnamefont {Bachtold}}, \bibinfo {author}
  {\bibfnamefont {A.~H.}\ \bibnamefont {MacDonald}}, \ and\ \bibinfo {author}
  {\bibfnamefont {D.~K.}\ \bibnamefont {Efetov}},\ }\href {\doibase
  10.1038/s41586-019-1695-0} {\bibfield  {journal} {\bibinfo  {journal}
  {Nature}\ }\textbf {\bibinfo {volume} {574}},\ \bibinfo {pages} {653}
  (\bibinfo {year} {2019})}\BibitemShut {NoStop}%
\bibitem [{\citenamefont {Yankowitz}\ \emph {et~al.}(2019)\citenamefont
  {Yankowitz}, \citenamefont {Chen}, \citenamefont {Polshyn}, \citenamefont
  {Zhang}, \citenamefont {Watanabe}, \citenamefont {Taniguchi}, \citenamefont
  {Graf}, \citenamefont {Young},\ and\ \citenamefont
  {Dean}}]{tBLG_Yankowitz_2019}%
  \BibitemOpen
  \bibfield  {author} {\bibinfo {author} {\bibfnamefont {M.}~\bibnamefont
  {Yankowitz}}, \bibinfo {author} {\bibfnamefont {S.}~\bibnamefont {Chen}},
  \bibinfo {author} {\bibfnamefont {H.}~\bibnamefont {Polshyn}}, \bibinfo
  {author} {\bibfnamefont {Y.}~\bibnamefont {Zhang}}, \bibinfo {author}
  {\bibfnamefont {K.}~\bibnamefont {Watanabe}}, \bibinfo {author}
  {\bibfnamefont {T.}~\bibnamefont {Taniguchi}}, \bibinfo {author}
  {\bibfnamefont {D.}~\bibnamefont {Graf}}, \bibinfo {author} {\bibfnamefont
  {A.~F.}\ \bibnamefont {Young}}, \ and\ \bibinfo {author} {\bibfnamefont
  {C.~R.}\ \bibnamefont {Dean}},\ }\href {\doibase 10.1126/science.aav1910}
  {\bibfield  {journal} {\bibinfo  {journal} {Science}\ }\textbf {\bibinfo
  {volume} {363}},\ \bibinfo {pages} {1059} (\bibinfo {year}
  {2019})}\BibitemShut {NoStop}%
\bibitem [{\citenamefont {Sharpe}\ \emph {et~al.}(2019)\citenamefont {Sharpe},
  \citenamefont {Fox}, \citenamefont {Barnard}, \citenamefont {Finney},
  \citenamefont {Watanabe}, \citenamefont {Taniguchi}, \citenamefont
  {Kastner},\ and\ \citenamefont {Goldhaber-Gordon}}]{tBLG_Sharpe_2019}%
  \BibitemOpen
  \bibfield  {author} {\bibinfo {author} {\bibfnamefont {A.~L.}\ \bibnamefont
  {Sharpe}}, \bibinfo {author} {\bibfnamefont {E.~J.}\ \bibnamefont {Fox}},
  \bibinfo {author} {\bibfnamefont {A.~W.}\ \bibnamefont {Barnard}}, \bibinfo
  {author} {\bibfnamefont {J.}~\bibnamefont {Finney}}, \bibinfo {author}
  {\bibfnamefont {K.}~\bibnamefont {Watanabe}}, \bibinfo {author}
  {\bibfnamefont {T.}~\bibnamefont {Taniguchi}}, \bibinfo {author}
  {\bibfnamefont {M.~A.}\ \bibnamefont {Kastner}}, \ and\ \bibinfo {author}
  {\bibfnamefont {D.}~\bibnamefont {Goldhaber-Gordon}},\ }\href {\doibase
  10.1126/science.aaw3780} {\bibfield  {journal} {\bibinfo  {journal}
  {Science}\ }\textbf {\bibinfo {volume} {365}},\ \bibinfo {pages} {605}
  (\bibinfo {year} {2019})}\BibitemShut {NoStop}%
\bibitem [{\citenamefont {Serlin}\ \emph {et~al.}(2020)\citenamefont {Serlin},
  \citenamefont {Tschirhart}, \citenamefont {Polshyn}, \citenamefont {Zhang},
  \citenamefont {Zhu}, \citenamefont {Watanabe}, \citenamefont {Taniguchi},
  \citenamefont {Balents},\ and\ \citenamefont {Young}}]{tBLG_Serlin_2020}%
  \BibitemOpen
  \bibfield  {author} {\bibinfo {author} {\bibfnamefont {M.}~\bibnamefont
  {Serlin}}, \bibinfo {author} {\bibfnamefont {C.~L.}\ \bibnamefont
  {Tschirhart}}, \bibinfo {author} {\bibfnamefont {H.}~\bibnamefont {Polshyn}},
  \bibinfo {author} {\bibfnamefont {Y.}~\bibnamefont {Zhang}}, \bibinfo
  {author} {\bibfnamefont {J.}~\bibnamefont {Zhu}}, \bibinfo {author}
  {\bibfnamefont {K.}~\bibnamefont {Watanabe}}, \bibinfo {author}
  {\bibfnamefont {T.}~\bibnamefont {Taniguchi}}, \bibinfo {author}
  {\bibfnamefont {L.}~\bibnamefont {Balents}}, \ and\ \bibinfo {author}
  {\bibfnamefont {A.~F.}\ \bibnamefont {Young}},\ }\href {\doibase
  10.1126/science.aay5533} {\bibfield  {journal} {\bibinfo  {journal}
  {Science}\ }\textbf {\bibinfo {volume} {367}},\ \bibinfo {pages} {900}
  (\bibinfo {year} {2020})}\BibitemShut {NoStop}%
\bibitem [{\citenamefont {Guinea}\ and\ \citenamefont
  {Walet}(2018)}]{tBLG_Guinea_2018}%
  \BibitemOpen
  \bibfield  {author} {\bibinfo {author} {\bibfnamefont {F.}~\bibnamefont
  {Guinea}}\ and\ \bibinfo {author} {\bibfnamefont {N.~R.}\ \bibnamefont
  {Walet}},\ }\href {\doibase 10.1073/pnas.1810947115} {\bibfield  {journal}
  {\bibinfo  {journal} {Proceedings of the National Academy of Sciences}\
  }\textbf {\bibinfo {volume} {115}},\ \bibinfo {pages} {13174} (\bibinfo
  {year} {2018})}\BibitemShut {NoStop}%
\bibitem [{\citenamefont {Xie}\ and\ \citenamefont
  {MacDonald}(2020)}]{tBLG_MingHF}%
  \BibitemOpen
  \bibfield  {author} {\bibinfo {author} {\bibfnamefont {M.}~\bibnamefont
  {Xie}}\ and\ \bibinfo {author} {\bibfnamefont {A.~H.}\ \bibnamefont
  {MacDonald}},\ }\href {\doibase 10.1103/PhysRevLett.124.097601} {\bibfield
  {journal} {\bibinfo  {journal} {Phys. Rev. Lett.}\ }\textbf {\bibinfo
  {volume} {124}},\ \bibinfo {pages} {097601} (\bibinfo {year}
  {2020})}\BibitemShut {NoStop}%
\bibitem [{\citenamefont {Xie}\ and\ \citenamefont
  {MacDonald}()}]{tBLG_Ming_weakfield}%
  \BibitemOpen
  \bibfield  {author} {\bibinfo {author} {\bibfnamefont {M.}~\bibnamefont
  {Xie}}\ and\ \bibinfo {author} {\bibfnamefont {A.~H.}\ \bibnamefont
  {MacDonald}},\ }\href {\doibase arXiv:2010.07928} {\
  arXiv:2010.07928}\BibitemShut {NoStop}%
\bibitem [{\citenamefont {Po}\ \emph {et~al.}(2018)\citenamefont {Po},
  \citenamefont {Zou}, \citenamefont {Vishwanath},\ and\ \citenamefont
  {Senthil}}]{tBLG_Po_2018}%
  \BibitemOpen
  \bibfield  {author} {\bibinfo {author} {\bibfnamefont {H.~C.}\ \bibnamefont
  {Po}}, \bibinfo {author} {\bibfnamefont {L.}~\bibnamefont {Zou}}, \bibinfo
  {author} {\bibfnamefont {A.}~\bibnamefont {Vishwanath}}, \ and\ \bibinfo
  {author} {\bibfnamefont {T.}~\bibnamefont {Senthil}},\ }\href {\doibase
  10.1103/PhysRevX.8.031089} {\bibfield  {journal} {\bibinfo  {journal} {Phys.
  Rev. X}\ }\textbf {\bibinfo {volume} {8}},\ \bibinfo {pages} {031089}
  (\bibinfo {year} {2018})}\BibitemShut {NoStop}%
\bibitem [{\citenamefont {Bultinck}\ \emph {et~al.}()\citenamefont {Bultinck},
  \citenamefont {Khalaf}, \citenamefont {Liu}, \citenamefont {Chatterjee},
  \citenamefont {Vishwanath},\ and\ \citenamefont
  {Zaletel}}]{tBLG_Bultinck_2019}%
  \BibitemOpen
  \bibfield  {author} {\bibinfo {author} {\bibfnamefont {N.}~\bibnamefont
  {Bultinck}}, \bibinfo {author} {\bibfnamefont {E.}~\bibnamefont {Khalaf}},
  \bibinfo {author} {\bibfnamefont {S.}~\bibnamefont {Liu}}, \bibinfo {author}
  {\bibfnamefont {S.}~\bibnamefont {Chatterjee}}, \bibinfo {author}
  {\bibfnamefont {A.}~\bibnamefont {Vishwanath}}, \ and\ \bibinfo {author}
  {\bibfnamefont {M.~P.}\ \bibnamefont {Zaletel}},\ }\href {\doibase
  arXiv:1911.02045} {\ arXiv:1911.02045}\BibitemShut {NoStop}%
\bibitem [{\citenamefont {You}\ and\ \citenamefont
  {Vishwanath}(2019)}]{tBLG_You_2019}%
  \BibitemOpen
  \bibfield  {author} {\bibinfo {author} {\bibfnamefont {Y.-Z.}\ \bibnamefont
  {You}}\ and\ \bibinfo {author} {\bibfnamefont {A.}~\bibnamefont
  {Vishwanath}},\ }\href {\doibase 10.1038/s41535-019-0153-4} {\bibfield
  {journal} {\bibinfo  {journal} {npj Quantum Materials}\ }\textbf {\bibinfo
  {volume} {4}},\ \bibinfo {pages} {16} (\bibinfo {year} {2019})}\BibitemShut
  {NoStop}%
\bibitem [{\citenamefont {Damascelli}\ \emph {et~al.}(2003)\citenamefont
  {Damascelli}, \citenamefont {Hussain},\ and\ \citenamefont
  {Shen}}]{ARPES_Damascelli_2003}%
  \BibitemOpen
  \bibfield  {author} {\bibinfo {author} {\bibfnamefont {A.}~\bibnamefont
  {Damascelli}}, \bibinfo {author} {\bibfnamefont {Z.}~\bibnamefont {Hussain}},
  \ and\ \bibinfo {author} {\bibfnamefont {Z.-X.}\ \bibnamefont {Shen}},\
  }\href {\doibase 10.1103/RevModPhys.75.473} {\bibfield  {journal} {\bibinfo
  {journal} {Rev. Mod. Phys.}\ }\textbf {\bibinfo {volume} {75}},\ \bibinfo
  {pages} {473} (\bibinfo {year} {2003})}\BibitemShut {NoStop}%
\bibitem [{\citenamefont {Vishik}\ \emph {et~al.}(2010)\citenamefont {Vishik},
  \citenamefont {Lee}, \citenamefont {He}, \citenamefont {Hashimoto},
  \citenamefont {Hussain}, \citenamefont {Devereaux},\ and\ \citenamefont
  {Shen}}]{ARPES_Vishik_2010}%
  \BibitemOpen
  \bibfield  {author} {\bibinfo {author} {\bibfnamefont {I.~M.}\ \bibnamefont
  {Vishik}}, \bibinfo {author} {\bibfnamefont {W.~S.}\ \bibnamefont {Lee}},
  \bibinfo {author} {\bibfnamefont {R.-H.}\ \bibnamefont {He}}, \bibinfo
  {author} {\bibfnamefont {M.}~\bibnamefont {Hashimoto}}, \bibinfo {author}
  {\bibfnamefont {Z.}~\bibnamefont {Hussain}}, \bibinfo {author} {\bibfnamefont
  {T.~P.}\ \bibnamefont {Devereaux}}, \ and\ \bibinfo {author} {\bibfnamefont
  {Z.-X.}\ \bibnamefont {Shen}},\ }\href {\doibase
  10.1088/1367-2630/12/10/105008} {\bibfield  {journal} {\bibinfo  {journal}
  {New Journal of Physics}\ }\textbf {\bibinfo {volume} {12}},\ \bibinfo
  {pages} {105008} (\bibinfo {year} {2010})}\BibitemShut {NoStop}%
\bibitem [{\citenamefont {Lu}\ \emph {et~al.}(2012)\citenamefont {Lu},
  \citenamefont {Vishik}, \citenamefont {Yi}, \citenamefont {Chen},
  \citenamefont {Moore},\ and\ \citenamefont {Shen}}]{ARPES_Lu_2012}%
  \BibitemOpen
  \bibfield  {author} {\bibinfo {author} {\bibfnamefont {D.}~\bibnamefont
  {Lu}}, \bibinfo {author} {\bibfnamefont {I.~M.}\ \bibnamefont {Vishik}},
  \bibinfo {author} {\bibfnamefont {M.}~\bibnamefont {Yi}}, \bibinfo {author}
  {\bibfnamefont {Y.}~\bibnamefont {Chen}}, \bibinfo {author} {\bibfnamefont
  {R.~G.}\ \bibnamefont {Moore}}, \ and\ \bibinfo {author} {\bibfnamefont
  {Z.-X.}\ \bibnamefont {Shen}},\ }\href {\doibase
  10.1146/annurev-conmatphys-020911-125027} {\bibfield  {journal} {\bibinfo
  {journal} {Annual Review of Condensed Matter Physics}\ }\textbf {\bibinfo
  {volume} {3}},\ \bibinfo {pages} {129} (\bibinfo {year} {2012})}\BibitemShut
  {NoStop}%
\bibitem [{\citenamefont {Xia}\ \emph {et~al.}(2009)\citenamefont {Xia},
  \citenamefont {Qian}, \citenamefont {Hsieh}, \citenamefont {Wray},
  \citenamefont {Pal}, \citenamefont {Lin}, \citenamefont {Bansil},
  \citenamefont {Grauer}, \citenamefont {Hor}, \citenamefont {Cava},\ and\
  \citenamefont {Hasan}}]{ARPES_3DTI_Xia_2009}%
  \BibitemOpen
  \bibfield  {author} {\bibinfo {author} {\bibfnamefont {Y.}~\bibnamefont
  {Xia}}, \bibinfo {author} {\bibfnamefont {D.}~\bibnamefont {Qian}}, \bibinfo
  {author} {\bibfnamefont {D.}~\bibnamefont {Hsieh}}, \bibinfo {author}
  {\bibfnamefont {L.}~\bibnamefont {Wray}}, \bibinfo {author} {\bibfnamefont
  {A.}~\bibnamefont {Pal}}, \bibinfo {author} {\bibfnamefont {H.}~\bibnamefont
  {Lin}}, \bibinfo {author} {\bibfnamefont {A.}~\bibnamefont {Bansil}},
  \bibinfo {author} {\bibfnamefont {D.}~\bibnamefont {Grauer}}, \bibinfo
  {author} {\bibfnamefont {Y.~S.}\ \bibnamefont {Hor}}, \bibinfo {author}
  {\bibfnamefont {R.~J.}\ \bibnamefont {Cava}}, \ and\ \bibinfo {author}
  {\bibfnamefont {M.~Z.}\ \bibnamefont {Hasan}},\ }\href
  {https://doi.org/10.1038/nphys1274} {\bibfield  {journal} {\bibinfo
  {journal} {Nature Physics}\ }\textbf {\bibinfo {volume} {5}},\ \bibinfo
  {pages} {398} (\bibinfo {year} {2009})}\BibitemShut {NoStop}%
\bibitem [{\citenamefont {Ohta}\ \emph {et~al.}(2006)\citenamefont {Ohta},
  \citenamefont {Bostwick}, \citenamefont {Seyller}, \citenamefont {Horn},\
  and\ \citenamefont {Rotenberg}}]{ARPES_BLG_Ohta_2006}%
  \BibitemOpen
  \bibfield  {author} {\bibinfo {author} {\bibfnamefont {T.}~\bibnamefont
  {Ohta}}, \bibinfo {author} {\bibfnamefont {A.}~\bibnamefont {Bostwick}},
  \bibinfo {author} {\bibfnamefont {T.}~\bibnamefont {Seyller}}, \bibinfo
  {author} {\bibfnamefont {K.}~\bibnamefont {Horn}}, \ and\ \bibinfo {author}
  {\bibfnamefont {E.}~\bibnamefont {Rotenberg}},\ }\href {\doibase
  10.1126/science.1130681} {\bibfield  {journal} {\bibinfo  {journal}
  {Science}\ }\textbf {\bibinfo {volume} {313}},\ \bibinfo {pages} {951}
  (\bibinfo {year} {2006})}\BibitemShut {NoStop}%
\bibitem [{\citenamefont {Ohta}\ \emph {et~al.}(2007)\citenamefont {Ohta},
  \citenamefont {Bostwick}, \citenamefont {McChesney}, \citenamefont {Seyller},
  \citenamefont {Horn},\ and\ \citenamefont {Rotenberg}}]{ARPES_Ohta_2007}%
  \BibitemOpen
  \bibfield  {author} {\bibinfo {author} {\bibfnamefont {T.}~\bibnamefont
  {Ohta}}, \bibinfo {author} {\bibfnamefont {A.}~\bibnamefont {Bostwick}},
  \bibinfo {author} {\bibfnamefont {J.~L.}\ \bibnamefont {McChesney}}, \bibinfo
  {author} {\bibfnamefont {T.}~\bibnamefont {Seyller}}, \bibinfo {author}
  {\bibfnamefont {K.}~\bibnamefont {Horn}}, \ and\ \bibinfo {author}
  {\bibfnamefont {E.}~\bibnamefont {Rotenberg}},\ }\href {\doibase
  10.1103/PhysRevLett.98.206802} {\bibfield  {journal} {\bibinfo  {journal}
  {Phys. Rev. Lett.}\ }\textbf {\bibinfo {volume} {98}},\ \bibinfo {pages}
  {206802} (\bibinfo {year} {2007})}\BibitemShut {NoStop}%
\bibitem [{\citenamefont {Sprinkle}\ \emph {et~al.}(2009)\citenamefont
  {Sprinkle}, \citenamefont {Siegel}, \citenamefont {Hu}, \citenamefont
  {Hicks}, \citenamefont {Tejeda}, \citenamefont {Taleb-Ibrahimi},
  \citenamefont {Le~F\`evre}, \citenamefont {Bertran}, \citenamefont {Vizzini},
  \citenamefont {Enriquez}, \citenamefont {Chiang}, \citenamefont
  {Soukiassian}, \citenamefont {Berger}, \citenamefont {de~Heer}, \citenamefont
  {Lanzara},\ and\ \citenamefont {Conrad}}]{ARPES_Sprinkle_2009}%
  \BibitemOpen
  \bibfield  {author} {\bibinfo {author} {\bibfnamefont {M.}~\bibnamefont
  {Sprinkle}}, \bibinfo {author} {\bibfnamefont {D.}~\bibnamefont {Siegel}},
  \bibinfo {author} {\bibfnamefont {Y.}~\bibnamefont {Hu}}, \bibinfo {author}
  {\bibfnamefont {J.}~\bibnamefont {Hicks}}, \bibinfo {author} {\bibfnamefont
  {A.}~\bibnamefont {Tejeda}}, \bibinfo {author} {\bibfnamefont
  {A.}~\bibnamefont {Taleb-Ibrahimi}}, \bibinfo {author} {\bibfnamefont
  {P.}~\bibnamefont {Le~F\`evre}}, \bibinfo {author} {\bibfnamefont
  {F.}~\bibnamefont {Bertran}}, \bibinfo {author} {\bibfnamefont
  {S.}~\bibnamefont {Vizzini}}, \bibinfo {author} {\bibfnamefont
  {H.}~\bibnamefont {Enriquez}}, \bibinfo {author} {\bibfnamefont
  {S.}~\bibnamefont {Chiang}}, \bibinfo {author} {\bibfnamefont
  {P.}~\bibnamefont {Soukiassian}}, \bibinfo {author} {\bibfnamefont
  {C.}~\bibnamefont {Berger}}, \bibinfo {author} {\bibfnamefont {W.~A.}\
  \bibnamefont {de~Heer}}, \bibinfo {author} {\bibfnamefont {A.}~\bibnamefont
  {Lanzara}}, \ and\ \bibinfo {author} {\bibfnamefont {E.~H.}\ \bibnamefont
  {Conrad}},\ }\href {\doibase 10.1103/PhysRevLett.103.226803} {\bibfield
  {journal} {\bibinfo  {journal} {Phys. Rev. Lett.}\ }\textbf {\bibinfo
  {volume} {103}},\ \bibinfo {pages} {226803} (\bibinfo {year}
  {2009})}\BibitemShut {NoStop}%
\bibitem [{\citenamefont {Bostwick}\ \emph
  {et~al.}(2007{\natexlab{a}})\citenamefont {Bostwick}, \citenamefont {Ohta},
  \citenamefont {Seyller}, \citenamefont {Horn},\ and\ \citenamefont
  {Rotenberg}}]{ARPES_Bostwick_2007}%
  \BibitemOpen
  \bibfield  {author} {\bibinfo {author} {\bibfnamefont {A.}~\bibnamefont
  {Bostwick}}, \bibinfo {author} {\bibfnamefont {T.}~\bibnamefont {Ohta}},
  \bibinfo {author} {\bibfnamefont {T.}~\bibnamefont {Seyller}}, \bibinfo
  {author} {\bibfnamefont {K.}~\bibnamefont {Horn}}, \ and\ \bibinfo {author}
  {\bibfnamefont {E.}~\bibnamefont {Rotenberg}},\ }\href {\doibase
  10.1038/nphys477} {\bibfield  {journal} {\bibinfo  {journal} {Nature
  Physics}\ }\textbf {\bibinfo {volume} {3}},\ \bibinfo {pages} {36} (\bibinfo
  {year} {2007}{\natexlab{a}})}\BibitemShut {NoStop}%
\bibitem [{\citenamefont {Bostwick}\ \emph
  {et~al.}(2007{\natexlab{b}})\citenamefont {Bostwick}, \citenamefont {Ohta},
  \citenamefont {McChesney}, \citenamefont {Emtsev}, \citenamefont {Seyller},
  \citenamefont {Horn},\ and\ \citenamefont {Rotenberg}}]{Bostwick_2007}%
  \BibitemOpen
  \bibfield  {author} {\bibinfo {author} {\bibfnamefont {A.}~\bibnamefont
  {Bostwick}}, \bibinfo {author} {\bibfnamefont {T.}~\bibnamefont {Ohta}},
  \bibinfo {author} {\bibfnamefont {J.~L.}\ \bibnamefont {McChesney}}, \bibinfo
  {author} {\bibfnamefont {K.~V.}\ \bibnamefont {Emtsev}}, \bibinfo {author}
  {\bibfnamefont {T.}~\bibnamefont {Seyller}}, \bibinfo {author} {\bibfnamefont
  {K.}~\bibnamefont {Horn}}, \ and\ \bibinfo {author} {\bibfnamefont
  {E.}~\bibnamefont {Rotenberg}},\ }\href {\doibase 10.1088/1367-2630/9/10/385}
  {\bibfield  {journal} {\bibinfo  {journal} {New Journal of Physics}\ }\textbf
  {\bibinfo {volume} {9}},\ \bibinfo {pages} {385} (\bibinfo {year}
  {2007}{\natexlab{b}})}\BibitemShut {NoStop}%
\bibitem [{\citenamefont {Zhou}\ \emph {et~al.}(2007)\citenamefont {Zhou},
  \citenamefont {Gweon}, \citenamefont {Fedorov}, \citenamefont {First},
  \citenamefont {de~Heer}, \citenamefont {Lee}, \citenamefont {Guinea},
  \citenamefont {Castro~Neto},\ and\ \citenamefont
  {Lanzara}}]{ARPES_Zhou_2007}%
  \BibitemOpen
  \bibfield  {author} {\bibinfo {author} {\bibfnamefont {S.~Y.}\ \bibnamefont
  {Zhou}}, \bibinfo {author} {\bibfnamefont {G.-H.}\ \bibnamefont {Gweon}},
  \bibinfo {author} {\bibfnamefont {A.~V.}\ \bibnamefont {Fedorov}}, \bibinfo
  {author} {\bibfnamefont {P.~N.}\ \bibnamefont {First}}, \bibinfo {author}
  {\bibfnamefont {W.~A.}\ \bibnamefont {de~Heer}}, \bibinfo {author}
  {\bibfnamefont {D.-H.}\ \bibnamefont {Lee}}, \bibinfo {author} {\bibfnamefont
  {F.}~\bibnamefont {Guinea}}, \bibinfo {author} {\bibfnamefont {A.~H.}\
  \bibnamefont {Castro~Neto}}, \ and\ \bibinfo {author} {\bibfnamefont
  {A.}~\bibnamefont {Lanzara}},\ }\href {\doibase 10.1038/nmat2003} {\bibfield
  {journal} {\bibinfo  {journal} {Nature Materials}\ }\textbf {\bibinfo
  {volume} {6}},\ \bibinfo {pages} {770} (\bibinfo {year} {2007})}\BibitemShut
  {NoStop}%
\bibitem [{\citenamefont {Bostwick}\ \emph {et~al.}()\citenamefont {Bostwick},
  \citenamefont {Ohta}, \citenamefont {Seyller}, \citenamefont {Horn},\ and\
  \citenamefont {Rotenberg}}]{ARPES_Bostwick_2006}%
  \BibitemOpen
  \bibfield  {author} {\bibinfo {author} {\bibfnamefont {A.}~\bibnamefont
  {Bostwick}}, \bibinfo {author} {\bibfnamefont {T.}~\bibnamefont {Ohta}},
  \bibinfo {author} {\bibfnamefont {T.}~\bibnamefont {Seyller}}, \bibinfo
  {author} {\bibfnamefont {K.}~\bibnamefont {Horn}}, \ and\ \bibinfo {author}
  {\bibfnamefont {E.}~\bibnamefont {Rotenberg}},\ }\href {\doibase
  arxiv.org/abs/cond-mat/0609660v1} {\
  arxiv.org/abs/cond-mat/0609660v1}\BibitemShut {NoStop}%
\bibitem [{\citenamefont {Coletti}\ \emph {et~al.}(2013)\citenamefont
  {Coletti}, \citenamefont {Forti}, \citenamefont {Principi}, \citenamefont
  {Emtsev}, \citenamefont {Zakharov}, \citenamefont {Daniels}, \citenamefont
  {Daas}, \citenamefont {Chandrashekhar}, \citenamefont {Ouisse}, \citenamefont
  {Chaussende}, \citenamefont {MacDonald}, \citenamefont {Polini},\ and\
  \citenamefont {Starke}}]{ARPES_Coletti_2013}%
  \BibitemOpen
  \bibfield  {author} {\bibinfo {author} {\bibfnamefont {C.}~\bibnamefont
  {Coletti}}, \bibinfo {author} {\bibfnamefont {S.}~\bibnamefont {Forti}},
  \bibinfo {author} {\bibfnamefont {A.}~\bibnamefont {Principi}}, \bibinfo
  {author} {\bibfnamefont {K.~V.}\ \bibnamefont {Emtsev}}, \bibinfo {author}
  {\bibfnamefont {A.~A.}\ \bibnamefont {Zakharov}}, \bibinfo {author}
  {\bibfnamefont {K.~M.}\ \bibnamefont {Daniels}}, \bibinfo {author}
  {\bibfnamefont {B.~K.}\ \bibnamefont {Daas}}, \bibinfo {author}
  {\bibfnamefont {M.~V.~S.}\ \bibnamefont {Chandrashekhar}}, \bibinfo {author}
  {\bibfnamefont {T.}~\bibnamefont {Ouisse}}, \bibinfo {author} {\bibfnamefont
  {D.}~\bibnamefont {Chaussende}}, \bibinfo {author} {\bibfnamefont {A.~H.}\
  \bibnamefont {MacDonald}}, \bibinfo {author} {\bibfnamefont {M.}~\bibnamefont
  {Polini}}, \ and\ \bibinfo {author} {\bibfnamefont {U.}~\bibnamefont
  {Starke}},\ }\href {\doibase 10.1103/PhysRevB.88.155439} {\bibfield
  {journal} {\bibinfo  {journal} {Phys. Rev. B}\ }\textbf {\bibinfo {volume}
  {88}},\ \bibinfo {pages} {155439} (\bibinfo {year} {2013})}\BibitemShut
  {NoStop}%
\bibitem [{\citenamefont {Hwang}\ \emph {et~al.}(2011)\citenamefont {Hwang},
  \citenamefont {Park}, \citenamefont {Siegel}, \citenamefont {Fedorov},
  \citenamefont {Louie},\ and\ \citenamefont {Lanzara}}]{ARPES_G_Hwang_2011}%
  \BibitemOpen
  \bibfield  {author} {\bibinfo {author} {\bibfnamefont {C.}~\bibnamefont
  {Hwang}}, \bibinfo {author} {\bibfnamefont {C.-H.}\ \bibnamefont {Park}},
  \bibinfo {author} {\bibfnamefont {D.~A.}\ \bibnamefont {Siegel}}, \bibinfo
  {author} {\bibfnamefont {A.~V.}\ \bibnamefont {Fedorov}}, \bibinfo {author}
  {\bibfnamefont {S.~G.}\ \bibnamefont {Louie}}, \ and\ \bibinfo {author}
  {\bibfnamefont {A.}~\bibnamefont {Lanzara}},\ }\href {\doibase
  10.1103/PhysRevB.84.125422} {\bibfield  {journal} {\bibinfo  {journal} {Phys.
  Rev. B}\ }\textbf {\bibinfo {volume} {84}},\ \bibinfo {pages} {125422}
  (\bibinfo {year} {2011})}\BibitemShut {NoStop}%
\bibitem [{\citenamefont {Gierz}\ \emph {et~al.}(2011)\citenamefont {Gierz},
  \citenamefont {Henk}, \citenamefont {H\"ochst}, \citenamefont {Ast},\ and\
  \citenamefont {Kern}}]{ARPES_G_Gierz_2011}%
  \BibitemOpen
  \bibfield  {author} {\bibinfo {author} {\bibfnamefont {I.}~\bibnamefont
  {Gierz}}, \bibinfo {author} {\bibfnamefont {J.}~\bibnamefont {Henk}},
  \bibinfo {author} {\bibfnamefont {H.}~\bibnamefont {H\"ochst}}, \bibinfo
  {author} {\bibfnamefont {C.~R.}\ \bibnamefont {Ast}}, \ and\ \bibinfo
  {author} {\bibfnamefont {K.}~\bibnamefont {Kern}},\ }\href {\doibase
  10.1103/PhysRevB.83.121408} {\bibfield  {journal} {\bibinfo  {journal} {Phys.
  Rev. B}\ }\textbf {\bibinfo {volume} {83}},\ \bibinfo {pages} {121408(R)}
  (\bibinfo {year} {2011})}\BibitemShut {NoStop}%
\bibitem [{\citenamefont {Ohta}\ \emph
  {et~al.}(2012{\natexlab{a}})\citenamefont {Ohta}, \citenamefont {Robinson},
  \citenamefont {Feibelman}, \citenamefont {Bostwick}, \citenamefont
  {Rotenberg},\ and\ \citenamefont {Beechem}}]{ARPES_TBGonSiC_Ohta}%
  \BibitemOpen
  \bibfield  {author} {\bibinfo {author} {\bibfnamefont {T.}~\bibnamefont
  {Ohta}}, \bibinfo {author} {\bibfnamefont {J.~T.}\ \bibnamefont {Robinson}},
  \bibinfo {author} {\bibfnamefont {P.~J.}\ \bibnamefont {Feibelman}}, \bibinfo
  {author} {\bibfnamefont {A.}~\bibnamefont {Bostwick}}, \bibinfo {author}
  {\bibfnamefont {E.}~\bibnamefont {Rotenberg}}, \ and\ \bibinfo {author}
  {\bibfnamefont {T.~E.}\ \bibnamefont {Beechem}},\ }\href {\doibase
  10.1103/PhysRevLett.109.186807} {\bibfield  {journal} {\bibinfo  {journal}
  {Phys. Rev. Lett.}\ }\textbf {\bibinfo {volume} {109}},\ \bibinfo {pages}
  {186807} (\bibinfo {year} {2012}{\natexlab{a}})}\BibitemShut {NoStop}%
\bibitem [{\citenamefont {Kim}\ \emph {et~al.}(2013)\citenamefont {Kim},
  \citenamefont {Walter}, \citenamefont {Moreschini}, \citenamefont {Seyller},
  \citenamefont {Horn}, \citenamefont {Rotenberg},\ and\ \citenamefont
  {Bostwick}}]{ARPES_BLG_Kim}%
  \BibitemOpen
  \bibfield  {author} {\bibinfo {author} {\bibfnamefont {K.~S.}\ \bibnamefont
  {Kim}}, \bibinfo {author} {\bibfnamefont {A.~L.}\ \bibnamefont {Walter}},
  \bibinfo {author} {\bibfnamefont {L.}~\bibnamefont {Moreschini}}, \bibinfo
  {author} {\bibfnamefont {T.}~\bibnamefont {Seyller}}, \bibinfo {author}
  {\bibfnamefont {K.}~\bibnamefont {Horn}}, \bibinfo {author} {\bibfnamefont
  {E.}~\bibnamefont {Rotenberg}}, \ and\ \bibinfo {author} {\bibfnamefont
  {A.}~\bibnamefont {Bostwick}},\ }\href {\doibase 10.1038/nmat3717} {\bibfield
   {journal} {\bibinfo  {journal} {Nature Materials}\ }\textbf {\bibinfo
  {volume} {12}},\ \bibinfo {pages} {887} (\bibinfo {year} {2013})}\BibitemShut
  {NoStop}%
\bibitem [{\citenamefont {Mo}(2017)}]{ARPES_2017}%
  \BibitemOpen
  \bibfield  {author} {\bibinfo {author} {\bibfnamefont {S.-K.}\ \bibnamefont
  {Mo}},\ }\href {https://doi.org/10.1186/s40580-017-0100-7} {\bibfield
  {journal} {\bibinfo  {journal} {Nano Convergence}\ }\textbf {\bibinfo
  {volume} {4}},\ \bibinfo {pages} {6} (\bibinfo {year} {2017})}\BibitemShut
  {NoStop}%
\bibitem [{\citenamefont {Dudin}\ \emph {et~al.}(2010)\citenamefont {Dudin},
  \citenamefont {Lacovig}, \citenamefont {Fava}, \citenamefont {Nicolini},
  \citenamefont {Bianco}, \citenamefont {Cautero},\ and\ \citenamefont
  {Barinov}}]{nanoARPES_Dudin_2010}%
  \BibitemOpen
  \bibfield  {author} {\bibinfo {author} {\bibfnamefont {P.}~\bibnamefont
  {Dudin}}, \bibinfo {author} {\bibfnamefont {P.}~\bibnamefont {Lacovig}},
  \bibinfo {author} {\bibfnamefont {C.}~\bibnamefont {Fava}}, \bibinfo {author}
  {\bibfnamefont {E.}~\bibnamefont {Nicolini}}, \bibinfo {author}
  {\bibfnamefont {A.}~\bibnamefont {Bianco}}, \bibinfo {author} {\bibfnamefont
  {G.}~\bibnamefont {Cautero}}, \ and\ \bibinfo {author} {\bibfnamefont
  {A.}~\bibnamefont {Barinov}},\ }\href {\doibase 10.1107/S0909049510013993}
  {\bibfield  {journal} {\bibinfo  {journal} {Journal of Synchrotron
  Radiation}\ }\textbf {\bibinfo {volume} {17}},\ \bibinfo {pages} {445}
  (\bibinfo {year} {2010})}\BibitemShut {NoStop}%
\bibitem [{\citenamefont {Bostwick}\ \emph {et~al.}(2012)\citenamefont
  {Bostwick}, \citenamefont {Rotenberg}, \citenamefont {Avila},\ and\
  \citenamefont {Asensio}}]{nanoARPES_Bostwick_2012}%
  \BibitemOpen
  \bibfield  {author} {\bibinfo {author} {\bibfnamefont {A.}~\bibnamefont
  {Bostwick}}, \bibinfo {author} {\bibfnamefont {E.}~\bibnamefont {Rotenberg}},
  \bibinfo {author} {\bibfnamefont {J.}~\bibnamefont {Avila}}, \ and\ \bibinfo
  {author} {\bibfnamefont {M.~C.}\ \bibnamefont {Asensio}},\ }\href {\doibase
  10.1080/08940886.2012.720162} {\bibfield  {journal} {\bibinfo  {journal}
  {Synchrotron Radiation News}\ }\textbf {\bibinfo {volume} {25}},\ \bibinfo
  {pages} {19} (\bibinfo {year} {2012})}\BibitemShut {NoStop}%
\bibitem [{\citenamefont {Avila}\ \emph
  {et~al.}(2013{\natexlab{a}})\citenamefont {Avila}, \citenamefont
  {Razado-Colambo}, \citenamefont {Lorcy}, \citenamefont {Giorgetta},
  \citenamefont {Polack},\ and\ \citenamefont
  {Asensio}}]{nanoARPES_Avila_2013}%
  \BibitemOpen
  \bibfield  {author} {\bibinfo {author} {\bibfnamefont {J.}~\bibnamefont
  {Avila}}, \bibinfo {author} {\bibfnamefont {I.}~\bibnamefont
  {Razado-Colambo}}, \bibinfo {author} {\bibfnamefont {S.}~\bibnamefont
  {Lorcy}}, \bibinfo {author} {\bibfnamefont {J.-L.}\ \bibnamefont
  {Giorgetta}}, \bibinfo {author} {\bibfnamefont {F.}~\bibnamefont {Polack}}, \
  and\ \bibinfo {author} {\bibfnamefont {M.~C.}\ \bibnamefont {Asensio}},\
  }\href {\doibase 10.1088/1742-6596/425/13/132013} {\bibfield  {journal}
  {\bibinfo  {journal} {Journal of Physics: Conference Series}\ }\textbf
  {\bibinfo {volume} {425}},\ \bibinfo {pages} {132013} (\bibinfo {year}
  {2013}{\natexlab{a}})}\BibitemShut {NoStop}%
\bibitem [{\citenamefont {Avila}\ \emph
  {et~al.}(2013{\natexlab{b}})\citenamefont {Avila}, \citenamefont
  {Razado-Colambo}, \citenamefont {Lorcy}, \citenamefont {Lagarde},
  \citenamefont {Giorgetta}, \citenamefont {Polack},\ and\ \citenamefont
  {Asensio}}]{nanoARPES_Avila_2013_2}%
  \BibitemOpen
  \bibfield  {author} {\bibinfo {author} {\bibfnamefont {J.}~\bibnamefont
  {Avila}}, \bibinfo {author} {\bibfnamefont {I.}~\bibnamefont
  {Razado-Colambo}}, \bibinfo {author} {\bibfnamefont {S.}~\bibnamefont
  {Lorcy}}, \bibinfo {author} {\bibfnamefont {B.}~\bibnamefont {Lagarde}},
  \bibinfo {author} {\bibfnamefont {J.-L.}\ \bibnamefont {Giorgetta}}, \bibinfo
  {author} {\bibfnamefont {F.}~\bibnamefont {Polack}}, \ and\ \bibinfo {author}
  {\bibfnamefont {M.~C.}\ \bibnamefont {Asensio}},\ }\href {\doibase
  10.1088/1742-6596/425/19/192023} {\bibfield  {journal} {\bibinfo  {journal}
  {Journal of Physics: Conference Series}\ }\textbf {\bibinfo {volume} {425}},\
  \bibinfo {pages} {192023} (\bibinfo {year} {2013}{\natexlab{b}})}\BibitemShut
  {NoStop}%
\bibitem [{\citenamefont {Avila}\ \emph
  {et~al.}(2013{\natexlab{c}})\citenamefont {Avila}, \citenamefont {Razado},
  \citenamefont {Lorcy}, \citenamefont {Fleurier}, \citenamefont {Pichonat},
  \citenamefont {Vignaud}, \citenamefont {Wallart},\ and\ \citenamefont
  {Asensio}}]{nanoARPES_Avila_2013_3}%
  \BibitemOpen
  \bibfield  {author} {\bibinfo {author} {\bibfnamefont {J.}~\bibnamefont
  {Avila}}, \bibinfo {author} {\bibfnamefont {I.}~\bibnamefont {Razado}},
  \bibinfo {author} {\bibfnamefont {S.}~\bibnamefont {Lorcy}}, \bibinfo
  {author} {\bibfnamefont {R.}~\bibnamefont {Fleurier}}, \bibinfo {author}
  {\bibfnamefont {E.}~\bibnamefont {Pichonat}}, \bibinfo {author}
  {\bibfnamefont {D.}~\bibnamefont {Vignaud}}, \bibinfo {author} {\bibfnamefont
  {X.}~\bibnamefont {Wallart}}, \ and\ \bibinfo {author} {\bibfnamefont
  {M.~C.}\ \bibnamefont {Asensio}},\ }\href {https://doi.org/10.1038/srep02439}
  {\bibfield  {journal} {\bibinfo  {journal} {Scientific Reports}\ }\textbf
  {\bibinfo {volume} {3}},\ \bibinfo {pages} {2439} (\bibinfo {year}
  {2013}{\natexlab{c}})}\BibitemShut {NoStop}%
\bibitem [{\citenamefont {Avila}\ and\ \citenamefont
  {Asensio}(2014)}]{nanoARPES_Avila_2014}%
  \BibitemOpen
  \bibfield  {author} {\bibinfo {author} {\bibfnamefont {J.}~\bibnamefont
  {Avila}}\ and\ \bibinfo {author} {\bibfnamefont {M.~C.}\ \bibnamefont
  {Asensio}},\ }\href {\doibase 10.1080/08940886.2014.889549} {\bibfield
  {journal} {\bibinfo  {journal} {Synchrotron Radiation News}\ }\textbf
  {\bibinfo {volume} {27}},\ \bibinfo {pages} {24} (\bibinfo {year}
  {2014})}\BibitemShut {NoStop}%
\bibitem [{\citenamefont {Coy~Diaz}\ \emph {et~al.}(2015)\citenamefont
  {Coy~Diaz}, \citenamefont {Avila}, \citenamefont {Chen}, \citenamefont
  {Addou}, \citenamefont {Asensio},\ and\ \citenamefont
  {Batzill}}]{nanoARPES_Coy_2015}%
  \BibitemOpen
  \bibfield  {author} {\bibinfo {author} {\bibfnamefont {H.}~\bibnamefont
  {Coy~Diaz}}, \bibinfo {author} {\bibfnamefont {J.}~\bibnamefont {Avila}},
  \bibinfo {author} {\bibfnamefont {C.}~\bibnamefont {Chen}}, \bibinfo {author}
  {\bibfnamefont {R.}~\bibnamefont {Addou}}, \bibinfo {author} {\bibfnamefont
  {M.~C.}\ \bibnamefont {Asensio}}, \ and\ \bibinfo {author} {\bibfnamefont
  {M.}~\bibnamefont {Batzill}},\ }\href {\doibase 10.1021/nl504167y} {\bibfield
   {journal} {\bibinfo  {journal} {Nano Letters}\ }\textbf {\bibinfo {volume}
  {15}},\ \bibinfo {pages} {1135} (\bibinfo {year} {2015})}\BibitemShut
  {NoStop}%
\bibitem [{\citenamefont {Joucken}\ \emph
  {et~al.}(2019{\natexlab{a}})\citenamefont {Joucken}, \citenamefont
  {Quezada-L\'opez}, \citenamefont {Avila}, \citenamefont {Chen}, \citenamefont
  {Davenport}, \citenamefont {Chen}, \citenamefont {Watanabe}, \citenamefont
  {Taniguchi}, \citenamefont {Asensio},\ and\ \citenamefont
  {Velasco}}]{nanoARPES_BLG_Joucken}%
  \BibitemOpen
  \bibfield  {author} {\bibinfo {author} {\bibfnamefont {F.}~\bibnamefont
  {Joucken}}, \bibinfo {author} {\bibfnamefont {E.~A.}\ \bibnamefont
  {Quezada-L\'opez}}, \bibinfo {author} {\bibfnamefont {J.}~\bibnamefont
  {Avila}}, \bibinfo {author} {\bibfnamefont {C.}~\bibnamefont {Chen}},
  \bibinfo {author} {\bibfnamefont {J.~L.}\ \bibnamefont {Davenport}}, \bibinfo
  {author} {\bibfnamefont {H.}~\bibnamefont {Chen}}, \bibinfo {author}
  {\bibfnamefont {K.}~\bibnamefont {Watanabe}}, \bibinfo {author}
  {\bibfnamefont {T.}~\bibnamefont {Taniguchi}}, \bibinfo {author}
  {\bibfnamefont {M.~C.}\ \bibnamefont {Asensio}}, \ and\ \bibinfo {author}
  {\bibfnamefont {J.}~\bibnamefont {Velasco}},\ }\href {\doibase
  10.1103/PhysRevB.99.161406} {\bibfield  {journal} {\bibinfo  {journal} {Phys.
  Rev. B}\ }\textbf {\bibinfo {volume} {99}},\ \bibinfo {pages} {161406(R)}
  (\bibinfo {year} {2019}{\natexlab{a}})}\BibitemShut {NoStop}%
\bibitem [{\citenamefont {Chen}\ \emph {et~al.}(2018)\citenamefont {Chen},
  \citenamefont {Avila}, \citenamefont {Wang}, \citenamefont {Wang},
  \citenamefont {Mucha-Kruczyński}, \citenamefont {Shen}, \citenamefont
  {Yang}, \citenamefont {Nosarzewski}, \citenamefont {Devereaux}, \citenamefont
  {Zhang},\ and\ \citenamefont {Asensio}}]{nanoARPES_G_Chen}%
  \BibitemOpen
  \bibfield  {author} {\bibinfo {author} {\bibfnamefont {C.}~\bibnamefont
  {Chen}}, \bibinfo {author} {\bibfnamefont {J.}~\bibnamefont {Avila}},
  \bibinfo {author} {\bibfnamefont {S.}~\bibnamefont {Wang}}, \bibinfo {author}
  {\bibfnamefont {Y.}~\bibnamefont {Wang}}, \bibinfo {author} {\bibfnamefont
  {M.}~\bibnamefont {Mucha-Kruczyński}}, \bibinfo {author} {\bibfnamefont
  {C.}~\bibnamefont {Shen}}, \bibinfo {author} {\bibfnamefont {R.}~\bibnamefont
  {Yang}}, \bibinfo {author} {\bibfnamefont {B.}~\bibnamefont {Nosarzewski}},
  \bibinfo {author} {\bibfnamefont {T.~P.}\ \bibnamefont {Devereaux}}, \bibinfo
  {author} {\bibfnamefont {G.}~\bibnamefont {Zhang}}, \ and\ \bibinfo {author}
  {\bibfnamefont {M.~C.}\ \bibnamefont {Asensio}},\ }\href {\doibase
  10.1021/acs.nanolett.7b04604} {\bibfield  {journal} {\bibinfo  {journal}
  {Nano Lett.}\ }\textbf {\bibinfo {volume} {18}},\ \bibinfo {pages} {1082}
  (\bibinfo {year} {2018})}\BibitemShut {NoStop}%
\bibitem [{\citenamefont {Katoch}\ \emph {et~al.}(2018)\citenamefont {Katoch},
  \citenamefont {Ulstrup}, \citenamefont {Koch}, \citenamefont {Moser},
  \citenamefont {McCreary}, \citenamefont {Singh}, \citenamefont {Xu},
  \citenamefont {Jonker}, \citenamefont {Kawakami}, \citenamefont {Bostwick},
  \citenamefont {Rotenberg},\ and\ \citenamefont
  {Jozwiak}}]{nanoARPES_WS2_Katoch}%
  \BibitemOpen
  \bibfield  {author} {\bibinfo {author} {\bibfnamefont {J.}~\bibnamefont
  {Katoch}}, \bibinfo {author} {\bibfnamefont {S.}~\bibnamefont {Ulstrup}},
  \bibinfo {author} {\bibfnamefont {R.~J.}\ \bibnamefont {Koch}}, \bibinfo
  {author} {\bibfnamefont {S.}~\bibnamefont {Moser}}, \bibinfo {author}
  {\bibfnamefont {K.~M.}\ \bibnamefont {McCreary}}, \bibinfo {author}
  {\bibfnamefont {S.}~\bibnamefont {Singh}}, \bibinfo {author} {\bibfnamefont
  {J.}~\bibnamefont {Xu}}, \bibinfo {author} {\bibfnamefont {B.~T.}\
  \bibnamefont {Jonker}}, \bibinfo {author} {\bibfnamefont {R.~K.}\
  \bibnamefont {Kawakami}}, \bibinfo {author} {\bibfnamefont {A.}~\bibnamefont
  {Bostwick}}, \bibinfo {author} {\bibfnamefont {E.}~\bibnamefont {Rotenberg}},
  \ and\ \bibinfo {author} {\bibfnamefont {C.}~\bibnamefont {Jozwiak}},\ }\href
  {\doibase 10.1038/s41567-017-0033-4} {\bibfield  {journal} {\bibinfo
  {journal} {Nature Physics}\ }\textbf {\bibinfo {volume} {14}},\ \bibinfo
  {pages} {355} (\bibinfo {year} {2018})}\BibitemShut {NoStop}%
\bibitem [{\citenamefont {Joucken}\ \emph
  {et~al.}(2019{\natexlab{b}})\citenamefont {Joucken}, \citenamefont {Avila},
  \citenamefont {Ge}, \citenamefont {Quezada-Lopez}, \citenamefont {Yi},
  \citenamefont {Le~Goff}, \citenamefont {Baudin}, \citenamefont {Davenport},
  \citenamefont {Watanabe}, \citenamefont {Taniguchi}, \citenamefont
  {Asensio},\ and\ \citenamefont {Velasco}}]{nanoARPES_Joucken}%
  \BibitemOpen
  \bibfield  {author} {\bibinfo {author} {\bibfnamefont {F.}~\bibnamefont
  {Joucken}}, \bibinfo {author} {\bibfnamefont {J.}~\bibnamefont {Avila}},
  \bibinfo {author} {\bibfnamefont {Z.}~\bibnamefont {Ge}}, \bibinfo {author}
  {\bibfnamefont {E.~A.}\ \bibnamefont {Quezada-Lopez}}, \bibinfo {author}
  {\bibfnamefont {H.}~\bibnamefont {Yi}}, \bibinfo {author} {\bibfnamefont
  {R.}~\bibnamefont {Le~Goff}}, \bibinfo {author} {\bibfnamefont
  {E.}~\bibnamefont {Baudin}}, \bibinfo {author} {\bibfnamefont {J.~L.}\
  \bibnamefont {Davenport}}, \bibinfo {author} {\bibfnamefont {K.}~\bibnamefont
  {Watanabe}}, \bibinfo {author} {\bibfnamefont {T.}~\bibnamefont {Taniguchi}},
  \bibinfo {author} {\bibfnamefont {M.~C.}\ \bibnamefont {Asensio}}, \ and\
  \bibinfo {author} {\bibfnamefont {J.}~\bibnamefont {Velasco}},\ }\href
  {\doibase 10.1021/acs.nanolett.9b00649} {\bibfield  {journal} {\bibinfo
  {journal} {Nano Letters}\ }\textbf {\bibinfo {volume} {19}},\ \bibinfo
  {pages} {2682} (\bibinfo {year} {2019}{\natexlab{b}})}\BibitemShut {NoStop}%
\bibitem [{\citenamefont {Nguyen}\ \emph {et~al.}(2019)\citenamefont {Nguyen},
  \citenamefont {Teutsch}, \citenamefont {Wilson}, \citenamefont {Kahn},
  \citenamefont {Xia}, \citenamefont {Graham}, \citenamefont {Kandyba},
  \citenamefont {Giampietri}, \citenamefont {Barinov}, \citenamefont
  {Constantinescu}, \citenamefont {Yeung}, \citenamefont {Hine}, \citenamefont
  {Xu}, \citenamefont {Cobden},\ and\ \citenamefont
  {Wilson}}]{nanoARPES_Nguyen}%
  \BibitemOpen
  \bibfield  {author} {\bibinfo {author} {\bibfnamefont {P.~V.}\ \bibnamefont
  {Nguyen}}, \bibinfo {author} {\bibfnamefont {N.~C.}\ \bibnamefont {Teutsch}},
  \bibinfo {author} {\bibfnamefont {N.~P.}\ \bibnamefont {Wilson}}, \bibinfo
  {author} {\bibfnamefont {J.}~\bibnamefont {Kahn}}, \bibinfo {author}
  {\bibfnamefont {X.}~\bibnamefont {Xia}}, \bibinfo {author} {\bibfnamefont
  {A.~J.}\ \bibnamefont {Graham}}, \bibinfo {author} {\bibfnamefont
  {V.}~\bibnamefont {Kandyba}}, \bibinfo {author} {\bibfnamefont
  {A.}~\bibnamefont {Giampietri}}, \bibinfo {author} {\bibfnamefont
  {A.}~\bibnamefont {Barinov}}, \bibinfo {author} {\bibfnamefont {G.~C.}\
  \bibnamefont {Constantinescu}}, \bibinfo {author} {\bibfnamefont
  {N.}~\bibnamefont {Yeung}}, \bibinfo {author} {\bibfnamefont {N.~D.~M.}\
  \bibnamefont {Hine}}, \bibinfo {author} {\bibfnamefont {X.}~\bibnamefont
  {Xu}}, \bibinfo {author} {\bibfnamefont {D.~H.}\ \bibnamefont {Cobden}}, \
  and\ \bibinfo {author} {\bibfnamefont {N.~R.}\ \bibnamefont {Wilson}},\
  }\href {\doibase 10.1038/s41586-019-1402-1} {\bibfield  {journal} {\bibinfo
  {journal} {Nature}\ }\textbf {\bibinfo {volume} {572}},\ \bibinfo {pages}
  {220} (\bibinfo {year} {2019})}\BibitemShut {NoStop}%
\bibitem [{\citenamefont {Wang}\ \emph
  {et~al.}(2016{\natexlab{a}})\citenamefont {Wang}, \citenamefont {Chen},
  \citenamefont {Wan}, \citenamefont {Lu}, \citenamefont {Chen}, \citenamefont
  {Avila}, \citenamefont {Fedorov}, \citenamefont {Zhang}, \citenamefont
  {Asensio}, \citenamefont {Zhang},\ and\ \citenamefont
  {Zhou}}]{nanoARPES_Wang}%
  \BibitemOpen
  \bibfield  {author} {\bibinfo {author} {\bibfnamefont {E.}~\bibnamefont
  {Wang}}, \bibinfo {author} {\bibfnamefont {G.}~\bibnamefont {Chen}}, \bibinfo
  {author} {\bibfnamefont {G.}~\bibnamefont {Wan}}, \bibinfo {author}
  {\bibfnamefont {X.}~\bibnamefont {Lu}}, \bibinfo {author} {\bibfnamefont
  {C.}~\bibnamefont {Chen}}, \bibinfo {author} {\bibfnamefont {J.}~\bibnamefont
  {Avila}}, \bibinfo {author} {\bibfnamefont {A.~V.}\ \bibnamefont {Fedorov}},
  \bibinfo {author} {\bibfnamefont {G.}~\bibnamefont {Zhang}}, \bibinfo
  {author} {\bibfnamefont {M.~C.}\ \bibnamefont {Asensio}}, \bibinfo {author}
  {\bibfnamefont {Y.}~\bibnamefont {Zhang}}, \ and\ \bibinfo {author}
  {\bibfnamefont {S.}~\bibnamefont {Zhou}},\ }\href {\doibase
  10.1088/0953-8984/28/44/444002} {\bibfield  {journal} {\bibinfo  {journal}
  {Journal of Physics: Condensed Matter}\ }\textbf {\bibinfo {volume} {28}},\
  \bibinfo {pages} {444002} (\bibinfo {year} {2016}{\natexlab{a}})}\BibitemShut
  {NoStop}%
\bibitem [{\citenamefont {Lisi}\ \emph {et~al.}()\citenamefont {Lisi},
  \citenamefont {Lu}, \citenamefont {Benschop}, \citenamefont {de~Jong},
  \citenamefont {Stepanov}, \citenamefont {Duran}, \citenamefont {Margot},
  \citenamefont {Cucchi}, \citenamefont {Cappelli}, \citenamefont {Hunter},
  \citenamefont {Tamai}, \citenamefont {Kandyba}, \citenamefont {Giampietri},
  \citenamefont {Barinov}, \citenamefont {Jobst}, \citenamefont {Stalman},
  \citenamefont {Leeuwenhoek}, \citenamefont {Watanabe}, \citenamefont
  {Taniguchi}, \citenamefont {Rademaker}, \citenamefont {van~der Molen},
  \citenamefont {Allan}, \citenamefont {Efetov},\ and\ \citenamefont
  {Baumberger}}]{nanoARPES_tBLG_Simone}%
  \BibitemOpen
  \bibfield  {author} {\bibinfo {author} {\bibfnamefont {S.}~\bibnamefont
  {Lisi}}, \bibinfo {author} {\bibfnamefont {X.}~\bibnamefont {Lu}}, \bibinfo
  {author} {\bibfnamefont {T.}~\bibnamefont {Benschop}}, \bibinfo {author}
  {\bibfnamefont {T.~A.}\ \bibnamefont {de~Jong}}, \bibinfo {author}
  {\bibfnamefont {P.}~\bibnamefont {Stepanov}}, \bibinfo {author}
  {\bibfnamefont {J.~R.}\ \bibnamefont {Duran}}, \bibinfo {author}
  {\bibfnamefont {F.}~\bibnamefont {Margot}}, \bibinfo {author} {\bibfnamefont
  {I.}~\bibnamefont {Cucchi}}, \bibinfo {author} {\bibfnamefont
  {E.}~\bibnamefont {Cappelli}}, \bibinfo {author} {\bibfnamefont
  {A.}~\bibnamefont {Hunter}}, \bibinfo {author} {\bibfnamefont
  {A.}~\bibnamefont {Tamai}}, \bibinfo {author} {\bibfnamefont
  {V.}~\bibnamefont {Kandyba}}, \bibinfo {author} {\bibfnamefont
  {A.}~\bibnamefont {Giampietri}}, \bibinfo {author} {\bibfnamefont
  {A.}~\bibnamefont {Barinov}}, \bibinfo {author} {\bibfnamefont
  {J.}~\bibnamefont {Jobst}}, \bibinfo {author} {\bibfnamefont
  {V.}~\bibnamefont {Stalman}}, \bibinfo {author} {\bibfnamefont
  {M.}~\bibnamefont {Leeuwenhoek}}, \bibinfo {author} {\bibfnamefont
  {K.}~\bibnamefont {Watanabe}}, \bibinfo {author} {\bibfnamefont
  {T.}~\bibnamefont {Taniguchi}}, \bibinfo {author} {\bibfnamefont
  {L.}~\bibnamefont {Rademaker}}, \bibinfo {author} {\bibfnamefont {S.~J.}\
  \bibnamefont {van~der Molen}}, \bibinfo {author} {\bibfnamefont
  {M.}~\bibnamefont {Allan}}, \bibinfo {author} {\bibfnamefont {D.~K.}\
  \bibnamefont {Efetov}}, \ and\ \bibinfo {author} {\bibfnamefont
  {F.}~\bibnamefont {Baumberger}},\ }\href {\doibase arXiv:2002.02289} {\
  arXiv:2002.02289}\BibitemShut {NoStop}%
\bibitem [{\citenamefont {Utama}\ \emph {et~al.}()\citenamefont {Utama},
  \citenamefont {Koch}, \citenamefont {Lee}, \citenamefont {Leconte},
  \citenamefont {Li}, \citenamefont {Zhao}, \citenamefont {Jiang},
  \citenamefont {Zhu}, \citenamefont {Watanabe}, \citenamefont {Taniguchi},
  \citenamefont {Ashby}, \citenamefont {Weber-Bargioni}, \citenamefont {Zettl},
  \citenamefont {Jozwiak}, \citenamefont {Jung}, \citenamefont {Rotenberg},
  \citenamefont {Bostwick},\ and\ \citenamefont {Wang}}]{nanoARPES_tBLG_Utama}%
  \BibitemOpen
  \bibfield  {author} {\bibinfo {author} {\bibfnamefont {M.~I.~B.}\
  \bibnamefont {Utama}}, \bibinfo {author} {\bibfnamefont {R.~J.}\ \bibnamefont
  {Koch}}, \bibinfo {author} {\bibfnamefont {K.}~\bibnamefont {Lee}}, \bibinfo
  {author} {\bibfnamefont {N.}~\bibnamefont {Leconte}}, \bibinfo {author}
  {\bibfnamefont {H.}~\bibnamefont {Li}}, \bibinfo {author} {\bibfnamefont
  {S.}~\bibnamefont {Zhao}}, \bibinfo {author} {\bibfnamefont {L.}~\bibnamefont
  {Jiang}}, \bibinfo {author} {\bibfnamefont {J.}~\bibnamefont {Zhu}}, \bibinfo
  {author} {\bibfnamefont {K.}~\bibnamefont {Watanabe}}, \bibinfo {author}
  {\bibfnamefont {T.}~\bibnamefont {Taniguchi}}, \bibinfo {author}
  {\bibfnamefont {P.~D.}\ \bibnamefont {Ashby}}, \bibinfo {author}
  {\bibfnamefont {A.}~\bibnamefont {Weber-Bargioni}}, \bibinfo {author}
  {\bibfnamefont {A.}~\bibnamefont {Zettl}}, \bibinfo {author} {\bibfnamefont
  {C.}~\bibnamefont {Jozwiak}}, \bibinfo {author} {\bibfnamefont
  {J.}~\bibnamefont {Jung}}, \bibinfo {author} {\bibfnamefont {E.}~\bibnamefont
  {Rotenberg}}, \bibinfo {author} {\bibfnamefont {A.}~\bibnamefont {Bostwick}},
  \ and\ \bibinfo {author} {\bibfnamefont {F.}~\bibnamefont {Wang}},\ }\href
  {\doibase arXiv:1912.00587} {\ arXiv:1912.00587}\BibitemShut {NoStop}%
\bibitem [{\citenamefont {Razado-Colambo}\ \emph {et~al.}(2016)\citenamefont
  {Razado-Colambo}, \citenamefont {Avila}, \citenamefont {Nys}, \citenamefont
  {Chen}, \citenamefont {Wallart}, \citenamefont {Asensio},\ and\ \citenamefont
  {Vignaud}}]{nanoARPES_Razado}%
  \BibitemOpen
  \bibfield  {author} {\bibinfo {author} {\bibfnamefont {I.}~\bibnamefont
  {Razado-Colambo}}, \bibinfo {author} {\bibfnamefont {J.}~\bibnamefont
  {Avila}}, \bibinfo {author} {\bibfnamefont {J.-P.}\ \bibnamefont {Nys}},
  \bibinfo {author} {\bibfnamefont {C.}~\bibnamefont {Chen}}, \bibinfo {author}
  {\bibfnamefont {X.}~\bibnamefont {Wallart}}, \bibinfo {author} {\bibfnamefont
  {M.-C.}\ \bibnamefont {Asensio}}, \ and\ \bibinfo {author} {\bibfnamefont
  {D.}~\bibnamefont {Vignaud}},\ }\href {https://doi.org/10.1038/srep27261}
  {\bibfield  {journal} {\bibinfo  {journal} {Scientific Reports}\ }\textbf
  {\bibinfo {volume} {6}},\ \bibinfo {pages} {27261} (\bibinfo {year}
  {2016})}\BibitemShut {NoStop}%
\bibitem [{\citenamefont {Ohta}\ \emph
  {et~al.}(2012{\natexlab{b}})\citenamefont {Ohta}, \citenamefont {Robinson},
  \citenamefont {Feibelman}, \citenamefont {Bostwick}, \citenamefont
  {Rotenberg},\ and\ \citenamefont {Beechem}}]{ARPES_tBLG_Ohta}%
  \BibitemOpen
  \bibfield  {author} {\bibinfo {author} {\bibfnamefont {T.}~\bibnamefont
  {Ohta}}, \bibinfo {author} {\bibfnamefont {J.~T.}\ \bibnamefont {Robinson}},
  \bibinfo {author} {\bibfnamefont {P.~J.}\ \bibnamefont {Feibelman}}, \bibinfo
  {author} {\bibfnamefont {A.}~\bibnamefont {Bostwick}}, \bibinfo {author}
  {\bibfnamefont {E.}~\bibnamefont {Rotenberg}}, \ and\ \bibinfo {author}
  {\bibfnamefont {T.~E.}\ \bibnamefont {Beechem}},\ }\href {\doibase
  10.1103/PhysRevLett.109.186807} {\bibfield  {journal} {\bibinfo  {journal}
  {Phys. Rev. Lett.}\ }\textbf {\bibinfo {volume} {109}},\ \bibinfo {pages}
  {186807} (\bibinfo {year} {2012}{\natexlab{b}})}\BibitemShut {NoStop}%
\bibitem [{\citenamefont {Amorim}(2018)}]{ARPES_theo_Amorim}%
  \BibitemOpen
  \bibfield  {author} {\bibinfo {author} {\bibfnamefont {B.}~\bibnamefont
  {Amorim}},\ }\href {\doibase 10.1103/PhysRevB.97.165414} {\bibfield
  {journal} {\bibinfo  {journal} {Phys. Rev. B}\ }\textbf {\bibinfo {volume}
  {97}},\ \bibinfo {pages} {165414} (\bibinfo {year} {2018})}\BibitemShut
  {NoStop}%
\bibitem [{\citenamefont {Amorim}\ and\ \citenamefont
  {Castro}()}]{ARPES_theo_tTL_Amorim}%
  \BibitemOpen
  \bibfield  {author} {\bibinfo {author} {\bibfnamefont {B.}~\bibnamefont
  {Amorim}}\ and\ \bibinfo {author} {\bibfnamefont {E.~V.}\ \bibnamefont
  {Castro}},\ }\href {\doibase arXiv:1807.11909} {\
  arXiv:1807.11909}\BibitemShut {NoStop}%
\bibitem [{\citenamefont {Pal}\ and\ \citenamefont
  {Mele}(2013)}]{ARPES_theo_Anshuman}%
  \BibitemOpen
  \bibfield  {author} {\bibinfo {author} {\bibfnamefont {A.}~\bibnamefont
  {Pal}}\ and\ \bibinfo {author} {\bibfnamefont {E.~J.}\ \bibnamefont {Mele}},\
  }\href {\doibase 10.1103/PhysRevB.87.205444} {\bibfield  {journal} {\bibinfo
  {journal} {Phys. Rev. B}\ }\textbf {\bibinfo {volume} {87}},\ \bibinfo
  {pages} {205444} (\bibinfo {year} {2013})}\BibitemShut {NoStop}%
\bibitem [{\citenamefont {Mucha-Kruczy\ifmmode~\acute{n}\else \'{n}\fi{}ski}\
  \emph {et~al.}(2016{\natexlab{a}})\citenamefont
  {Mucha-Kruczy\ifmmode~\acute{n}\else \'{n}\fi{}ski}, \citenamefont
  {Wallbank},\ and\ \citenamefont {Fal'ko}}]{ARPES_theo_GhBN_Falko}%
  \BibitemOpen
  \bibfield  {author} {\bibinfo {author} {\bibfnamefont {M.}~\bibnamefont
  {Mucha-Kruczy\ifmmode~\acute{n}\else \'{n}\fi{}ski}}, \bibinfo {author}
  {\bibfnamefont {J.~R.}\ \bibnamefont {Wallbank}}, \ and\ \bibinfo {author}
  {\bibfnamefont {V.~I.}\ \bibnamefont {Fal'ko}},\ }\href {\doibase
  10.1103/PhysRevB.93.085409} {\bibfield  {journal} {\bibinfo  {journal} {Phys.
  Rev. B}\ }\textbf {\bibinfo {volume} {93}},\ \bibinfo {pages} {085409}
  (\bibinfo {year} {2016}{\natexlab{a}})}\BibitemShut {NoStop}%
\bibitem [{\citenamefont {Jung}\ \emph {et~al.}(2014)\citenamefont {Jung},
  \citenamefont {Raoux}, \citenamefont {Qiao},\ and\ \citenamefont
  {MacDonald}}]{GhBN_Jeil_2014}%
  \BibitemOpen
  \bibfield  {author} {\bibinfo {author} {\bibfnamefont {J.}~\bibnamefont
  {Jung}}, \bibinfo {author} {\bibfnamefont {A.}~\bibnamefont {Raoux}},
  \bibinfo {author} {\bibfnamefont {Z.}~\bibnamefont {Qiao}}, \ and\ \bibinfo
  {author} {\bibfnamefont {A.~H.}\ \bibnamefont {MacDonald}},\ }\href {\doibase
  10.1103/PhysRevB.89.205414} {\bibfield  {journal} {\bibinfo  {journal} {Phys.
  Rev. B}\ }\textbf {\bibinfo {volume} {89}},\ \bibinfo {pages} {205414}
  (\bibinfo {year} {2014})}\BibitemShut {NoStop}%
\bibitem [{\citenamefont {Jung}\ \emph {et~al.}(2017)\citenamefont {Jung},
  \citenamefont {Laksono}, \citenamefont {DaSilva}, \citenamefont {MacDonald},
  \citenamefont {Mucha-Kruczy\ifmmode~\acute{n}\else \'{n}\fi{}ski},\ and\
  \citenamefont {Adam}}]{GhBN_Jeil_2017}%
  \BibitemOpen
  \bibfield  {author} {\bibinfo {author} {\bibfnamefont {J.}~\bibnamefont
  {Jung}}, \bibinfo {author} {\bibfnamefont {E.}~\bibnamefont {Laksono}},
  \bibinfo {author} {\bibfnamefont {A.~M.}\ \bibnamefont {DaSilva}}, \bibinfo
  {author} {\bibfnamefont {A.~H.}\ \bibnamefont {MacDonald}}, \bibinfo {author}
  {\bibfnamefont {M.}~\bibnamefont {Mucha-Kruczy\ifmmode~\acute{n}\else
  \'{n}\fi{}ski}}, \ and\ \bibinfo {author} {\bibfnamefont {S.}~\bibnamefont
  {Adam}},\ }\href {\doibase 10.1103/PhysRevB.96.085442} {\bibfield  {journal}
  {\bibinfo  {journal} {Phys. Rev. B}\ }\textbf {\bibinfo {volume} {96}},\
  \bibinfo {pages} {085442} (\bibinfo {year} {2017})}\BibitemShut {NoStop}%
\bibitem [{\citenamefont {Jung}\ \emph {et~al.}(2015)\citenamefont {Jung},
  \citenamefont {DaSilva}, \citenamefont {MacDonald},\ and\ \citenamefont
  {Adam}}]{GhBN_Jeil_2015}%
  \BibitemOpen
  \bibfield  {author} {\bibinfo {author} {\bibfnamefont {J.}~\bibnamefont
  {Jung}}, \bibinfo {author} {\bibfnamefont {A.~M.}\ \bibnamefont {DaSilva}},
  \bibinfo {author} {\bibfnamefont {A.~H.}\ \bibnamefont {MacDonald}}, \ and\
  \bibinfo {author} {\bibfnamefont {S.}~\bibnamefont {Adam}},\ }\href
  {https://doi.org/10.1038/ncomms7308} {\bibfield  {journal} {\bibinfo
  {journal} {Nature Communications}\ }\textbf {\bibinfo {volume} {6}},\
  \bibinfo {pages} {6308 EP } (\bibinfo {year} {2015})}\BibitemShut {NoStop}%
\bibitem [{\citenamefont {Strocov}\ \emph {et~al.}(2012)\citenamefont
  {Strocov}, \citenamefont {Shi}, \citenamefont {Kobayashi}, \citenamefont
  {Monney}, \citenamefont {Wang}, \citenamefont {Krempasky}, \citenamefont
  {Schmitt}, \citenamefont {Patthey}, \citenamefont {Berger},\ and\
  \citenamefont {Blaha}}]{final_state_Strocov}%
  \BibitemOpen
  \bibfield  {author} {\bibinfo {author} {\bibfnamefont {V.~N.}\ \bibnamefont
  {Strocov}}, \bibinfo {author} {\bibfnamefont {M.}~\bibnamefont {Shi}},
  \bibinfo {author} {\bibfnamefont {M.}~\bibnamefont {Kobayashi}}, \bibinfo
  {author} {\bibfnamefont {C.}~\bibnamefont {Monney}}, \bibinfo {author}
  {\bibfnamefont {X.}~\bibnamefont {Wang}}, \bibinfo {author} {\bibfnamefont
  {J.}~\bibnamefont {Krempasky}}, \bibinfo {author} {\bibfnamefont
  {T.}~\bibnamefont {Schmitt}}, \bibinfo {author} {\bibfnamefont
  {L.}~\bibnamefont {Patthey}}, \bibinfo {author} {\bibfnamefont
  {H.}~\bibnamefont {Berger}}, \ and\ \bibinfo {author} {\bibfnamefont
  {P.}~\bibnamefont {Blaha}},\ }\href {\doibase 10.1103/PhysRevLett.109.086401}
  {\bibfield  {journal} {\bibinfo  {journal} {Phys. Rev. Lett.}\ }\textbf
  {\bibinfo {volume} {109}},\ \bibinfo {pages} {086401} (\bibinfo {year}
  {2012})}\BibitemShut {NoStop}%
\bibitem [{\citenamefont {Damascelli}(2004)}]{Damascelli_ARPES_2004}%
  \BibitemOpen
  \bibfield  {author} {\bibinfo {author} {\bibfnamefont {A.}~\bibnamefont
  {Damascelli}},\ }\href {\doibase 10.1238/physica.topical.109a00061}
  {\bibfield  {journal} {\bibinfo  {journal} {Physica Scripta}\ }\textbf
  {\bibinfo {volume} {T109}},\ \bibinfo {pages} {61} (\bibinfo {year}
  {2004})}\BibitemShut {NoStop}%
\bibitem [{\citenamefont {Puschnig}\ \emph {et~al.}(2009)\citenamefont
  {Puschnig}, \citenamefont {Berkebile}, \citenamefont {Fleming}, \citenamefont
  {Koller}, \citenamefont {Emtsev}, \citenamefont {Seyller}, \citenamefont
  {Riley}, \citenamefont {Ambrosch-Draxl}, \citenamefont {Netzer},\ and\
  \citenamefont {Ramsey}}]{Puschnig_2009}%
  \BibitemOpen
  \bibfield  {author} {\bibinfo {author} {\bibfnamefont {P.}~\bibnamefont
  {Puschnig}}, \bibinfo {author} {\bibfnamefont {S.}~\bibnamefont {Berkebile}},
  \bibinfo {author} {\bibfnamefont {A.~J.}\ \bibnamefont {Fleming}}, \bibinfo
  {author} {\bibfnamefont {G.}~\bibnamefont {Koller}}, \bibinfo {author}
  {\bibfnamefont {K.}~\bibnamefont {Emtsev}}, \bibinfo {author} {\bibfnamefont
  {T.}~\bibnamefont {Seyller}}, \bibinfo {author} {\bibfnamefont {J.~D.}\
  \bibnamefont {Riley}}, \bibinfo {author} {\bibfnamefont {C.}~\bibnamefont
  {Ambrosch-Draxl}}, \bibinfo {author} {\bibfnamefont {F.~P.}\ \bibnamefont
  {Netzer}}, \ and\ \bibinfo {author} {\bibfnamefont {M.~G.}\ \bibnamefont
  {Ramsey}},\ }\href {\doibase 10.1126/science.1176105} {\bibfield  {journal}
  {\bibinfo  {journal} {Science}\ }\textbf {\bibinfo {volume} {326}},\ \bibinfo
  {pages} {702} (\bibinfo {year} {2009})}\BibitemShut {NoStop}%
\bibitem [{\citenamefont {Medjanik}\ \emph {et~al.}(2017)\citenamefont
  {Medjanik}, \citenamefont {Fedchenko}, \citenamefont {Chernov}, \citenamefont
  {Kutnyakhov}, \citenamefont {Ellguth}, \citenamefont {Oelsner}, \citenamefont
  {Schönhense}, \citenamefont {Peixoto}, \citenamefont {Lutz}, \citenamefont
  {Min}, \citenamefont {Reinert}, \citenamefont {Däster}, \citenamefont
  {Acremann}, \citenamefont {Viefhaus}, \citenamefont {Wurth}, \citenamefont
  {Elmers},\ and\ \citenamefont {Schönhense}}]{Medjanik_2017}%
  \BibitemOpen
  \bibfield  {author} {\bibinfo {author} {\bibfnamefont {K.}~\bibnamefont
  {Medjanik}}, \bibinfo {author} {\bibfnamefont {O.}~\bibnamefont {Fedchenko}},
  \bibinfo {author} {\bibfnamefont {S.}~\bibnamefont {Chernov}}, \bibinfo
  {author} {\bibfnamefont {D.}~\bibnamefont {Kutnyakhov}}, \bibinfo {author}
  {\bibfnamefont {M.}~\bibnamefont {Ellguth}}, \bibinfo {author} {\bibfnamefont
  {A.}~\bibnamefont {Oelsner}}, \bibinfo {author} {\bibfnamefont
  {B.}~\bibnamefont {Schönhense}}, \bibinfo {author} {\bibfnamefont
  {T.~R.~F.}\ \bibnamefont {Peixoto}}, \bibinfo {author} {\bibfnamefont
  {P.}~\bibnamefont {Lutz}}, \bibinfo {author} {\bibfnamefont {C.~H.}\
  \bibnamefont {Min}}, \bibinfo {author} {\bibfnamefont {F.}~\bibnamefont
  {Reinert}}, \bibinfo {author} {\bibfnamefont {S.}~\bibnamefont {Däster}},
  \bibinfo {author} {\bibfnamefont {Y.}~\bibnamefont {Acremann}}, \bibinfo
  {author} {\bibfnamefont {J.}~\bibnamefont {Viefhaus}}, \bibinfo {author}
  {\bibfnamefont {W.}~\bibnamefont {Wurth}}, \bibinfo {author} {\bibfnamefont
  {H.~J.}\ \bibnamefont {Elmers}}, \ and\ \bibinfo {author} {\bibfnamefont
  {G.}~\bibnamefont {Schönhense}},\ }\href {\doibase 10.1038/nmat4875}
  {\bibfield  {journal} {\bibinfo  {journal} {Nature Materials}\ }\textbf
  {\bibinfo {volume} {16}},\ \bibinfo {pages} {615} (\bibinfo {year}
  {2017})}\BibitemShut {NoStop}%
\bibitem [{\citenamefont {Puschnig}\ and\ \citenamefont
  {Lüftner}(2015)}]{ARPES_Puschnig}%
  \BibitemOpen
  \bibfield  {author} {\bibinfo {author} {\bibfnamefont {P.}~\bibnamefont
  {Puschnig}}\ and\ \bibinfo {author} {\bibfnamefont {D.}~\bibnamefont
  {Lüftner}},\ }\href {\doibase https://doi.org/10.1016/j.elspec.2015.06.003}
  {\bibfield  {journal} {\bibinfo  {journal} {Journal of Electron Spectroscopy
  and Related Phenomena}\ }\textbf {\bibinfo {volume} {200}},\ \bibinfo {pages}
  {193 } (\bibinfo {year} {2015})},\ \bibinfo {note} {special Anniversary
  Issue: Volume 200}\BibitemShut {NoStop}%
\bibitem [{\citenamefont {Shirley}\ \emph {et~al.}(1995)\citenamefont
  {Shirley}, \citenamefont {Terminello}, \citenamefont {Santoni},\ and\
  \citenamefont {Himpsel}}]{ARPES_Shirley}%
  \BibitemOpen
  \bibfield  {author} {\bibinfo {author} {\bibfnamefont {E.~L.}\ \bibnamefont
  {Shirley}}, \bibinfo {author} {\bibfnamefont {L.~J.}\ \bibnamefont
  {Terminello}}, \bibinfo {author} {\bibfnamefont {A.}~\bibnamefont {Santoni}},
  \ and\ \bibinfo {author} {\bibfnamefont {F.~J.}\ \bibnamefont {Himpsel}},\
  }\href {\doibase 10.1103/PhysRevB.51.13614} {\bibfield  {journal} {\bibinfo
  {journal} {Phys. Rev. B}\ }\textbf {\bibinfo {volume} {51}},\ \bibinfo
  {pages} {13614} (\bibinfo {year} {1995})}\BibitemShut {NoStop}%
\bibitem [{\citenamefont {Ayria}\ \emph {et~al.}(2015)\citenamefont {Ayria},
  \citenamefont {Nugraha}, \citenamefont {Hasdeo}, \citenamefont {Czank},
  \citenamefont {Tanaka},\ and\ \citenamefont {Saito}}]{final_state_Ayria}%
  \BibitemOpen
  \bibfield  {author} {\bibinfo {author} {\bibfnamefont {P.}~\bibnamefont
  {Ayria}}, \bibinfo {author} {\bibfnamefont {A.~R.~T.}\ \bibnamefont
  {Nugraha}}, \bibinfo {author} {\bibfnamefont {E.~H.}\ \bibnamefont {Hasdeo}},
  \bibinfo {author} {\bibfnamefont {T.~R.}\ \bibnamefont {Czank}}, \bibinfo
  {author} {\bibfnamefont {S.-i.}\ \bibnamefont {Tanaka}}, \ and\ \bibinfo
  {author} {\bibfnamefont {R.}~\bibnamefont {Saito}},\ }\href {\doibase
  10.1103/PhysRevB.92.195148} {\bibfield  {journal} {\bibinfo  {journal} {Phys.
  Rev. B}\ }\textbf {\bibinfo {volume} {92}},\ \bibinfo {pages} {195148}
  (\bibinfo {year} {2015})}\BibitemShut {NoStop}%
\bibitem [{\citenamefont {Barrett}\ \emph {et~al.}(2005)\citenamefont
  {Barrett}, \citenamefont {Krasovskii}, \citenamefont {Themlin},\ and\
  \citenamefont {Strocov}}]{final_state_Barrett}%
  \BibitemOpen
  \bibfield  {author} {\bibinfo {author} {\bibfnamefont {N.}~\bibnamefont
  {Barrett}}, \bibinfo {author} {\bibfnamefont {E.~E.}\ \bibnamefont
  {Krasovskii}}, \bibinfo {author} {\bibfnamefont {J.-M.}\ \bibnamefont
  {Themlin}}, \ and\ \bibinfo {author} {\bibfnamefont {V.~N.}\ \bibnamefont
  {Strocov}},\ }\href {\doibase 10.1103/PhysRevB.71.035427} {\bibfield
  {journal} {\bibinfo  {journal} {Phys. Rev. B}\ }\textbf {\bibinfo {volume}
  {71}},\ \bibinfo {pages} {035427} (\bibinfo {year} {2005})}\BibitemShut
  {NoStop}%
\bibitem [{\citenamefont {Strocov}\ \emph {et~al.}(2006)\citenamefont
  {Strocov}, \citenamefont {Krasovskii}, \citenamefont {Schattke},
  \citenamefont {Barrett}, \citenamefont {Berger}, \citenamefont {Schrupp},\
  and\ \citenamefont {Claessen}}]{final_state_Strocov_2006}%
  \BibitemOpen
  \bibfield  {author} {\bibinfo {author} {\bibfnamefont {V.~N.}\ \bibnamefont
  {Strocov}}, \bibinfo {author} {\bibfnamefont {E.~E.}\ \bibnamefont
  {Krasovskii}}, \bibinfo {author} {\bibfnamefont {W.}~\bibnamefont
  {Schattke}}, \bibinfo {author} {\bibfnamefont {N.}~\bibnamefont {Barrett}},
  \bibinfo {author} {\bibfnamefont {H.}~\bibnamefont {Berger}}, \bibinfo
  {author} {\bibfnamefont {D.}~\bibnamefont {Schrupp}}, \ and\ \bibinfo
  {author} {\bibfnamefont {R.}~\bibnamefont {Claessen}},\ }\href {\doibase
  10.1103/PhysRevB.74.195125} {\bibfield  {journal} {\bibinfo  {journal} {Phys.
  Rev. B}\ }\textbf {\bibinfo {volume} {74}},\ \bibinfo {pages} {195125}
  (\bibinfo {year} {2006})}\BibitemShut {NoStop}%
\bibitem [{\citenamefont {Mucha-Kruczy\ifmmode~\acute{n}\else \'{n}\fi{}ski}\
  \emph {et~al.}(2008)\citenamefont {Mucha-Kruczy\ifmmode~\acute{n}\else
  \'{n}\fi{}ski}, \citenamefont {Tsyplyatyev}, \citenamefont {Grishin},
  \citenamefont {McCann}, \citenamefont {Fal'ko}, \citenamefont {Bostwick},\
  and\ \citenamefont {Rotenberg}}]{ARPES_Eli2008}%
  \BibitemOpen
  \bibfield  {author} {\bibinfo {author} {\bibfnamefont {M.}~\bibnamefont
  {Mucha-Kruczy\ifmmode~\acute{n}\else \'{n}\fi{}ski}}, \bibinfo {author}
  {\bibfnamefont {O.}~\bibnamefont {Tsyplyatyev}}, \bibinfo {author}
  {\bibfnamefont {A.}~\bibnamefont {Grishin}}, \bibinfo {author} {\bibfnamefont
  {E.}~\bibnamefont {McCann}}, \bibinfo {author} {\bibfnamefont {V.~I.}\
  \bibnamefont {Fal'ko}}, \bibinfo {author} {\bibfnamefont {A.}~\bibnamefont
  {Bostwick}}, \ and\ \bibinfo {author} {\bibfnamefont {E.}~\bibnamefont
  {Rotenberg}},\ }\href {\doibase 10.1103/PhysRevB.77.195403} {\bibfield
  {journal} {\bibinfo  {journal} {Phys. Rev. B}\ }\textbf {\bibinfo {volume}
  {77}},\ \bibinfo {pages} {195403} (\bibinfo {year} {2008})}\BibitemShut
  {NoStop}%
\bibitem [{\citenamefont {Ismail-Beigi}\ \emph {et~al.}(2001)\citenamefont
  {Ismail-Beigi}, \citenamefont {Chang},\ and\ \citenamefont
  {Louie}}]{Adotv_Ismail}%
  \BibitemOpen
  \bibfield  {author} {\bibinfo {author} {\bibfnamefont {S.}~\bibnamefont
  {Ismail-Beigi}}, \bibinfo {author} {\bibfnamefont {E.~K.}\ \bibnamefont
  {Chang}}, \ and\ \bibinfo {author} {\bibfnamefont {S.~G.}\ \bibnamefont
  {Louie}},\ }\href {\doibase 10.1103/PhysRevLett.87.087402} {\bibfield
  {journal} {\bibinfo  {journal} {Phys. Rev. Lett.}\ }\textbf {\bibinfo
  {volume} {87}},\ \bibinfo {pages} {087402} (\bibinfo {year}
  {2001})}\BibitemShut {NoStop}%
\bibitem [{\citenamefont {Hunt}\ \emph {et~al.}(2013)\citenamefont {Hunt},
  \citenamefont {Sanchez-Yamagishi}, \citenamefont {Young}, \citenamefont
  {Yankowitz}, \citenamefont {LeRoy}, \citenamefont {Watanabe}, \citenamefont
  {Taniguchi}, \citenamefont {Moon}, \citenamefont {Koshino}, \citenamefont
  {Jarillo-Herrero},\ and\ \citenamefont {Ashoori}}]{GhBN_gap_Hunt}%
  \BibitemOpen
  \bibfield  {author} {\bibinfo {author} {\bibfnamefont {B.}~\bibnamefont
  {Hunt}}, \bibinfo {author} {\bibfnamefont {J.~D.}\ \bibnamefont
  {Sanchez-Yamagishi}}, \bibinfo {author} {\bibfnamefont {A.~F.}\ \bibnamefont
  {Young}}, \bibinfo {author} {\bibfnamefont {M.}~\bibnamefont {Yankowitz}},
  \bibinfo {author} {\bibfnamefont {B.~J.}\ \bibnamefont {LeRoy}}, \bibinfo
  {author} {\bibfnamefont {K.}~\bibnamefont {Watanabe}}, \bibinfo {author}
  {\bibfnamefont {T.}~\bibnamefont {Taniguchi}}, \bibinfo {author}
  {\bibfnamefont {P.}~\bibnamefont {Moon}}, \bibinfo {author} {\bibfnamefont
  {M.}~\bibnamefont {Koshino}}, \bibinfo {author} {\bibfnamefont
  {P.}~\bibnamefont {Jarillo-Herrero}}, \ and\ \bibinfo {author} {\bibfnamefont
  {R.~C.}\ \bibnamefont {Ashoori}},\ }\href {\doibase 10.1126/science.1237240}
  {\bibfield  {journal} {\bibinfo  {journal} {Science}\ }\textbf {\bibinfo
  {volume} {340}},\ \bibinfo {pages} {1427} (\bibinfo {year}
  {2013})}\BibitemShut {NoStop}%
\bibitem [{\citenamefont {Woods}\ \emph {et~al.}(2014)\citenamefont {Woods},
  \citenamefont {Britnell}, \citenamefont {Eckmann}, \citenamefont {Ma},
  \citenamefont {Lu}, \citenamefont {Guo}, \citenamefont {Lin}, \citenamefont
  {Yu}, \citenamefont {Cao}, \citenamefont {Gorbachev}, \citenamefont
  {Kretinin}, \citenamefont {Park}, \citenamefont {Ponomarenko}, \citenamefont
  {Katsnelson}, \citenamefont {Gornostyrev}, \citenamefont {Watanabe},
  \citenamefont {Taniguchi}, \citenamefont {Casiraghi}, \citenamefont {Gao},
  \citenamefont {Geim},\ and\ \citenamefont {Novoselov}}]{GhBN_gap_Woods}%
  \BibitemOpen
  \bibfield  {author} {\bibinfo {author} {\bibfnamefont {C.~R.}\ \bibnamefont
  {Woods}}, \bibinfo {author} {\bibfnamefont {L.}~\bibnamefont {Britnell}},
  \bibinfo {author} {\bibfnamefont {A.}~\bibnamefont {Eckmann}}, \bibinfo
  {author} {\bibfnamefont {R.~S.}\ \bibnamefont {Ma}}, \bibinfo {author}
  {\bibfnamefont {J.~C.}\ \bibnamefont {Lu}}, \bibinfo {author} {\bibfnamefont
  {H.~M.}\ \bibnamefont {Guo}}, \bibinfo {author} {\bibfnamefont
  {X.}~\bibnamefont {Lin}}, \bibinfo {author} {\bibfnamefont {G.~L.}\
  \bibnamefont {Yu}}, \bibinfo {author} {\bibfnamefont {Y.}~\bibnamefont
  {Cao}}, \bibinfo {author} {\bibfnamefont {R.}~\bibnamefont {Gorbachev}},
  \bibinfo {author} {\bibfnamefont {A.~V.}\ \bibnamefont {Kretinin}}, \bibinfo
  {author} {\bibfnamefont {J.}~\bibnamefont {Park}}, \bibinfo {author}
  {\bibfnamefont {L.~A.}\ \bibnamefont {Ponomarenko}}, \bibinfo {author}
  {\bibfnamefont {M.~I.}\ \bibnamefont {Katsnelson}}, \bibinfo {author}
  {\bibfnamefont {Y.}~\bibnamefont {Gornostyrev}}, \bibinfo {author}
  {\bibfnamefont {K.}~\bibnamefont {Watanabe}}, \bibinfo {author}
  {\bibfnamefont {T.}~\bibnamefont {Taniguchi}}, \bibinfo {author}
  {\bibfnamefont {C.}~\bibnamefont {Casiraghi}}, \bibinfo {author}
  {\bibfnamefont {H.-J.}\ \bibnamefont {Gao}}, \bibinfo {author} {\bibfnamefont
  {A.~K.}\ \bibnamefont {Geim}}, \ and\ \bibinfo {author} {\bibfnamefont
  {K.}~\bibnamefont {Novoselov}},\ }\href {\doibase 10.1038/nphys2954}
  {\bibfield  {journal} {\bibinfo  {journal} {Nature Physics}\ }\textbf
  {\bibinfo {volume} {10}},\ \bibinfo {pages} {451} (\bibinfo {year}
  {2014})}\BibitemShut {NoStop}%
\bibitem [{\citenamefont {Ponomarenko}\ \emph
  {et~al.}(2013{\natexlab{a}})\citenamefont {Ponomarenko}, \citenamefont
  {Gorbachev}, \citenamefont {Yu}, \citenamefont {Elias}, \citenamefont
  {Jalil}, \citenamefont {Patel}, \citenamefont {Mishchenko}, \citenamefont
  {Mayorov}, \citenamefont {Woods}, \citenamefont {Wallbank}, \citenamefont
  {Mucha-Kruczynski}, \citenamefont {Piot}, \citenamefont {Potemski},
  \citenamefont {Grigorieva}, \citenamefont {Novoselov}, \citenamefont
  {Guinea}, \citenamefont {Fal’ko},\ and\ \citenamefont
  {Geim}}]{SDC_Ponomarenko}%
  \BibitemOpen
  \bibfield  {author} {\bibinfo {author} {\bibfnamefont {L.~A.}\ \bibnamefont
  {Ponomarenko}}, \bibinfo {author} {\bibfnamefont {R.~V.}\ \bibnamefont
  {Gorbachev}}, \bibinfo {author} {\bibfnamefont {G.~L.}\ \bibnamefont {Yu}},
  \bibinfo {author} {\bibfnamefont {D.~C.}\ \bibnamefont {Elias}}, \bibinfo
  {author} {\bibfnamefont {R.}~\bibnamefont {Jalil}}, \bibinfo {author}
  {\bibfnamefont {A.~A.}\ \bibnamefont {Patel}}, \bibinfo {author}
  {\bibfnamefont {A.}~\bibnamefont {Mishchenko}}, \bibinfo {author}
  {\bibfnamefont {A.~S.}\ \bibnamefont {Mayorov}}, \bibinfo {author}
  {\bibfnamefont {C.~R.}\ \bibnamefont {Woods}}, \bibinfo {author}
  {\bibfnamefont {J.~R.}\ \bibnamefont {Wallbank}}, \bibinfo {author}
  {\bibfnamefont {M.}~\bibnamefont {Mucha-Kruczynski}}, \bibinfo {author}
  {\bibfnamefont {B.~A.}\ \bibnamefont {Piot}}, \bibinfo {author}
  {\bibfnamefont {M.}~\bibnamefont {Potemski}}, \bibinfo {author}
  {\bibfnamefont {I.~V.}\ \bibnamefont {Grigorieva}}, \bibinfo {author}
  {\bibfnamefont {K.~S.}\ \bibnamefont {Novoselov}}, \bibinfo {author}
  {\bibfnamefont {F.}~\bibnamefont {Guinea}}, \bibinfo {author} {\bibfnamefont
  {V.~I.}\ \bibnamefont {Fal’ko}}, \ and\ \bibinfo {author} {\bibfnamefont
  {A.~K.}\ \bibnamefont {Geim}},\ }\href {\doibase 10.1038/nature12187}
  {\bibfield  {journal} {\bibinfo  {journal} {Nature}\ }\textbf {\bibinfo
  {volume} {497}},\ \bibinfo {pages} {594} (\bibinfo {year}
  {2013}{\natexlab{a}})}\BibitemShut {NoStop}%
\bibitem [{\citenamefont {Park}\ \emph {et~al.}(2008)\citenamefont {Park},
  \citenamefont {Yang}, \citenamefont {Son}, \citenamefont {Cohen},\ and\
  \citenamefont {Louie}}]{SDC_Park}%
  \BibitemOpen
  \bibfield  {author} {\bibinfo {author} {\bibfnamefont {C.-H.}\ \bibnamefont
  {Park}}, \bibinfo {author} {\bibfnamefont {L.}~\bibnamefont {Yang}}, \bibinfo
  {author} {\bibfnamefont {Y.-W.}\ \bibnamefont {Son}}, \bibinfo {author}
  {\bibfnamefont {M.~L.}\ \bibnamefont {Cohen}}, \ and\ \bibinfo {author}
  {\bibfnamefont {S.~G.}\ \bibnamefont {Louie}},\ }\href {\doibase
  10.1103/PhysRevLett.101.126804} {\bibfield  {journal} {\bibinfo  {journal}
  {Phys. Rev. Lett.}\ }\textbf {\bibinfo {volume} {101}},\ \bibinfo {pages}
  {126804} (\bibinfo {year} {2008})}\BibitemShut {NoStop}%
\bibitem [{\citenamefont {DaSilva}\ \emph {et~al.}(2015)\citenamefont
  {DaSilva}, \citenamefont {Jung}, \citenamefont {Adam},\ and\ \citenamefont
  {MacDonald}}]{SDC_Jung}%
  \BibitemOpen
  \bibfield  {author} {\bibinfo {author} {\bibfnamefont {A.~M.}\ \bibnamefont
  {DaSilva}}, \bibinfo {author} {\bibfnamefont {J.}~\bibnamefont {Jung}},
  \bibinfo {author} {\bibfnamefont {S.}~\bibnamefont {Adam}}, \ and\ \bibinfo
  {author} {\bibfnamefont {A.~H.}\ \bibnamefont {MacDonald}},\ }\href {\doibase
  10.1103/PhysRevB.91.245422} {\bibfield  {journal} {\bibinfo  {journal} {Phys.
  Rev. B}\ }\textbf {\bibinfo {volume} {91}},\ \bibinfo {pages} {245422}
  (\bibinfo {year} {2015})}\BibitemShut {NoStop}%
\bibitem [{\citenamefont {Ortix}\ \emph {et~al.}(2012)\citenamefont {Ortix},
  \citenamefont {Yang},\ and\ \citenamefont {van~den Brink}}]{SDC_Ortix}%
  \BibitemOpen
  \bibfield  {author} {\bibinfo {author} {\bibfnamefont {C.}~\bibnamefont
  {Ortix}}, \bibinfo {author} {\bibfnamefont {L.}~\bibnamefont {Yang}}, \ and\
  \bibinfo {author} {\bibfnamefont {J.}~\bibnamefont {van~den Brink}},\ }\href
  {\doibase 10.1103/PhysRevB.86.081405} {\bibfield  {journal} {\bibinfo
  {journal} {Phys. Rev. B}\ }\textbf {\bibinfo {volume} {86}},\ \bibinfo
  {pages} {081405(R)} (\bibinfo {year} {2012})}\BibitemShut {NoStop}%
\bibitem [{\citenamefont {Wallbank}\ \emph {et~al.}(2013)\citenamefont
  {Wallbank}, \citenamefont {Patel}, \citenamefont
  {Mucha-Kruczy\ifmmode~\acute{n}\else \'{n}\fi{}ski}, \citenamefont {Geim},\
  and\ \citenamefont {Fal'ko}}]{SDC_Wallbank}%
  \BibitemOpen
  \bibfield  {author} {\bibinfo {author} {\bibfnamefont {J.~R.}\ \bibnamefont
  {Wallbank}}, \bibinfo {author} {\bibfnamefont {A.~A.}\ \bibnamefont {Patel}},
  \bibinfo {author} {\bibfnamefont {M.}~\bibnamefont
  {Mucha-Kruczy\ifmmode~\acute{n}\else \'{n}\fi{}ski}}, \bibinfo {author}
  {\bibfnamefont {A.~K.}\ \bibnamefont {Geim}}, \ and\ \bibinfo {author}
  {\bibfnamefont {V.~I.}\ \bibnamefont {Fal'ko}},\ }\href {\doibase
  10.1103/PhysRevB.87.245408} {\bibfield  {journal} {\bibinfo  {journal} {Phys.
  Rev. B}\ }\textbf {\bibinfo {volume} {87}},\ \bibinfo {pages} {245408}
  (\bibinfo {year} {2013})}\BibitemShut {NoStop}%
\bibitem [{\citenamefont {Wang}\ \emph
  {et~al.}(2016{\natexlab{b}})\citenamefont {Wang}, \citenamefont {Lu},
  \citenamefont {Ding}, \citenamefont {Yao}, \citenamefont {Yan}, \citenamefont
  {Wan}, \citenamefont {Deng}, \citenamefont {Wang}, \citenamefont {Chen},
  \citenamefont {Ma}, \citenamefont {Jung}, \citenamefont {Fedorov},
  \citenamefont {Zhang}, \citenamefont {Zhang},\ and\ \citenamefont
  {Zhou}}]{G_hBN_SDC_2016}%
  \BibitemOpen
  \bibfield  {author} {\bibinfo {author} {\bibfnamefont {E.}~\bibnamefont
  {Wang}}, \bibinfo {author} {\bibfnamefont {X.}~\bibnamefont {Lu}}, \bibinfo
  {author} {\bibfnamefont {S.}~\bibnamefont {Ding}}, \bibinfo {author}
  {\bibfnamefont {W.}~\bibnamefont {Yao}}, \bibinfo {author} {\bibfnamefont
  {M.}~\bibnamefont {Yan}}, \bibinfo {author} {\bibfnamefont {G.}~\bibnamefont
  {Wan}}, \bibinfo {author} {\bibfnamefont {K.}~\bibnamefont {Deng}}, \bibinfo
  {author} {\bibfnamefont {S.}~\bibnamefont {Wang}}, \bibinfo {author}
  {\bibfnamefont {G.}~\bibnamefont {Chen}}, \bibinfo {author} {\bibfnamefont
  {L.}~\bibnamefont {Ma}}, \bibinfo {author} {\bibfnamefont {J.}~\bibnamefont
  {Jung}}, \bibinfo {author} {\bibfnamefont {A.~V.}\ \bibnamefont {Fedorov}},
  \bibinfo {author} {\bibfnamefont {Y.}~\bibnamefont {Zhang}}, \bibinfo
  {author} {\bibfnamefont {G.}~\bibnamefont {Zhang}}, \ and\ \bibinfo {author}
  {\bibfnamefont {S.}~\bibnamefont {Zhou}},\ }\href {\doibase
  10.1038/nphys3856} {\bibfield  {journal} {\bibinfo  {journal} {Nature
  Physics}\ }\textbf {\bibinfo {volume} {12}},\ \bibinfo {pages} {1111}
  (\bibinfo {year} {2016}{\natexlab{b}})}\BibitemShut {NoStop}%
\bibitem [{\citenamefont {Yankowitz}\ \emph
  {et~al.}(2012{\natexlab{a}})\citenamefont {Yankowitz}, \citenamefont {Xue},
  \citenamefont {Cormode}, \citenamefont {Sanchez-Yamagishi}, \citenamefont
  {Watanabe}, \citenamefont {Taniguchi}, \citenamefont {Jarillo-Herrero},
  \citenamefont {Jacquod},\ and\ \citenamefont {LeRoy}}]{SDC_Yankowitz}%
  \BibitemOpen
  \bibfield  {author} {\bibinfo {author} {\bibfnamefont {M.}~\bibnamefont
  {Yankowitz}}, \bibinfo {author} {\bibfnamefont {J.}~\bibnamefont {Xue}},
  \bibinfo {author} {\bibfnamefont {D.}~\bibnamefont {Cormode}}, \bibinfo
  {author} {\bibfnamefont {J.~D.}\ \bibnamefont {Sanchez-Yamagishi}}, \bibinfo
  {author} {\bibfnamefont {K.}~\bibnamefont {Watanabe}}, \bibinfo {author}
  {\bibfnamefont {T.}~\bibnamefont {Taniguchi}}, \bibinfo {author}
  {\bibfnamefont {P.}~\bibnamefont {Jarillo-Herrero}}, \bibinfo {author}
  {\bibfnamefont {P.}~\bibnamefont {Jacquod}}, \ and\ \bibinfo {author}
  {\bibfnamefont {B.~J.}\ \bibnamefont {LeRoy}},\ }\href {\doibase
  10.1038/nphys2272} {\bibfield  {journal} {\bibinfo  {journal} {Nature
  Physics}\ }\textbf {\bibinfo {volume} {8}},\ \bibinfo {pages} {382} (\bibinfo
  {year} {2012}{\natexlab{a}})}\BibitemShut {NoStop}%
\bibitem [{\citenamefont {Mucha-Kruczy\ifmmode~\acute{n}\else \'{n}\fi{}ski}\
  \emph {et~al.}(2016{\natexlab{b}})\citenamefont
  {Mucha-Kruczy\ifmmode~\acute{n}\else \'{n}\fi{}ski}, \citenamefont
  {Wallbank},\ and\ \citenamefont {Fal'ko}}]{ARPES_theo_Falko}%
  \BibitemOpen
  \bibfield  {author} {\bibinfo {author} {\bibfnamefont {M.}~\bibnamefont
  {Mucha-Kruczy\ifmmode~\acute{n}\else \'{n}\fi{}ski}}, \bibinfo {author}
  {\bibfnamefont {J.~R.}\ \bibnamefont {Wallbank}}, \ and\ \bibinfo {author}
  {\bibfnamefont {V.~I.}\ \bibnamefont {Fal'ko}},\ }\href {\doibase
  10.1103/PhysRevB.93.085409} {\bibfield  {journal} {\bibinfo  {journal} {Phys.
  Rev. B}\ }\textbf {\bibinfo {volume} {93}},\ \bibinfo {pages} {085409}
  (\bibinfo {year} {2016}{\natexlab{b}})}\BibitemShut {NoStop}%
\bibitem [{\citenamefont {Ulstrup}\ \emph {et~al.}(2015)\citenamefont
  {Ulstrup}, \citenamefont {Johannsen}, \citenamefont {Crepaldi}, \citenamefont
  {Cilento}, \citenamefont {Zacchigna}, \citenamefont {Cacho}, \citenamefont
  {Chapman}, \citenamefont {Springate}, \citenamefont {Fromm}, \citenamefont
  {Raidel}, \citenamefont {Seyller}, \citenamefont {Parmigiani}, \citenamefont
  {Grioni},\ and\ \citenamefont {Hofmann}}]{ARPES_G_Soren}%
  \BibitemOpen
  \bibfield  {author} {\bibinfo {author} {\bibfnamefont {S.}~\bibnamefont
  {Ulstrup}}, \bibinfo {author} {\bibfnamefont {J.~C.}\ \bibnamefont
  {Johannsen}}, \bibinfo {author} {\bibfnamefont {A.}~\bibnamefont {Crepaldi}},
  \bibinfo {author} {\bibfnamefont {F.}~\bibnamefont {Cilento}}, \bibinfo
  {author} {\bibfnamefont {M.}~\bibnamefont {Zacchigna}}, \bibinfo {author}
  {\bibfnamefont {C.}~\bibnamefont {Cacho}}, \bibinfo {author} {\bibfnamefont
  {R.~T.}\ \bibnamefont {Chapman}}, \bibinfo {author} {\bibfnamefont
  {E.}~\bibnamefont {Springate}}, \bibinfo {author} {\bibfnamefont
  {F.}~\bibnamefont {Fromm}}, \bibinfo {author} {\bibfnamefont
  {C.}~\bibnamefont {Raidel}}, \bibinfo {author} {\bibfnamefont
  {T.}~\bibnamefont {Seyller}}, \bibinfo {author} {\bibfnamefont
  {F.}~\bibnamefont {Parmigiani}}, \bibinfo {author} {\bibfnamefont
  {M.}~\bibnamefont {Grioni}}, \ and\ \bibinfo {author} {\bibfnamefont
  {P.}~\bibnamefont {Hofmann}},\ }\href {\doibase
  10.1088/0953-8984/27/16/164206} {\bibfield  {journal} {\bibinfo  {journal}
  {Journal of Physics: Condensed Matter}\ }\textbf {\bibinfo {volume} {27}},\
  \bibinfo {pages} {164206} (\bibinfo {year} {2015})}\BibitemShut {NoStop}%
\bibitem [{\citenamefont {Liu}\ \emph {et~al.}(2011)\citenamefont {Liu},
  \citenamefont {Bian}, \citenamefont {Miller},\ and\ \citenamefont
  {Chiang}}]{ARPES_polarization_Liu}%
  \BibitemOpen
  \bibfield  {author} {\bibinfo {author} {\bibfnamefont {Y.}~\bibnamefont
  {Liu}}, \bibinfo {author} {\bibfnamefont {G.}~\bibnamefont {Bian}}, \bibinfo
  {author} {\bibfnamefont {T.}~\bibnamefont {Miller}}, \ and\ \bibinfo {author}
  {\bibfnamefont {T.-C.}\ \bibnamefont {Chiang}},\ }\href {\doibase
  10.1103/PhysRevLett.107.166803} {\bibfield  {journal} {\bibinfo  {journal}
  {Phys. Rev. Lett.}\ }\textbf {\bibinfo {volume} {107}},\ \bibinfo {pages}
  {166803} (\bibinfo {year} {2011})}\BibitemShut {NoStop}%
\bibitem [{\citenamefont {Moser}(2017)}]{ARPES_review_Moser}%
  \BibitemOpen
  \bibfield  {author} {\bibinfo {author} {\bibfnamefont {S.}~\bibnamefont
  {Moser}},\ }\href {\doibase https://doi.org/10.1016/j.elspec.2016.11.007}
  {\bibfield  {journal} {\bibinfo  {journal} {Journal of Electron Spectroscopy
  and Related Phenomena}\ }\textbf {\bibinfo {volume} {214}},\ \bibinfo {pages}
  {29 } (\bibinfo {year} {2017})}\BibitemShut {NoStop}%
\bibitem [{\citenamefont {Gierz}\ \emph {et~al.}(2012)\citenamefont {Gierz},
  \citenamefont {Lindroos}, \citenamefont {Höchst}, \citenamefont {Ast},\ and\
  \citenamefont {Kern}}]{Gierz_G_subsym}%
  \BibitemOpen
  \bibfield  {author} {\bibinfo {author} {\bibfnamefont {I.}~\bibnamefont
  {Gierz}}, \bibinfo {author} {\bibfnamefont {M.}~\bibnamefont {Lindroos}},
  \bibinfo {author} {\bibfnamefont {H.}~\bibnamefont {Höchst}}, \bibinfo
  {author} {\bibfnamefont {C.~R.}\ \bibnamefont {Ast}}, \ and\ \bibinfo
  {author} {\bibfnamefont {K.}~\bibnamefont {Kern}},\ }\href {\doibase
  10.1021/nl300512q} {\bibfield  {journal} {\bibinfo  {journal} {Nano Letters}\
  }\textbf {\bibinfo {volume} {12}},\ \bibinfo {pages} {3900} (\bibinfo {year}
  {2012})},\ \bibinfo {note} {pMID: 22784029}\BibitemShut {NoStop}%
\bibitem [{\citenamefont {Polshyn}\ \emph {et~al.}()\citenamefont {Polshyn},
  \citenamefont {Zhu}, \citenamefont {Kumar}, \citenamefont {Zhang},
  \citenamefont {Yang}, \citenamefont {Tschirhart}, \citenamefont {Serlin},
  \citenamefont {Watanabe}, \citenamefont {Taniguchi}, \citenamefont
  {MacDonald},\ and\ \citenamefont {Young}}]{QAHE_Polshyn}%
  \BibitemOpen
  \bibfield  {author} {\bibinfo {author} {\bibfnamefont {H.}~\bibnamefont
  {Polshyn}}, \bibinfo {author} {\bibfnamefont {J.}~\bibnamefont {Zhu}},
  \bibinfo {author} {\bibfnamefont {M.~A.}\ \bibnamefont {Kumar}}, \bibinfo
  {author} {\bibfnamefont {Y.}~\bibnamefont {Zhang}}, \bibinfo {author}
  {\bibfnamefont {F.}~\bibnamefont {Yang}}, \bibinfo {author} {\bibfnamefont
  {C.~L.}\ \bibnamefont {Tschirhart}}, \bibinfo {author} {\bibfnamefont
  {M.}~\bibnamefont {Serlin}}, \bibinfo {author} {\bibfnamefont
  {K.}~\bibnamefont {Watanabe}}, \bibinfo {author} {\bibfnamefont
  {T.}~\bibnamefont {Taniguchi}}, \bibinfo {author} {\bibfnamefont {A.~H.}\
  \bibnamefont {MacDonald}}, \ and\ \bibinfo {author} {\bibfnamefont {A.~F.}\
  \bibnamefont {Young}},\ }\href {\doibase arXiv:2004.11353} {\
  arXiv:2004.11353}\BibitemShut {NoStop}%
\bibitem [{\citenamefont {Stepanov}\ \emph {et~al.}()\citenamefont {Stepanov},
  \citenamefont {Das}, \citenamefont {Lu}, \citenamefont {Fahimniya},
  \citenamefont {Watanabe}, \citenamefont {Taniguchi}, \citenamefont {Koppens},
  \citenamefont {Johannes~Lischner},\ and\ \citenamefont
  {Efetov}}]{QAHE_Stepanov}%
  \BibitemOpen
  \bibfield  {author} {\bibinfo {author} {\bibfnamefont {P.}~\bibnamefont
  {Stepanov}}, \bibinfo {author} {\bibfnamefont {I.}~\bibnamefont {Das}},
  \bibinfo {author} {\bibfnamefont {X.}~\bibnamefont {Lu}}, \bibinfo {author}
  {\bibfnamefont {A.}~\bibnamefont {Fahimniya}}, \bibinfo {author}
  {\bibfnamefont {K.}~\bibnamefont {Watanabe}}, \bibinfo {author}
  {\bibfnamefont {T.}~\bibnamefont {Taniguchi}}, \bibinfo {author}
  {\bibfnamefont {F.~H.~L.}\ \bibnamefont {Koppens}}, \bibinfo {author}
  {\bibfnamefont {L.~L.}\ \bibnamefont {Johannes~Lischner}}, \ and\ \bibinfo
  {author} {\bibfnamefont {D.~K.}\ \bibnamefont {Efetov}},\ }\href {\doibase
  arXiv:1911.09198} {\ arXiv:1911.09198}\BibitemShut {NoStop}%
\bibitem [{\citenamefont {Bultinck}\ \emph {et~al.}(2020)\citenamefont
  {Bultinck}, \citenamefont {Chatterjee},\ and\ \citenamefont
  {Zaletel}}]{QAHE_theo_Bultinck}%
  \BibitemOpen
  \bibfield  {author} {\bibinfo {author} {\bibfnamefont {N.}~\bibnamefont
  {Bultinck}}, \bibinfo {author} {\bibfnamefont {S.}~\bibnamefont
  {Chatterjee}}, \ and\ \bibinfo {author} {\bibfnamefont {M.~P.}\ \bibnamefont
  {Zaletel}},\ }\href {\doibase 10.1103/PhysRevLett.124.166601} {\bibfield
  {journal} {\bibinfo  {journal} {Phys. Rev. Lett.}\ }\textbf {\bibinfo
  {volume} {124}},\ \bibinfo {pages} {166601} (\bibinfo {year}
  {2020})}\BibitemShut {NoStop}%
\bibitem [{\citenamefont {Zhang}\ \emph
  {et~al.}(2019{\natexlab{a}})\citenamefont {Zhang}, \citenamefont {Mao},
  \citenamefont {Cao}, \citenamefont {Jarillo-Herrero},\ and\ \citenamefont
  {Senthil}}]{QAHE_theo_Ya-Hui}%
  \BibitemOpen
  \bibfield  {author} {\bibinfo {author} {\bibfnamefont {Y.-H.}\ \bibnamefont
  {Zhang}}, \bibinfo {author} {\bibfnamefont {D.}~\bibnamefont {Mao}}, \bibinfo
  {author} {\bibfnamefont {Y.}~\bibnamefont {Cao}}, \bibinfo {author}
  {\bibfnamefont {P.}~\bibnamefont {Jarillo-Herrero}}, \ and\ \bibinfo {author}
  {\bibfnamefont {T.}~\bibnamefont {Senthil}},\ }\href {\doibase
  10.1103/PhysRevB.99.075127} {\bibfield  {journal} {\bibinfo  {journal} {Phys.
  Rev. B}\ }\textbf {\bibinfo {volume} {99}},\ \bibinfo {pages} {075127}
  (\bibinfo {year} {2019}{\natexlab{a}})}\BibitemShut {NoStop}%
\bibitem [{\citenamefont {Zhang}\ \emph
  {et~al.}(2019{\natexlab{b}})\citenamefont {Zhang}, \citenamefont {Mao},\ and\
  \citenamefont {Senthil}}]{QAHE_theo_Ya-Hui_PRR}%
  \BibitemOpen
  \bibfield  {author} {\bibinfo {author} {\bibfnamefont {Y.-H.}\ \bibnamefont
  {Zhang}}, \bibinfo {author} {\bibfnamefont {D.}~\bibnamefont {Mao}}, \ and\
  \bibinfo {author} {\bibfnamefont {T.}~\bibnamefont {Senthil}},\ }\href
  {\doibase 10.1103/PhysRevResearch.1.033126} {\bibfield  {journal} {\bibinfo
  {journal} {Phys. Rev. Research}\ }\textbf {\bibinfo {volume} {1}},\ \bibinfo
  {pages} {033126} (\bibinfo {year} {2019}{\natexlab{b}})}\BibitemShut
  {NoStop}%
\bibitem [{\citenamefont {Song}\ \emph {et~al.}(2015)\citenamefont {Song},
  \citenamefont {Samutpraphoot},\ and\ \citenamefont
  {Levitov}}]{QAHE_theo_Song}%
  \BibitemOpen
  \bibfield  {author} {\bibinfo {author} {\bibfnamefont {J.~C.~W.}\
  \bibnamefont {Song}}, \bibinfo {author} {\bibfnamefont {P.}~\bibnamefont
  {Samutpraphoot}}, \ and\ \bibinfo {author} {\bibfnamefont {L.~S.}\
  \bibnamefont {Levitov}},\ }\href {\doibase 10.1073/pnas.1424760112}
  {\bibfield  {journal} {\bibinfo  {journal} {Proceedings of the National
  Academy of Sciences}\ }\textbf {\bibinfo {volume} {112}},\ \bibinfo {pages}
  {10879} (\bibinfo {year} {2015})},\ \Eprint
  {http://arxiv.org/abs/https://www.pnas.org/content/112/35/10879.full.pdf}
  {https://www.pnas.org/content/112/35/10879.full.pdf} \BibitemShut {NoStop}%
\bibitem [{\citenamefont {Xiao}\ \emph {et~al.}(2007)\citenamefont {Xiao},
  \citenamefont {Yao},\ and\ \citenamefont {Niu}}]{QAHE_theo_XiaoDi}%
  \BibitemOpen
  \bibfield  {author} {\bibinfo {author} {\bibfnamefont {D.}~\bibnamefont
  {Xiao}}, \bibinfo {author} {\bibfnamefont {W.}~\bibnamefont {Yao}}, \ and\
  \bibinfo {author} {\bibfnamefont {Q.}~\bibnamefont {Niu}},\ }\href {\doibase
  10.1103/PhysRevLett.99.236809} {\bibfield  {journal} {\bibinfo  {journal}
  {Phys. Rev. Lett.}\ }\textbf {\bibinfo {volume} {99}},\ \bibinfo {pages}
  {236809} (\bibinfo {year} {2007})}\BibitemShut {NoStop}%
\bibitem [{\citenamefont {Wolf}\ \emph {et~al.}(2018)\citenamefont {Wolf},
  \citenamefont {Zilberberg}, \citenamefont {Levkivskyi},\ and\ \citenamefont
  {Blatter}}]{QAHE_theo_Wolf}%
  \BibitemOpen
  \bibfield  {author} {\bibinfo {author} {\bibfnamefont {T.~M.~R.}\
  \bibnamefont {Wolf}}, \bibinfo {author} {\bibfnamefont {O.}~\bibnamefont
  {Zilberberg}}, \bibinfo {author} {\bibfnamefont {I.}~\bibnamefont
  {Levkivskyi}}, \ and\ \bibinfo {author} {\bibfnamefont {G.}~\bibnamefont
  {Blatter}},\ }\href {\doibase 10.1103/PhysRevB.98.125408} {\bibfield
  {journal} {\bibinfo  {journal} {Phys. Rev. B}\ }\textbf {\bibinfo {volume}
  {98}},\ \bibinfo {pages} {125408} (\bibinfo {year} {2018})}\BibitemShut
  {NoStop}%
\bibitem [{\citenamefont {Liu}\ and\ \citenamefont
  {Dai}()}]{QAHE_theo_Jianpeng}%
  \BibitemOpen
  \bibfield  {author} {\bibinfo {author} {\bibfnamefont {J.}~\bibnamefont
  {Liu}}\ and\ \bibinfo {author} {\bibfnamefont {X.}~\bibnamefont {Dai}},\
  }\href {\doibase arXiv:1911.03760} {\ arXiv:1911.03760}\BibitemShut {NoStop}%
\bibitem [{\citenamefont {Repellin}\ \emph {et~al.}(2020)\citenamefont
  {Repellin}, \citenamefont {Dong}, \citenamefont {Zhang},\ and\ \citenamefont
  {Senthil}}]{QAHE_theo_Repellin}%
  \BibitemOpen
  \bibfield  {author} {\bibinfo {author} {\bibfnamefont {C.}~\bibnamefont
  {Repellin}}, \bibinfo {author} {\bibfnamefont {Z.}~\bibnamefont {Dong}},
  \bibinfo {author} {\bibfnamefont {Y.-H.}\ \bibnamefont {Zhang}}, \ and\
  \bibinfo {author} {\bibfnamefont {T.}~\bibnamefont {Senthil}},\ }\href
  {\doibase 10.1103/PhysRevLett.124.187601} {\bibfield  {journal} {\bibinfo
  {journal} {Phys. Rev. Lett.}\ }\textbf {\bibinfo {volume} {124}},\ \bibinfo
  {pages} {187601} (\bibinfo {year} {2020})}\BibitemShut {NoStop}%
\bibitem [{\citenamefont {Wu}\ and\ \citenamefont
  {Das~Sarma}(2020)}]{QAHE_theo_Fengcheng}%
  \BibitemOpen
  \bibfield  {author} {\bibinfo {author} {\bibfnamefont {F.}~\bibnamefont
  {Wu}}\ and\ \bibinfo {author} {\bibfnamefont {S.}~\bibnamefont {Das~Sarma}},\
  }\href {\doibase 10.1103/PhysRevLett.124.046403} {\bibfield  {journal}
  {\bibinfo  {journal} {Phys. Rev. Lett.}\ }\textbf {\bibinfo {volume} {124}},\
  \bibinfo {pages} {046403} (\bibinfo {year} {2020})}\BibitemShut {NoStop}%
\bibitem [{\citenamefont {Uchida}\ \emph {et~al.}(2014)\citenamefont {Uchida},
  \citenamefont {Furuya}, \citenamefont {Iwata},\ and\ \citenamefont
  {Oshiyama}}]{Uchida_tBLG_corrugation}%
  \BibitemOpen
  \bibfield  {author} {\bibinfo {author} {\bibfnamefont {K.}~\bibnamefont
  {Uchida}}, \bibinfo {author} {\bibfnamefont {S.}~\bibnamefont {Furuya}},
  \bibinfo {author} {\bibfnamefont {J.-I.}\ \bibnamefont {Iwata}}, \ and\
  \bibinfo {author} {\bibfnamefont {A.}~\bibnamefont {Oshiyama}},\ }\href
  {\doibase 10.1103/PhysRevB.90.155451} {\bibfield  {journal} {\bibinfo
  {journal} {Phys. Rev. B}\ }\textbf {\bibinfo {volume} {90}},\ \bibinfo
  {pages} {155451} (\bibinfo {year} {2014})}\BibitemShut {NoStop}%
\bibitem [{\citenamefont {Koshino}\ \emph {et~al.}(2018)\citenamefont
  {Koshino}, \citenamefont {Yuan}, \citenamefont {Koretsune}, \citenamefont
  {Ochi}, \citenamefont {Kuroki},\ and\ \citenamefont
  {Fu}}]{Koshino_tBLG_corrugation}%
  \BibitemOpen
  \bibfield  {author} {\bibinfo {author} {\bibfnamefont {M.}~\bibnamefont
  {Koshino}}, \bibinfo {author} {\bibfnamefont {N.~F.~Q.}\ \bibnamefont
  {Yuan}}, \bibinfo {author} {\bibfnamefont {T.}~\bibnamefont {Koretsune}},
  \bibinfo {author} {\bibfnamefont {M.}~\bibnamefont {Ochi}}, \bibinfo {author}
  {\bibfnamefont {K.}~\bibnamefont {Kuroki}}, \ and\ \bibinfo {author}
  {\bibfnamefont {L.}~\bibnamefont {Fu}},\ }\href {\doibase
  10.1103/PhysRevX.8.031087} {\bibfield  {journal} {\bibinfo  {journal} {Phys.
  Rev. X}\ }\textbf {\bibinfo {volume} {8}},\ \bibinfo {pages} {031087}
  (\bibinfo {year} {2018})}\BibitemShut {NoStop}%
\bibitem [{\citenamefont {van Wijk}\ \emph {et~al.}(2015)\citenamefont {van
  Wijk}, \citenamefont {Schuring}, \citenamefont {Katsnelson},\ and\
  \citenamefont {Fasolino}}]{tBLG_corrugation_2015}%
  \BibitemOpen
  \bibfield  {author} {\bibinfo {author} {\bibfnamefont {M.~M.}\ \bibnamefont
  {van Wijk}}, \bibinfo {author} {\bibfnamefont {A.}~\bibnamefont {Schuring}},
  \bibinfo {author} {\bibfnamefont {M.~I.}\ \bibnamefont {Katsnelson}}, \ and\
  \bibinfo {author} {\bibfnamefont {A.}~\bibnamefont {Fasolino}},\ }\href
  {\doibase 10.1088/2053-1583/2/3/034010} {\bibfield  {journal} {\bibinfo
  {journal} {2D Materials}\ }\textbf {\bibinfo {volume} {2}},\ \bibinfo {pages}
  {034010} (\bibinfo {year} {2015})}\BibitemShut {NoStop}%
\bibitem [{\citenamefont {Dai}\ \emph {et~al.}(2016)\citenamefont {Dai},
  \citenamefont {Xiang},\ and\ \citenamefont
  {Srolovitz}}]{tBLG_corrugation_2016}%
  \BibitemOpen
  \bibfield  {author} {\bibinfo {author} {\bibfnamefont {S.}~\bibnamefont
  {Dai}}, \bibinfo {author} {\bibfnamefont {Y.}~\bibnamefont {Xiang}}, \ and\
  \bibinfo {author} {\bibfnamefont {D.~J.}\ \bibnamefont {Srolovitz}},\ }\href
  {\doibase 10.1021/acs.nanolett.6b02870} {\bibfield  {journal} {\bibinfo
  {journal} {Nano Letters}\ }\textbf {\bibinfo {volume} {16}},\ \bibinfo
  {pages} {5923} (\bibinfo {year} {2016})},\ \bibinfo {note} {pMID:
  27533089}\BibitemShut {NoStop}%
\bibitem [{\citenamefont {Jain}\ \emph {et~al.}(2016)\citenamefont {Jain},
  \citenamefont {Juri{\v{c}}i{\'{c}}},\ and\ \citenamefont
  {Barkema}}]{tBLG_corrugation_Jain}%
  \BibitemOpen
  \bibfield  {author} {\bibinfo {author} {\bibfnamefont {S.~K.}\ \bibnamefont
  {Jain}}, \bibinfo {author} {\bibfnamefont {V.}~\bibnamefont
  {Juri{\v{c}}i{\'{c}}}}, \ and\ \bibinfo {author} {\bibfnamefont {G.~T.}\
  \bibnamefont {Barkema}},\ }\href {\doibase 10.1088/2053-1583/4/1/015018}
  {\bibfield  {journal} {\bibinfo  {journal} {2D Materials}\ }\textbf {\bibinfo
  {volume} {4}},\ \bibinfo {pages} {015018} (\bibinfo {year}
  {2016})}\BibitemShut {NoStop}%
\bibitem [{\citenamefont {Carr}\ \emph {et~al.}(2019)\citenamefont {Carr},
  \citenamefont {Fang}, \citenamefont {Zhu},\ and\ \citenamefont
  {Kaxiras}}]{tBLG_relaxation_Kaxiras}%
  \BibitemOpen
  \bibfield  {author} {\bibinfo {author} {\bibfnamefont {S.}~\bibnamefont
  {Carr}}, \bibinfo {author} {\bibfnamefont {S.}~\bibnamefont {Fang}}, \bibinfo
  {author} {\bibfnamefont {Z.}~\bibnamefont {Zhu}}, \ and\ \bibinfo {author}
  {\bibfnamefont {E.}~\bibnamefont {Kaxiras}},\ }\href {\doibase
  10.1103/PhysRevResearch.1.013001} {\bibfield  {journal} {\bibinfo  {journal}
  {Phys. Rev. Research}\ }\textbf {\bibinfo {volume} {1}},\ \bibinfo {pages}
  {013001} (\bibinfo {year} {2019})}\BibitemShut {NoStop}%
\bibitem [{\citenamefont {Nam}\ and\ \citenamefont
  {Koshino}(2017)}]{tBLG_relaxation_Koshino}%
  \BibitemOpen
  \bibfield  {author} {\bibinfo {author} {\bibfnamefont {N.~N.~T.}\
  \bibnamefont {Nam}}\ and\ \bibinfo {author} {\bibfnamefont {M.}~\bibnamefont
  {Koshino}},\ }\href {\doibase 10.1103/PhysRevB.96.075311} {\bibfield
  {journal} {\bibinfo  {journal} {Phys. Rev. B}\ }\textbf {\bibinfo {volume}
  {96}},\ \bibinfo {pages} {075311} (\bibinfo {year} {2017})}\BibitemShut
  {NoStop}%
\bibitem [{\citenamefont {Yankowitz}\ \emph
  {et~al.}(2012{\natexlab{b}})\citenamefont {Yankowitz}, \citenamefont {Xue},
  \citenamefont {Cormode}, \citenamefont {Sanchez-Yamagishi}, \citenamefont
  {Watanabe}, \citenamefont {Taniguchi}, \citenamefont {Jarillo-Herrero},
  \citenamefont {Jacquod},\ and\ \citenamefont
  {LeRoy}}]{second_Dirac_point2012}%
  \BibitemOpen
  \bibfield  {author} {\bibinfo {author} {\bibfnamefont {M.}~\bibnamefont
  {Yankowitz}}, \bibinfo {author} {\bibfnamefont {J.}~\bibnamefont {Xue}},
  \bibinfo {author} {\bibfnamefont {D.}~\bibnamefont {Cormode}}, \bibinfo
  {author} {\bibfnamefont {J.~D.}\ \bibnamefont {Sanchez-Yamagishi}}, \bibinfo
  {author} {\bibfnamefont {K.}~\bibnamefont {Watanabe}}, \bibinfo {author}
  {\bibfnamefont {T.}~\bibnamefont {Taniguchi}}, \bibinfo {author}
  {\bibfnamefont {P.}~\bibnamefont {Jarillo-Herrero}}, \bibinfo {author}
  {\bibfnamefont {P.}~\bibnamefont {Jacquod}}, \ and\ \bibinfo {author}
  {\bibfnamefont {B.~J.}\ \bibnamefont {LeRoy}},\ }\href
  {https://doi.org/10.1038/nphys2272} {\bibfield  {journal} {\bibinfo
  {journal} {Nature Physics}\ }\textbf {\bibinfo {volume} {8}},\ \bibinfo
  {pages} {382 EP } (\bibinfo {year} {2012}{\natexlab{b}})}\BibitemShut
  {NoStop}%
\bibitem [{\citenamefont {Ponomarenko}\ \emph
  {et~al.}(2013{\natexlab{b}})\citenamefont {Ponomarenko}, \citenamefont
  {Gorbachev}, \citenamefont {Yu}, \citenamefont {Elias}, \citenamefont
  {Jalil}, \citenamefont {Patel}, \citenamefont {Mishchenko}, \citenamefont
  {Mayorov}, \citenamefont {Woods}, \citenamefont {Wallbank}, \citenamefont
  {Mucha-Kruczynski}, \citenamefont {Piot}, \citenamefont {Potemski},
  \citenamefont {Grigorieva}, \citenamefont {Novoselov}, \citenamefont
  {Guinea}, \citenamefont {Fal’ko},\ and\ \citenamefont
  {Geim}}]{second_Dirac_point2013}%
  \BibitemOpen
  \bibfield  {author} {\bibinfo {author} {\bibfnamefont {L.~A.}\ \bibnamefont
  {Ponomarenko}}, \bibinfo {author} {\bibfnamefont {R.~V.}\ \bibnamefont
  {Gorbachev}}, \bibinfo {author} {\bibfnamefont {G.~L.}\ \bibnamefont {Yu}},
  \bibinfo {author} {\bibfnamefont {D.~C.}\ \bibnamefont {Elias}}, \bibinfo
  {author} {\bibfnamefont {R.}~\bibnamefont {Jalil}}, \bibinfo {author}
  {\bibfnamefont {A.~A.}\ \bibnamefont {Patel}}, \bibinfo {author}
  {\bibfnamefont {A.}~\bibnamefont {Mishchenko}}, \bibinfo {author}
  {\bibfnamefont {A.~S.}\ \bibnamefont {Mayorov}}, \bibinfo {author}
  {\bibfnamefont {C.~R.}\ \bibnamefont {Woods}}, \bibinfo {author}
  {\bibfnamefont {J.~R.}\ \bibnamefont {Wallbank}}, \bibinfo {author}
  {\bibfnamefont {M.}~\bibnamefont {Mucha-Kruczynski}}, \bibinfo {author}
  {\bibfnamefont {B.~A.}\ \bibnamefont {Piot}}, \bibinfo {author}
  {\bibfnamefont {M.}~\bibnamefont {Potemski}}, \bibinfo {author}
  {\bibfnamefont {I.~V.}\ \bibnamefont {Grigorieva}}, \bibinfo {author}
  {\bibfnamefont {K.~S.}\ \bibnamefont {Novoselov}}, \bibinfo {author}
  {\bibfnamefont {F.}~\bibnamefont {Guinea}}, \bibinfo {author} {\bibfnamefont
  {V.~I.}\ \bibnamefont {Fal’ko}}, \ and\ \bibinfo {author} {\bibfnamefont
  {A.~K.}\ \bibnamefont {Geim}},\ }\href {https://doi.org/10.1038/nature12187}
  {\bibfield  {journal} {\bibinfo  {journal} {Nature}\ }\textbf {\bibinfo
  {volume} {497}},\ \bibinfo {pages} {594 EP } (\bibinfo {year}
  {2013}{\natexlab{b}})}\BibitemShut {NoStop}%
\bibitem [{\citenamefont {Bi}\ \emph {et~al.}(2019)\citenamefont {Bi},
  \citenamefont {Yuan},\ and\ \citenamefont {Fu}}]{C3_strain_Bi}%
  \BibitemOpen
  \bibfield  {author} {\bibinfo {author} {\bibfnamefont {Z.}~\bibnamefont
  {Bi}}, \bibinfo {author} {\bibfnamefont {N.~F.~Q.}\ \bibnamefont {Yuan}}, \
  and\ \bibinfo {author} {\bibfnamefont {L.}~\bibnamefont {Fu}},\ }\href
  {\doibase 10.1103/PhysRevB.100.035448} {\bibfield  {journal} {\bibinfo
  {journal} {Phys. Rev. B}\ }\textbf {\bibinfo {volume} {100}},\ \bibinfo
  {pages} {035448} (\bibinfo {year} {2019})}\BibitemShut {NoStop}%
\bibitem [{\citenamefont {Liu}\ \emph {et~al.}()\citenamefont {Liu},
  \citenamefont {Khalaf}, \citenamefont {Lee},\ and\ \citenamefont
  {Vishwanath}}]{C3_Shang}%
  \BibitemOpen
  \bibfield  {author} {\bibinfo {author} {\bibfnamefont {S.}~\bibnamefont
  {Liu}}, \bibinfo {author} {\bibfnamefont {E.}~\bibnamefont {Khalaf}},
  \bibinfo {author} {\bibfnamefont {J.~Y.}\ \bibnamefont {Lee}}, \ and\
  \bibinfo {author} {\bibfnamefont {A.}~\bibnamefont {Vishwanath}},\ }\href
  {\doibase arXiv:1905.07409} {\ arXiv:1905.07409}\BibitemShut {NoStop}%
\bibitem [{\citenamefont {Klug}()}]{C3_Markus}%
  \BibitemOpen
  \bibfield  {author} {\bibinfo {author} {\bibfnamefont {M.~J.}\ \bibnamefont
  {Klug}},\ }\href {\doibase arXiv:1909.03074} {\ arXiv:1909.03074}\BibitemShut
  {NoStop}%
\bibitem [{\citenamefont {Jiang}\ \emph {et~al.}(2019)\citenamefont {Jiang},
  \citenamefont {Lai}, \citenamefont {Watanabe}, \citenamefont {Taniguchi},
  \citenamefont {Haule}, \citenamefont {Mao},\ and\ \citenamefont
  {Andrei}}]{TBG_STM_Jiang}%
  \BibitemOpen
  \bibfield  {author} {\bibinfo {author} {\bibfnamefont {Y.}~\bibnamefont
  {Jiang}}, \bibinfo {author} {\bibfnamefont {X.}~\bibnamefont {Lai}}, \bibinfo
  {author} {\bibfnamefont {K.}~\bibnamefont {Watanabe}}, \bibinfo {author}
  {\bibfnamefont {T.}~\bibnamefont {Taniguchi}}, \bibinfo {author}
  {\bibfnamefont {K.}~\bibnamefont {Haule}}, \bibinfo {author} {\bibfnamefont
  {J.}~\bibnamefont {Mao}}, \ and\ \bibinfo {author} {\bibfnamefont {E.~Y.}\
  \bibnamefont {Andrei}},\ }\href {\doibase 10.1038/s41586-019-1460-4}
  {\bibfield  {journal} {\bibinfo  {journal} {Nature}\ }\textbf {\bibinfo
  {volume} {573}},\ \bibinfo {pages} {91} (\bibinfo {year} {2019})}\BibitemShut
  {NoStop}%
\bibitem [{\citenamefont {Kerelsky}\ \emph {et~al.}(2019)\citenamefont
  {Kerelsky}, \citenamefont {McGilly}, \citenamefont {Kennes}, \citenamefont
  {Xian}, \citenamefont {Yankowitz}, \citenamefont {Chen}, \citenamefont
  {Watanabe}, \citenamefont {Taniguchi}, \citenamefont {Hone}, \citenamefont
  {Dean}, \citenamefont {Rubio},\ and\ \citenamefont
  {Pasupathy}}]{TBG_STM_Kerelsky}%
  \BibitemOpen
  \bibfield  {author} {\bibinfo {author} {\bibfnamefont {A.}~\bibnamefont
  {Kerelsky}}, \bibinfo {author} {\bibfnamefont {L.~J.}\ \bibnamefont
  {McGilly}}, \bibinfo {author} {\bibfnamefont {D.~M.}\ \bibnamefont {Kennes}},
  \bibinfo {author} {\bibfnamefont {L.}~\bibnamefont {Xian}}, \bibinfo {author}
  {\bibfnamefont {M.}~\bibnamefont {Yankowitz}}, \bibinfo {author}
  {\bibfnamefont {S.}~\bibnamefont {Chen}}, \bibinfo {author} {\bibfnamefont
  {K.}~\bibnamefont {Watanabe}}, \bibinfo {author} {\bibfnamefont
  {T.}~\bibnamefont {Taniguchi}}, \bibinfo {author} {\bibfnamefont
  {J.}~\bibnamefont {Hone}}, \bibinfo {author} {\bibfnamefont {C.}~\bibnamefont
  {Dean}}, \bibinfo {author} {\bibfnamefont {A.}~\bibnamefont {Rubio}}, \ and\
  \bibinfo {author} {\bibfnamefont {A.~N.}\ \bibnamefont {Pasupathy}},\ }\href
  {\doibase 10.1038/s41586-019-1431-9} {\bibfield  {journal} {\bibinfo
  {journal} {Nature}\ }\textbf {\bibinfo {volume} {572}},\ \bibinfo {pages}
  {95} (\bibinfo {year} {2019})}\BibitemShut {NoStop}%
\bibitem [{\citenamefont {Choi}\ \emph {et~al.}(2019)\citenamefont {Choi},
  \citenamefont {Kemmer}, \citenamefont {Peng}, \citenamefont {Thomson},
  \citenamefont {Arora}, \citenamefont {Polski}, \citenamefont {Zhang},
  \citenamefont {Ren}, \citenamefont {Alicea}, \citenamefont {Refael},
  \citenamefont {von Oppen}, \citenamefont {Watanabe}, \citenamefont
  {Taniguchi},\ and\ \citenamefont {Nadj-Perge}}]{TBG_STM_Choi}%
  \BibitemOpen
  \bibfield  {author} {\bibinfo {author} {\bibfnamefont {Y.}~\bibnamefont
  {Choi}}, \bibinfo {author} {\bibfnamefont {J.}~\bibnamefont {Kemmer}},
  \bibinfo {author} {\bibfnamefont {Y.}~\bibnamefont {Peng}}, \bibinfo {author}
  {\bibfnamefont {A.}~\bibnamefont {Thomson}}, \bibinfo {author} {\bibfnamefont
  {H.}~\bibnamefont {Arora}}, \bibinfo {author} {\bibfnamefont
  {R.}~\bibnamefont {Polski}}, \bibinfo {author} {\bibfnamefont
  {Y.}~\bibnamefont {Zhang}}, \bibinfo {author} {\bibfnamefont
  {H.}~\bibnamefont {Ren}}, \bibinfo {author} {\bibfnamefont {J.}~\bibnamefont
  {Alicea}}, \bibinfo {author} {\bibfnamefont {G.}~\bibnamefont {Refael}},
  \bibinfo {author} {\bibfnamefont {F.}~\bibnamefont {von Oppen}}, \bibinfo
  {author} {\bibfnamefont {K.}~\bibnamefont {Watanabe}}, \bibinfo {author}
  {\bibfnamefont {T.}~\bibnamefont {Taniguchi}}, \ and\ \bibinfo {author}
  {\bibfnamefont {S.}~\bibnamefont {Nadj-Perge}},\ }\href {\doibase
  10.1038/s41567-019-0606-5} {\bibfield  {journal} {\bibinfo  {journal} {Nature
  Physics}\ }\textbf {\bibinfo {volume} {15}},\ \bibinfo {pages} {1174}
  (\bibinfo {year} {2019})}\BibitemShut {NoStop}%
\bibitem [{\citenamefont {Xie}\ \emph {et~al.}(2019)\citenamefont {Xie},
  \citenamefont {Lian}, \citenamefont {Jäck}, \citenamefont {Liu},
  \citenamefont {Chiu}, \citenamefont {Watanabe}, \citenamefont {Taniguchi},
  \citenamefont {Bernevig},\ and\ \citenamefont {Yazdani}}]{TBG_STM_Xie}%
  \BibitemOpen
  \bibfield  {author} {\bibinfo {author} {\bibfnamefont {Y.}~\bibnamefont
  {Xie}}, \bibinfo {author} {\bibfnamefont {B.}~\bibnamefont {Lian}}, \bibinfo
  {author} {\bibfnamefont {B.}~\bibnamefont {Jäck}}, \bibinfo {author}
  {\bibfnamefont {X.}~\bibnamefont {Liu}}, \bibinfo {author} {\bibfnamefont
  {C.-L.}\ \bibnamefont {Chiu}}, \bibinfo {author} {\bibfnamefont
  {K.}~\bibnamefont {Watanabe}}, \bibinfo {author} {\bibfnamefont
  {T.}~\bibnamefont {Taniguchi}}, \bibinfo {author} {\bibfnamefont {B.~A.}\
  \bibnamefont {Bernevig}}, \ and\ \bibinfo {author} {\bibfnamefont
  {A.}~\bibnamefont {Yazdani}},\ }\href {\doibase 10.1038/s41586-019-1422-x}
  {\bibfield  {journal} {\bibinfo  {journal} {Nature}\ }\textbf {\bibinfo
  {volume} {572}},\ \bibinfo {pages} {101} (\bibinfo {year}
  {2019})}\BibitemShut {NoStop}%
\bibitem [{\citenamefont {Wong}\ \emph {et~al.}()\citenamefont {Wong},
  \citenamefont {Nuckolls}, \citenamefont {Oh}, \citenamefont {Lian},
  \citenamefont {Xie}, \citenamefont {Jeon}, \citenamefont {Watanabe},
  \citenamefont {Taniguchi}, \citenamefont {Bernevig},\ and\ \citenamefont
  {Yazdani}}]{TBG_STM_Dillon}%
  \BibitemOpen
  \bibfield  {author} {\bibinfo {author} {\bibfnamefont {D.}~\bibnamefont
  {Wong}}, \bibinfo {author} {\bibfnamefont {K.~P.}\ \bibnamefont {Nuckolls}},
  \bibinfo {author} {\bibfnamefont {M.}~\bibnamefont {Oh}}, \bibinfo {author}
  {\bibfnamefont {B.}~\bibnamefont {Lian}}, \bibinfo {author} {\bibfnamefont
  {Y.}~\bibnamefont {Xie}}, \bibinfo {author} {\bibfnamefont {S.}~\bibnamefont
  {Jeon}}, \bibinfo {author} {\bibfnamefont {K.}~\bibnamefont {Watanabe}},
  \bibinfo {author} {\bibfnamefont {T.}~\bibnamefont {Taniguchi}}, \bibinfo
  {author} {\bibfnamefont {B.~A.}\ \bibnamefont {Bernevig}}, \ and\ \bibinfo
  {author} {\bibfnamefont {A.}~\bibnamefont {Yazdani}},\ }\href {\doibase
  arXiv:1912.06145} {\ arXiv:1912.06145}\BibitemShut {NoStop}%
\bibitem [{\citenamefont {Cea}\ \emph {et~al.}(2019)\citenamefont {Cea},
  \citenamefont {Walet},\ and\ \citenamefont {Guinea}}]{TBG_VHS}%
  \BibitemOpen
  \bibfield  {author} {\bibinfo {author} {\bibfnamefont {T.}~\bibnamefont
  {Cea}}, \bibinfo {author} {\bibfnamefont {N.~R.}\ \bibnamefont {Walet}}, \
  and\ \bibinfo {author} {\bibfnamefont {F.}~\bibnamefont {Guinea}},\ }\href
  {\doibase 10.1103/PhysRevB.100.205113} {\bibfield  {journal} {\bibinfo
  {journal} {Phys. Rev. B}\ }\textbf {\bibinfo {volume} {100}},\ \bibinfo
  {pages} {205113} (\bibinfo {year} {2019})}\BibitemShut {NoStop}%
\bibitem [{\citenamefont {Wu}\ \emph {et~al.}(2021)\citenamefont {Wu},
  \citenamefont {Zhang}, \citenamefont {Watanabe}, \citenamefont {Taniguchi},\
  and\ \citenamefont {Andrei}}]{VHS_Wu2021}%
  \BibitemOpen
  \bibfield  {author} {\bibinfo {author} {\bibfnamefont {S.}~\bibnamefont
  {Wu}}, \bibinfo {author} {\bibfnamefont {Z.}~\bibnamefont {Zhang}}, \bibinfo
  {author} {\bibfnamefont {K.}~\bibnamefont {Watanabe}}, \bibinfo {author}
  {\bibfnamefont {T.}~\bibnamefont {Taniguchi}}, \ and\ \bibinfo {author}
  {\bibfnamefont {E.~Y.}\ \bibnamefont {Andrei}},\ }\href {\doibase
  10.1038/s41563-020-00911-2} {\bibfield  {journal} {\bibinfo  {journal}
  {Nature Materials}\ } (\bibinfo {year} {2021}),\
  10.1038/s41563-020-00911-2}\BibitemShut {NoStop}%
\bibitem [{\citenamefont {Kohn}\ and\ \citenamefont
  {Luttinger}(1965)}]{VHS_kohn_1965}%
  \BibitemOpen
  \bibfield  {author} {\bibinfo {author} {\bibfnamefont {W.}~\bibnamefont
  {Kohn}}\ and\ \bibinfo {author} {\bibfnamefont {J.~M.}\ \bibnamefont
  {Luttinger}},\ }\href {\doibase 10.1103/PhysRevLett.15.524} {\bibfield
  {journal} {\bibinfo  {journal} {Phys. Rev. Lett.}\ }\textbf {\bibinfo
  {volume} {15}},\ \bibinfo {pages} {524} (\bibinfo {year} {1965})}\BibitemShut
  {NoStop}%
\bibitem [{\citenamefont {Markiewicz}(1997)}]{VHS_Markiewicz_1997}%
  \BibitemOpen
  \bibfield  {author} {\bibinfo {author} {\bibfnamefont {R.}~\bibnamefont
  {Markiewicz}},\ }\href {\doibase
  https://doi.org/10.1016/S0022-3697(97)00025-5} {\bibfield  {journal}
  {\bibinfo  {journal} {Journal of Physics and Chemistry of Solids}\ }\textbf
  {\bibinfo {volume} {58}},\ \bibinfo {pages} {1179 } (\bibinfo {year}
  {1997})}\BibitemShut {NoStop}%
\bibitem [{\citenamefont {Gonz\'alez}(2008)}]{VHS_Gonzalez_2008}%
  \BibitemOpen
  \bibfield  {author} {\bibinfo {author} {\bibfnamefont {J.}~\bibnamefont
  {Gonz\'alez}},\ }\href {\doibase 10.1103/PhysRevB.78.205431} {\bibfield
  {journal} {\bibinfo  {journal} {Phys. Rev. B}\ }\textbf {\bibinfo {volume}
  {78}},\ \bibinfo {pages} {205431} (\bibinfo {year} {2008})}\BibitemShut
  {NoStop}%
\bibitem [{\citenamefont {Fleck}\ \emph {et~al.}(1997)\citenamefont {Fleck},
  \citenamefont {Ole\ifmmode~\acute{s}\else \'{s}\fi{}},\ and\ \citenamefont
  {Hedin}}]{VHS_Fleck_1997}%
  \BibitemOpen
  \bibfield  {author} {\bibinfo {author} {\bibfnamefont {M.}~\bibnamefont
  {Fleck}}, \bibinfo {author} {\bibfnamefont {A.~M.}\ \bibnamefont
  {Ole\ifmmode~\acute{s}\else \'{s}\fi{}}}, \ and\ \bibinfo {author}
  {\bibfnamefont {L.}~\bibnamefont {Hedin}},\ }\href {\doibase
  10.1103/PhysRevB.56.3159} {\bibfield  {journal} {\bibinfo  {journal} {Phys.
  Rev. B}\ }\textbf {\bibinfo {volume} {56}},\ \bibinfo {pages} {3159}
  (\bibinfo {year} {1997})}\BibitemShut {NoStop}%
\bibitem [{\citenamefont {Rice}\ and\ \citenamefont
  {Scott}(1975)}]{VHS_Rice_1975}%
  \BibitemOpen
  \bibfield  {author} {\bibinfo {author} {\bibfnamefont {T.~M.}\ \bibnamefont
  {Rice}}\ and\ \bibinfo {author} {\bibfnamefont {G.~K.}\ \bibnamefont
  {Scott}},\ }\href {\doibase 10.1103/PhysRevLett.35.120} {\bibfield  {journal}
  {\bibinfo  {journal} {Phys. Rev. Lett.}\ }\textbf {\bibinfo {volume} {35}},\
  \bibinfo {pages} {120} (\bibinfo {year} {1975})}\BibitemShut {NoStop}%
\bibitem [{\citenamefont {Valenzuela}\ and\ \citenamefont
  {Vozmediano}(2008)}]{VHS_Valenzuela_2008}%
  \BibitemOpen
  \bibfield  {author} {\bibinfo {author} {\bibfnamefont {B.}~\bibnamefont
  {Valenzuela}}\ and\ \bibinfo {author} {\bibfnamefont {M.~A.~H.}\ \bibnamefont
  {Vozmediano}},\ }\href {\doibase 10.1088/1367-2630/10/11/113009} {\bibfield
  {journal} {\bibinfo  {journal} {New Journal of Physics}\ }\textbf {\bibinfo
  {volume} {10}},\ \bibinfo {pages} {113009} (\bibinfo {year}
  {2008})}\BibitemShut {NoStop}%
\bibitem [{\citenamefont {Makogon}\ \emph {et~al.}(2011)\citenamefont
  {Makogon}, \citenamefont {van Gelderen}, \citenamefont {Rold\'an},\ and\
  \citenamefont {Smith}}]{VHS_Makogon_2011}%
  \BibitemOpen
  \bibfield  {author} {\bibinfo {author} {\bibfnamefont {D.}~\bibnamefont
  {Makogon}}, \bibinfo {author} {\bibfnamefont {R.}~\bibnamefont {van
  Gelderen}}, \bibinfo {author} {\bibfnamefont {R.}~\bibnamefont {Rold\'an}}, \
  and\ \bibinfo {author} {\bibfnamefont {C.~M.}\ \bibnamefont {Smith}},\ }\href
  {\doibase 10.1103/PhysRevB.84.125404} {\bibfield  {journal} {\bibinfo
  {journal} {Phys. Rev. B}\ }\textbf {\bibinfo {volume} {84}},\ \bibinfo
  {pages} {125404} (\bibinfo {year} {2011})}\BibitemShut {NoStop}%
\bibitem [{\citenamefont {Nandkishore}\ \emph {et~al.}(2012)\citenamefont
  {Nandkishore}, \citenamefont {Levitov},\ and\ \citenamefont
  {Chubukov}}]{VHS_Nandkishore_2012}%
  \BibitemOpen
  \bibfield  {author} {\bibinfo {author} {\bibfnamefont {R.}~\bibnamefont
  {Nandkishore}}, \bibinfo {author} {\bibfnamefont {L.~S.}\ \bibnamefont
  {Levitov}}, \ and\ \bibinfo {author} {\bibfnamefont {A.~V.}\ \bibnamefont
  {Chubukov}},\ }\href {\doibase 10.1038/nphys2208} {\bibfield  {journal}
  {\bibinfo  {journal} {Nature Physics}\ }\textbf {\bibinfo {volume} {8}},\
  \bibinfo {pages} {158} (\bibinfo {year} {2012})}\BibitemShut {NoStop}%
\bibitem [{\citenamefont {Li}(2012)}]{VHS_Li_2012}%
  \BibitemOpen
  \bibfield  {author} {\bibinfo {author} {\bibfnamefont {T.}~\bibnamefont
  {Li}},\ }\href {\doibase 10.1209/0295-5075/97/37001} {\bibfield  {journal}
  {\bibinfo  {journal} {{EPL} (Europhysics Letters)}\ }\textbf {\bibinfo
  {volume} {97}},\ \bibinfo {pages} {37001} (\bibinfo {year}
  {2012})}\BibitemShut {NoStop}%
\bibitem [{\citenamefont {Yuan}\ \emph {et~al.}(2019)\citenamefont {Yuan},
  \citenamefont {Isobe},\ and\ \citenamefont {Fu}}]{LiangFu_VHS}%
  \BibitemOpen
  \bibfield  {author} {\bibinfo {author} {\bibfnamefont {N.~F.~Q.}\
  \bibnamefont {Yuan}}, \bibinfo {author} {\bibfnamefont {H.}~\bibnamefont
  {Isobe}}, \ and\ \bibinfo {author} {\bibfnamefont {L.}~\bibnamefont {Fu}},\
  }\href {\doibase 10.1038/s41467-019-13670-9} {\bibfield  {journal} {\bibinfo
  {journal} {Nature Communications}\ }\textbf {\bibinfo {volume} {10}},\
  \bibinfo {pages} {5769} (\bibinfo {year} {2019})}\BibitemShut {NoStop}%
\bibitem [{\citenamefont {McCann}\ and\ \citenamefont
  {Koshino}(2013)}]{BLG_spectrum_Koshino}%
  \BibitemOpen
  \bibfield  {author} {\bibinfo {author} {\bibfnamefont {E.}~\bibnamefont
  {McCann}}\ and\ \bibinfo {author} {\bibfnamefont {M.}~\bibnamefont
  {Koshino}},\ }\href {\doibase 10.1088/0034-4885/76/5/056503} {\bibfield
  {journal} {\bibinfo  {journal} {Reports on Progress in Physics}\ }\textbf
  {\bibinfo {volume} {76}},\ \bibinfo {pages} {056503} (\bibinfo {year}
  {2013})}\BibitemShut {NoStop}%
\bibitem [{\citenamefont {Iorsh}\ \emph {et~al.}(2017)\citenamefont {Iorsh},
  \citenamefont {Dini}, \citenamefont {Kibis},\ and\ \citenamefont
  {Shelykh}}]{BLG_t3_Iorsh}%
  \BibitemOpen
  \bibfield  {author} {\bibinfo {author} {\bibfnamefont {I.~V.}\ \bibnamefont
  {Iorsh}}, \bibinfo {author} {\bibfnamefont {K.}~\bibnamefont {Dini}},
  \bibinfo {author} {\bibfnamefont {O.~V.}\ \bibnamefont {Kibis}}, \ and\
  \bibinfo {author} {\bibfnamefont {I.~A.}\ \bibnamefont {Shelykh}},\ }\href
  {\doibase 10.1103/PhysRevB.96.155432} {\bibfield  {journal} {\bibinfo
  {journal} {Phys. Rev. B}\ }\textbf {\bibinfo {volume} {96}},\ \bibinfo
  {pages} {155432} (\bibinfo {year} {2017})}\BibitemShut {NoStop}%
\bibitem [{\citenamefont {Park}(2012)}]{BLG_t3_Park}%
  \BibitemOpen
  \bibfield  {author} {\bibinfo {author} {\bibfnamefont {C.-S.}\ \bibnamefont
  {Park}},\ }\href {\doibase https://doi.org/10.1016/j.ssc.2012.08.014}
  {\bibfield  {journal} {\bibinfo  {journal} {Solid State Communications}\
  }\textbf {\bibinfo {volume} {152}},\ \bibinfo {pages} {2018 } (\bibinfo
  {year} {2012})}\BibitemShut {NoStop}%
\bibitem [{\citenamefont {Gr\"uneis}\ \emph {et~al.}(2008)\citenamefont
  {Gr\"uneis}, \citenamefont {Attaccalite}, \citenamefont {Wirtz},
  \citenamefont {Shiozawa}, \citenamefont {Saito}, \citenamefont {Pichler},\
  and\ \citenamefont {Rubio}}]{BLG_t3_Gruneis}%
  \BibitemOpen
  \bibfield  {author} {\bibinfo {author} {\bibfnamefont {A.}~\bibnamefont
  {Gr\"uneis}}, \bibinfo {author} {\bibfnamefont {C.}~\bibnamefont
  {Attaccalite}}, \bibinfo {author} {\bibfnamefont {L.}~\bibnamefont {Wirtz}},
  \bibinfo {author} {\bibfnamefont {H.}~\bibnamefont {Shiozawa}}, \bibinfo
  {author} {\bibfnamefont {R.}~\bibnamefont {Saito}}, \bibinfo {author}
  {\bibfnamefont {T.}~\bibnamefont {Pichler}}, \ and\ \bibinfo {author}
  {\bibfnamefont {A.}~\bibnamefont {Rubio}},\ }\href {\doibase
  10.1103/PhysRevB.78.205425} {\bibfield  {journal} {\bibinfo  {journal} {Phys.
  Rev. B}\ }\textbf {\bibinfo {volume} {78}},\ \bibinfo {pages} {205425}
  (\bibinfo {year} {2008})}\BibitemShut {NoStop}%
\bibitem [{\citenamefont {Jolie}\ \emph {et~al.}(2018)\citenamefont {Jolie},
  \citenamefont {Lux}, \citenamefont {P\"ortner}, \citenamefont {Dombrowski},
  \citenamefont {Herbig}, \citenamefont {Knispel}, \citenamefont {Simon},
  \citenamefont {Michely}, \citenamefont {Rosch},\ and\ \citenamefont
  {Busse}}]{BLG_t3_Jolie}%
  \BibitemOpen
  \bibfield  {author} {\bibinfo {author} {\bibfnamefont {W.}~\bibnamefont
  {Jolie}}, \bibinfo {author} {\bibfnamefont {J.}~\bibnamefont {Lux}}, \bibinfo
  {author} {\bibfnamefont {M.}~\bibnamefont {P\"ortner}}, \bibinfo {author}
  {\bibfnamefont {D.}~\bibnamefont {Dombrowski}}, \bibinfo {author}
  {\bibfnamefont {C.}~\bibnamefont {Herbig}}, \bibinfo {author} {\bibfnamefont
  {T.}~\bibnamefont {Knispel}}, \bibinfo {author} {\bibfnamefont
  {S.}~\bibnamefont {Simon}}, \bibinfo {author} {\bibfnamefont
  {T.}~\bibnamefont {Michely}}, \bibinfo {author} {\bibfnamefont
  {A.}~\bibnamefont {Rosch}}, \ and\ \bibinfo {author} {\bibfnamefont
  {C.}~\bibnamefont {Busse}},\ }\href {\doibase 10.1103/PhysRevLett.120.106801}
  {\bibfield  {journal} {\bibinfo  {journal} {Phys. Rev. Lett.}\ }\textbf
  {\bibinfo {volume} {120}},\ \bibinfo {pages} {106801} (\bibinfo {year}
  {2018})}\BibitemShut {NoStop}%
\bibitem [{\citenamefont {McCann}\ and\ \citenamefont
  {Fal'ko}(2006)}]{BLG_t3_McCann}%
  \BibitemOpen
  \bibfield  {author} {\bibinfo {author} {\bibfnamefont {E.}~\bibnamefont
  {McCann}}\ and\ \bibinfo {author} {\bibfnamefont {V.~I.}\ \bibnamefont
  {Fal'ko}},\ }\href {\doibase 10.1103/PhysRevLett.96.086805} {\bibfield
  {journal} {\bibinfo  {journal} {Phys. Rev. Lett.}\ }\textbf {\bibinfo
  {volume} {96}},\ \bibinfo {pages} {086805} (\bibinfo {year}
  {2006})}\BibitemShut {NoStop}%
\bibitem [{\citenamefont {Cserti}\ \emph {et~al.}(2007)\citenamefont {Cserti},
  \citenamefont {Csord\'as},\ and\ \citenamefont {D\'avid}}]{BLG_t3_Cserti}%
  \BibitemOpen
  \bibfield  {author} {\bibinfo {author} {\bibfnamefont {J.}~\bibnamefont
  {Cserti}}, \bibinfo {author} {\bibfnamefont {A.}~\bibnamefont {Csord\'as}}, \
  and\ \bibinfo {author} {\bibfnamefont {G.}~\bibnamefont {D\'avid}},\ }\href
  {\doibase 10.1103/PhysRevLett.99.066802} {\bibfield  {journal} {\bibinfo
  {journal} {Phys. Rev. Lett.}\ }\textbf {\bibinfo {volume} {99}},\ \bibinfo
  {pages} {066802} (\bibinfo {year} {2007})}\BibitemShut {NoStop}%
\bibitem [{\citenamefont {Jung}\ \emph {et~al.}(2011)\citenamefont {Jung},
  \citenamefont {Zhang},\ and\ \citenamefont {MacDonald}}]{BLG_t3_Jung_2011}%
  \BibitemOpen
  \bibfield  {author} {\bibinfo {author} {\bibfnamefont {J.}~\bibnamefont
  {Jung}}, \bibinfo {author} {\bibfnamefont {F.}~\bibnamefont {Zhang}}, \ and\
  \bibinfo {author} {\bibfnamefont {A.~H.}\ \bibnamefont {MacDonald}},\ }\href
  {\doibase 10.1103/PhysRevB.83.115408} {\bibfield  {journal} {\bibinfo
  {journal} {Phys. Rev. B}\ }\textbf {\bibinfo {volume} {83}},\ \bibinfo
  {pages} {115408} (\bibinfo {year} {2011})}\BibitemShut {NoStop}%
\bibitem [{\citenamefont {Nilsson}\ \emph {et~al.}(2008)\citenamefont
  {Nilsson}, \citenamefont {Castro~Neto}, \citenamefont {Guinea},\ and\
  \citenamefont {Peres}}]{BLG_t3_Nilsson}%
  \BibitemOpen
  \bibfield  {author} {\bibinfo {author} {\bibfnamefont {J.}~\bibnamefont
  {Nilsson}}, \bibinfo {author} {\bibfnamefont {A.~H.}\ \bibnamefont
  {Castro~Neto}}, \bibinfo {author} {\bibfnamefont {F.}~\bibnamefont {Guinea}},
  \ and\ \bibinfo {author} {\bibfnamefont {N.~M.~R.}\ \bibnamefont {Peres}},\
  }\href {\doibase 10.1103/PhysRevB.78.045405} {\bibfield  {journal} {\bibinfo
  {journal} {Phys. Rev. B}\ }\textbf {\bibinfo {volume} {78}},\ \bibinfo
  {pages} {045405} (\bibinfo {year} {2008})}\BibitemShut {NoStop}%
\bibitem [{\citenamefont {Castro}\ \emph {et~al.}(2010)\citenamefont {Castro},
  \citenamefont {Novoselov}, \citenamefont {Morozov}, \citenamefont {Peres},
  \citenamefont {dos Santos}, \citenamefont {Nilsson}, \citenamefont {Guinea},
  \citenamefont {Geim},\ and\ \citenamefont {Neto}}]{BLG_t3_Castro_2010}%
  \BibitemOpen
  \bibfield  {author} {\bibinfo {author} {\bibfnamefont {E.~V.}\ \bibnamefont
  {Castro}}, \bibinfo {author} {\bibfnamefont {K.~S.}\ \bibnamefont
  {Novoselov}}, \bibinfo {author} {\bibfnamefont {S.~V.}\ \bibnamefont
  {Morozov}}, \bibinfo {author} {\bibfnamefont {N.~M.~R.}\ \bibnamefont
  {Peres}}, \bibinfo {author} {\bibfnamefont {J.~M. B.~L.}\ \bibnamefont {dos
  Santos}}, \bibinfo {author} {\bibfnamefont {J.}~\bibnamefont {Nilsson}},
  \bibinfo {author} {\bibfnamefont {F.}~\bibnamefont {Guinea}}, \bibinfo
  {author} {\bibfnamefont {A.~K.}\ \bibnamefont {Geim}}, \ and\ \bibinfo
  {author} {\bibfnamefont {A.~H.~C.}\ \bibnamefont {Neto}},\ }\href {\doibase
  10.1088/0953-8984/22/17/175503} {\bibfield  {journal} {\bibinfo  {journal}
  {Journal of Physics: Condensed Matter}\ }\textbf {\bibinfo {volume} {22}},\
  \bibinfo {pages} {175503} (\bibinfo {year} {2010})}\BibitemShut {NoStop}%
\bibitem [{\citenamefont {Jung}\ and\ \citenamefont
  {MacDonald}(2014)}]{BLG_sign_Jeil}%
  \BibitemOpen
  \bibfield  {author} {\bibinfo {author} {\bibfnamefont {J.}~\bibnamefont
  {Jung}}\ and\ \bibinfo {author} {\bibfnamefont {A.~H.}\ \bibnamefont
  {MacDonald}},\ }\href {\doibase 10.1103/PhysRevB.89.035405} {\bibfield
  {journal} {\bibinfo  {journal} {Phys. Rev. B}\ }\textbf {\bibinfo {volume}
  {89}},\ \bibinfo {pages} {035405} (\bibinfo {year} {2014})}\BibitemShut
  {NoStop}%
\bibitem [{\citenamefont {Joucken}\ \emph {et~al.}(2020)\citenamefont
  {Joucken}, \citenamefont {Ge}, \citenamefont {Quezada-L\'opez}, \citenamefont
  {Davenport}, \citenamefont {Watanabe}, \citenamefont {Taniguchi},\ and\
  \citenamefont {Velasco}}]{BLG_STM_Joucken}%
  \BibitemOpen
  \bibfield  {author} {\bibinfo {author} {\bibfnamefont {F.}~\bibnamefont
  {Joucken}}, \bibinfo {author} {\bibfnamefont {Z.}~\bibnamefont {Ge}},
  \bibinfo {author} {\bibfnamefont {E.~A.}\ \bibnamefont {Quezada-L\'opez}},
  \bibinfo {author} {\bibfnamefont {J.~L.}\ \bibnamefont {Davenport}}, \bibinfo
  {author} {\bibfnamefont {K.}~\bibnamefont {Watanabe}}, \bibinfo {author}
  {\bibfnamefont {T.}~\bibnamefont {Taniguchi}}, \ and\ \bibinfo {author}
  {\bibfnamefont {J.}~\bibnamefont {Velasco}},\ }\href {\doibase
  10.1103/PhysRevB.101.161103} {\bibfield  {journal} {\bibinfo  {journal}
  {Phys. Rev. B}\ }\textbf {\bibinfo {volume} {101}},\ \bibinfo {pages}
  {161103(R)} (\bibinfo {year} {2020})}\BibitemShut {NoStop}%
\end{thebibliography}%
\end{document}